\PassOptionsToPackage{unicode}{hyperref}
\PassOptionsToPackage{hyphens}{url}
\PassOptionsToPackage{dvipsnames,svgnames,x11names}{xcolor}
\documentclass[
  11pt,
  letterpaper,
  DIV=11,
  numbers=noendperiod]{scrartcl}

\usepackage{amsmath,amssymb}
\usepackage{iftex}
\ifPDFTeX
  \usepackage[T1]{fontenc}
  \usepackage[utf8]{inputenc}
  \usepackage{textcomp} 
\else 
  \usepackage{unicode-math}
  \defaultfontfeatures{Scale=MatchLowercase}
  \defaultfontfeatures[\rmfamily]{Ligatures=TeX,Scale=1}
\fi
\usepackage{lmodern}
\ifPDFTeX\else  
\fi
\IfFileExists{upquote.sty}{\usepackage{upquote}}{}
\IfFileExists{microtype.sty}{
  \usepackage[]{microtype}
  \UseMicrotypeSet[protrusion]{basicmath} 
}{}
\makeatletter
\@ifundefined{KOMAClassName}{
  \IfFileExists{parskip.sty}{%
    \usepackage{parskip}
  }{
    \setlength{\parindent}{0pt}
    \setlength{\parskip}{6pt plus 2pt minus 1pt}}
}{
  \KOMAoptions{parskip=half}}
\makeatother
\usepackage{xcolor}
\makeatletter
\ifx\paragraph\undefined\else
  \let\oldparagraph\paragraph
  \renewcommand{\paragraph}{
    \@ifstar
      \xxxParagraphStar
      \xxxParagraphNoStar
  }
  \newcommand{\xxxParagraphStar}[1]{\oldparagraph*{#1}\mbox{}}
  \newcommand{\xxxParagraphNoStar}[1]{\oldparagraph{#1}\mbox{}}
\fi
\ifx\subparagraph\undefined\else
  \let\oldsubparagraph\subparagraph
  \renewcommand{\subparagraph}{
    \@ifstar
      \xxxSubParagraphStar
      \xxxSubParagraphNoStar
  }
  \newcommand{\xxxSubParagraphStar}[1]{\oldsubparagraph*{#1}\mbox{}}
  \newcommand{\xxxSubParagraphNoStar}[1]{\oldsubparagraph{#1}\mbox{}}
\fi
\makeatother

\providecommand{\tightlist}{%
  \setlength{\itemsep}{0pt}\setlength{\parskip}{0pt}}\usepackage{longtable,booktabs,array}
\usepackage{calc} 
\usepackage{etoolbox}
\makeatletter
\patchcmd\longtable{\par}{\if@noskipsec\mbox{}\fi\par}{}{}
\makeatother
\IfFileExists{footnotehyper.sty}{\usepackage{footnotehyper}}{\usepackage{footnote}}
\makesavenoteenv{longtable}
\usepackage{graphicx}
\makeatletter
\newsavebox\pandoc@box
\newcommand*\pandocbounded[1]{
  \sbox\pandoc@box{#1}%
  \Gscale@div\@tempa{\textheight}{\dimexpr\ht\pandoc@box+\dp\pandoc@box\relax}%
  \Gscale@div\@tempb{\linewidth}{\wd\pandoc@box}%
  \ifdim\@tempb\p@<\@tempa\p@\let\@tempa\@tempb\fi
  \ifdim\@tempa\p@<\p@\scalebox{\@tempa}{\usebox\pandoc@box}%
  \else\usebox{\pandoc@box}%
  \fi%
}
\def\fps@figure{htbp}
\makeatother
\NewDocumentCommand\citeproctext{}{}

\makeatletter
 \let\@cite@ofmt\@firstofone
 \def\@biblabel#1{}
 \def\@cite#1#2{{#1\if@tempswa , #2\fi}}
\makeatother
\newlength{\cslhangindent}
\newlength{\csllabelwidth}
\newenvironment{CSLReferences}[2] 
 {\begin{list}{}{%
  \setlength{\itemindent}{0pt}
  \setlength{\leftmargin}{0pt}
  \setlength{\parsep}{0pt}
  \ifodd #1
   \setlength{\leftmargin}{\cslhangindent}
   \setlength{\itemindent}{-1\cslhangindent}
  \fi
  \setlength{\itemsep}{#2\baselineskip}}}
 {\end{list}}
\usepackage{calc}

\usepackage{enumitem}
\setlist[itemize]{label = $\bullet$}

\usepackage[ruled,vlined]{algorithm2e}

\AtBeginDocument{
  \setlength{\abovedisplayskip}{-0.5cm}
  \setlength{\belowdisplayskip}{0cm}
  \setlength{\abovedisplayshortskip}{-0.35cm}
  \setlength{\belowdisplayshortskip}{0.2cm}
}
\usepackage{booktabs}
\usepackage{longtable}
\usepackage{array}
\usepackage{multirow}
\usepackage{wrapfig}
\usepackage{float}
\usepackage{colortbl}
\usepackage{pdflscape}
\usepackage{tabu}
\usepackage{threeparttable}
\usepackage{threeparttablex}
\usepackage[normalem]{ulem}
\usepackage{makecell}
\usepackage{xcolor}
\KOMAoption{captions}{tableheading}
\makeatletter
\@ifpackageloaded{caption}{}{\usepackage{caption}}
\AtBeginDocument{%
\ifdefined\contentsname
  \renewcommand*\contentsname{Table of contents}
\else
  \newcommand\contentsname{Table of contents}
\fi
\ifdefined\listfigurename
  \renewcommand*\listfigurename{List of Figures}
\else
  \newcommand\listfigurename{List of Figures}
\fi
\ifdefined\listtablename
  \renewcommand*\listtablename{List of Tables}
\else
  \newcommand\listtablename{List of Tables}
\fi
\ifdefined\figurename
  \renewcommand*\figurename{Figure}
\else
  \newcommand\figurename{Figure}
\fi
\ifdefined\tablename
  \renewcommand*\tablename{Table}
\else
  \newcommand\tablename{Table}
\fi
}
\@ifpackageloaded{float}{}{\usepackage{float}}
\floatstyle{ruled}
\@ifundefined{c@chapter}{\newfloat{codelisting}{h}{lop}}{\newfloat{codelisting}{h}{lop}[chapter]}
\floatname{codelisting}{Listing}

\makeatother
\makeatletter
\@ifpackageloaded{caption}{}{\usepackage{caption}}
\@ifpackageloaded{subcaption}{}{\usepackage{subcaption}}
\makeatother

\ifLuaTeX
\usepackage[bidi=basic]{babel}
\else
\usepackage[bidi=default]{babel}
\fi
\babelprovide[main,import]{english}

\def\languageshorthands#1{}
\ifLuaTeX
  \usepackage[english]{selnolig} 
\fi
\usepackage{bookmark}

\IfFileExists{xurl.sty}{\usepackage{xurl}}{} 
\hypersetup{
  pdftitle={Whittaker-Henderson Smoothing Revisited: A Modern Statistical Framework for Practical Use},
  pdfauthor={Guillaume Biessy, PhD, LinkPact and Sorbonne Université},
  pdflang={en},
  colorlinks=true,
  linkcolor={blue},
  filecolor={Maroon},
  citecolor={Blue},
  urlcolor={Blue},
  pdfcreator={LaTeX via pandoc}}

\title{Whittaker-Henderson Smoothing Revisited: A Modern Statistical
Framework for Practical Use}
\author{Guillaume Biessy\footnote{guillaume.biessy78@gmail.com}, PhD,
LinkPact\footnote{LinkPact, 75015 Paris, France} and Sorbonne
Université\footnote{Sorbonne Université, CNRS, Laboratoire de
  Probabilités, Statistique et Modélisation, LPSM, 75005 Paris, France}}
\date{September 3, 2025}

\begin{document}
\maketitle
\begin{abstract}
Introduced over a century ago, Whittaker-Henderson smoothing remains
widely used by actuaries in constructing one-dimensional and
two-dimensional experience tables for mortality, disability and other
life insurance risks. In this paper, we reinterpret this smoothing
technique within a modern statistical framework and address six
practically relevant questions about its use.

First, we adopt a Bayesian perspective on this method to construct
credible intervals. Second, in the context of survival analysis, we
clarify how to choose the observation and weight vectors by linking the
smoothing technique to a maximum likelihood estimator. Third, we improve
accuracy by relaxing the method's reliance on an implicit normal
approximation. Fourth, we select the smoothing parameters by maximizing
a marginal likelihood function. Fifth, we improve computational
efficiency when dealing with numerous observation points and
consequently parameters. Finally, we develop an extrapolation procedure
that ensures consistency between estimated and predicted values through
constraints.
\end{abstract}

\renewcommand*\contentsname{Table of contents}
{
\hypersetup{linkcolor=}
\setcounter{tocdepth}{2}
\tableofcontents
}

\section*{Notations}\label{notations}

In this paper, vectors are denoted in boldface and matrix names in
uppercase letters. If \(\mathbf{y}\) is a vector and \(A\) is a matrix,
\(\text{Var}(\mathbf{y})\) denotes the variance-covariance matrix
associated with \(\mathbf{y}\), \(\textbf{diag}(A)\) represents the
diagonal of matrix \(A\), and \(\text{Diag}(\mathbf{y})\) is the
diagonal matrix such that
\(\textbf{diag}(\text{Diag}(\mathbf{y})) = \mathbf{y}\). The sum of the
diagonal elements of \(A\) is denoted as \(\text{tr}(A)\) and its
transpose as \(A^{T}\). In the case where \(A\) is invertible,
\(A^{- 1}\) denotes its inverse and \(|A|\) denotes the product of the
eigenvalues of \(A\). For a non-invertible matrix \(A\), \(A^{-}\)
refers to the Moore-Penrose pseudo-inverse of \(A\), and \(|A|_+\)
denotes the product of the non-zero eigenvalues of \(A\). By writing the
eigendecomposition as \(A = U\Sigma V^{T}\), where \(U\) and \(V\) are
orthogonal matrices and \(\Sigma\) is a diagonal matrix containing the
eigenvalues of \(A\), and by denoting \(\Sigma^-\) as the matrix
obtained by replacing the non-zero eigenvalues in \(\Sigma\) with their
inverses leaving the zero eigenvalues unchanged, the pseudo-inverse is
given by \(A^{-} = V\Sigma^{-}U^{T}\). The Kronecker product of two
matrices \(A\) and \(B\) is denoted as \(A \otimes B\), and their
Hadamard (element-wise) product, is denoted as \(A \odot B\).
\(\lfloor x\rfloor\) denotes the greatest integer less than or equal to
\(x\in\mathbb{R}\). Finally, the symbol \(\propto\) denotes
proportionality between the expressions on both sides.

\section{Introduction}\label{sec-intro}

Whittaker-Henderson (WH) smoothing is a graduation method designed to
mitigate the effects of sampling fluctuations in a vector of evenly
spaced discrete observations. Although this method was originally
proposed by Bohlmann (1899), it is named after Whittaker (1923), who
applied it to graduate mortality tables, and Henderson (1924), who
popularized it among actuaries in the United States. The method was
later extended to two dimensions by Knorr (1984). WH smoothing may be
used to build experience tables for a broad spectrum of life insurance
risks, such as mortality, disability, long-term care, lapse, mortgage
default and unemployment. We begin with a brief overview of the method
before outlining the structure and main contributions of the paper.

\subsection{A brief reminder of WH smoothing mathematical
formulation}\label{sec-brief}

\subsubsection{The one-dimensional case}\label{the-one-dimensional-case}

Let \(\mathbf{y}\) be a vector of observations and \(\mathbf{w}\) a
vector of positive weights, both of size \(n\). The estimator associated
with Whittaker-Henderson smoothing is given by:

\begin{equation}\phantomsection\label{eq-WH1}{
\hat{\mathbf{y}} = \underset{\boldsymbol{\theta}}{\text{argmin}}\{F(\mathbf{y},\mathbf{w},\boldsymbol{\theta}) + R_{\lambda,q}(\boldsymbol{\theta})\}
}\end{equation}

where:

\begin{itemize}
\item
  \(F(\mathbf{y},\mathbf{w},\boldsymbol{\theta}) = \underset{i = 1}{\overset{n}{\sum}} w_i(y_i - \theta_i)^2\)
  represents a fidelity criterion with respect to the observations,
\item
  \(R_{\lambda,q}(\boldsymbol{\theta}) = \lambda \underset{i = 1}{\overset{n - q}{\sum}} (\Delta^q\boldsymbol{\theta})_i^2\)
  represents a smoothness criterion.
\end{itemize}

In the latter expression, \(\lambda \ge 0\) is a smoothing parameter and
\(\Delta^q\) denotes the forward difference operator of order \(q\),
such that for any \(i\in\{1,\dots,n - q\}\):

\[
(\Delta^q\boldsymbol{\theta})_i = \underset{k = 0}{\overset{q}{\sum}} \begin{pmatrix}q \\ k\end{pmatrix}(- 1)^{q - k} \theta_{i + k}.
\]

Define \(W = \text{Diag}(\mathbf{w})\), the diagonal matrix of weights,
and \(D_{n,q}\) as the order \(q\) difference matrix of dimensions
\((n-q) \times n\), such that
\((D_{n,q}\boldsymbol{\theta})_i = (\Delta^q\boldsymbol{\theta})_i\) for
all \(i \in [1, n-q]\). The first- and second-order difference matrices
are given by:

\[
D_{n,1} = \begin{bmatrix}
-1 & 1 & 0 & \ldots & 0 \\
0 & -1 & 1 & \ddots & \vdots \\
\vdots & \ddots & \ddots & \ddots & 0 \\
0 & \ldots & 0 & -1 & 1 \\
\end{bmatrix}
\quad\text{and}\quad
D_{n,2} = \begin{bmatrix}
1 & -2 & 1 & 0 & \ldots & 0 \\
0 & 1 & -2 & 1 & \ddots & \vdots \\
\vdots & \ddots & \ddots & \ddots & \ddots & 0 \\
0 & \ldots & 0 & 1 & -2 & 1 \\
\end{bmatrix}.
\]

while higher-order difference matrices follow the recursive formula
\(D_{n,q} = D_{n - 1,q - 1}D_{n,1}\). The fidelity and smoothness
criteria can be rewritten with matrix notations as:

\[
F(\mathbf{y},\mathbf{w},\boldsymbol{\theta}) = (\mathbf{y} - \boldsymbol{\theta})^TW(\mathbf{y} - \boldsymbol{\theta}) \quad \text{and} \quad R_{\lambda,q}(\boldsymbol{\theta}) = \lambda\boldsymbol{\theta}^TD_{n,q}^TD_{n,q}\boldsymbol{\theta}
\]

and the WH smoothing estimator thus becomes:

\begin{equation}\phantomsection\label{eq-WH2}{
\hat{\mathbf{y}} = \underset{\boldsymbol{\theta}}{\text{argmin}} \left\lbrace(\mathbf{y} - \boldsymbol{\theta})^TW(\mathbf{y} - \boldsymbol{\theta}) + \boldsymbol{\theta}^TP_{\lambda}\boldsymbol{\theta}\right\rbrace
}\end{equation}

where \(P_{\lambda} = \lambda D_{n,q}^TD_{n,q}\).

\subsubsection{The two-dimensional case}\label{the-two-dimensional-case}

In the two-dimensional case, consider a matrix \(Y\) of observations and
a matrix \(\Omega\) of non-negative weights, both of dimensions
\(n_x \times n_z\). The WH smoothing estimator solves:

\[
\widehat{Y} = \underset{\Theta}{\text{argmin}}\{F(Y,\Omega, \Theta) + R_{\lambda,q}(\Theta)\}
\]

where:

\begin{itemize}
\item
  \(F(Y,\Omega, \Theta) = \sum_{i = 1}^{n_x}\sum_{j = 1}^{n_z} \Omega_{i,j}(Y_{i,j} - \Theta_{i,j})^2\)
  represents a fidelity criterion with respect to the observations,
\item
  \(R_{\lambda,q}(\Theta) = \lambda_x \sum_{j = 1}^{n_z}\sum_{i = 1}^{n_x - q_x} (\Delta^{q_x}\Theta_{\bullet,j})_i^2 + \lambda_z \sum_{i = 1}^{n_x}\sum_{j = 1}^{n_z - q_z} (\Delta^{q_z}\Theta_{i,\bullet})_j^2\)
  is a smoothness criterion with \(\lambda = (\lambda_x,\lambda_z)\).
\end{itemize}

This latter criterion adds row-wise and column-wise regularization
criteria to \(\Theta\), with respective orders \(q_x\) and \(q_z\),
weighted by non-negative smoothing parameters \(\lambda_x\) and
\(\lambda_z\). In matrix notation, let \(\mathbf{y} = \textbf{vec}(Y)\),
\(\mathbf{w} = \textbf{vec}(\Omega)\), and
\(\boldsymbol{\theta} = \textbf{vec}(\Theta)\) as the vectors obtained
by stacking the columns of the matrices \(Y\), \(\Omega\), and
\(\Theta\), respectively. Additionally, denote
\(W = \text{Diag}(\mathbf{w})\) and \(n = n_x \times n_z\). The fidelity
and smoothness criteria become:

\[
\begin{aligned}
F(\mathbf{y},\mathbf{w}, \boldsymbol{\theta}) &= (\mathbf{y} - \boldsymbol{\theta})^TW(\mathbf{y} - \boldsymbol{\theta}) \\
R_{\lambda,q}(\boldsymbol{\theta}) &= \boldsymbol{\theta}^{T}(\lambda_x I_{n_z} \otimes D_{n_x,q_x}^{T}D_{n_x,q_x} + \lambda_z D_{n_z,q_z}^{T}D_{n_z,q_z} \otimes I_{n_x}) \boldsymbol{\theta}
\end{aligned}
\]

and the associated estimator also takes the form of
Equation~\ref{eq-WH2} except in this case

\[P_{\lambda} = \lambda_x I_{n_z} \otimes D_{n_x,q_x}^{T}D_{n_x,q_x} + \lambda_z D_{n_z,q_z}^{T}D_{n_z,q_z} \otimes I_{n_x}.\]

Extension to higher dimensions is straightforward and not discussed
here.

\subsubsection{An explicit solution}\label{sec-explicit}

If \(W + P_{\lambda}\) is invertible, Equation~\ref{eq-WH2} admits the
closed-form solution:

\begin{equation}\phantomsection\label{eq-WH3}{\hat{\mathbf{y}} = (W + P_{\lambda})^{-1}W\mathbf{y}.}\end{equation}

Indeed, as a minimum, \(\hat{\mathbf{y}}\) satisfies:

\[0 = \left.\frac{\partial}{\partial \boldsymbol{\theta}}\right|_{\hat{\mathbf{y}}}\left\lbrace(\mathbf{y} - \boldsymbol{\theta})^{T}W(\mathbf{y} - \boldsymbol{\theta}) + \boldsymbol{\theta}^{T}P_{\lambda}\boldsymbol{\theta}\right\rbrace = - 2 W(y - \hat{\mathbf{y}}) +2P_{\lambda} \hat{\mathbf{y}}.\]

It follows that \((W + P_{\lambda})\hat{\mathbf{y}} = W\mathbf{y}\),
proving Equation~\ref{eq-WH3}. If \(\lambda \neq 0\),
\(W + P_{\lambda}\) is invertible as long as \(\mathbf{w}\) has \(q\)
non-zero elements in the one-dimensional case, and \(\Omega\) has at
least \(q_x \times q_z\) non-zero elements spread across \(q_x\)
different rows and \(q_z\) different columns in the two-dimensional
case. These conditions are always met in real datasets.

\subsection{Structure of the paper}\label{structure-of-the-paper}

Introduced a century ago, Whittaker-Henderson (WH) smoothing remains
widely used by actuaries, particularly in France and North America
(Canadian Institute of Actuaries 2017; Society of Actuaries 2018). Other
non-parametric smoothing methods have since emerged, notably
spline-based techniques (Reinsch 1967), which gained even greater
popularity with P-splines (Eilers and Marx 1996). A broader overview of
alternative smoothers is available in Wood (2017, chap. 5).

For evenly spaced discrete observations, WH smoothing may be considered
a particular case of P-splines with degree-zero splines and identity
model matrix. Its appeal lies in its simplicity: no selection of knots,
parameters equal to fitted values, and shape controlled solely via
penalization. However, it involves more parameters than low-rank
smoothers, making it more computationally intensive.

Originally proposed as an empirical alternative to polynomial regression
and weighted averages, WH smoothing offered key benefits noted by
Whittaker (1923): first \(q\) moment preservation, adjustable smoothing
parameters, and robustness at boundaries. While smoothing theory has
evolved---particularly via generalized additive models (Hastie and
Tibshirani 1990), use of WH smoothing by actuaries remains largely
unchanged. This paper reinterprets WH within modern statistical theory
to bridge that gap and address six practical questions, each discussed
in a dedicated section.

\subsubsection{How to measure uncertainty in smoothing
results?}\label{how-to-measure-uncertainty-in-smoothing-results}

We propose a method to quantify the uncertainty in WH smoothing based on
data volume, a topic that has received little attention in the
literature. In a Frequentist framework, the WH estimator is biased,
which complicates the construction of valid confidence intervals for
finite samples. However, under certain conditions, WH smoothing can be
viewed as a Bayesian model, enabling the derivation of credible
intervals. This Bayesian interpretation was originally suggested by
Whittaker (1923) as a justification for the method and formally
revisited decades later by Taylor (1992). In this section, we build on
that equivalence to derive credible intervals for WH smoothing.

\subsubsection{Which observation and weight vectors to
use?}\label{which-observation-and-weight-vectors-to-use}

For the Bayesian interpretation of WH smoothing discussed in
Section~\ref{sec-CI} to hold, it must be applied to a vector
\(\mathbf{y}\) of independent, normally distributed observations with
known variances. The weight vector \(\mathbf{w}\) should then contain
the inverse variances (up to a constant), as noted by Taylor (1992) and
Verrall (1993). We show that, under piecewise constant transition
intensities in duration models, the maximum likelihood estimator of
crude rates produces vectors \((\mathbf{y}, \mathbf{w})\) that
asymptotically meet these conditions. This, combined with the results
from the previous section, offers a statistical foundation for the use
of WH smoothing in constructing experience tables for life insurance
risks.

\subsubsection{How to improve the accuracy of smoothing with limited
data
volume?}\label{how-to-improve-the-accuracy-of-smoothing-with-limited-data-volume}

The standard approach applies WH smoothing to crude rate estimates,
assuming they are asymptotically normal. However, this assumption often
breaks down in practice when data are limited, making the method
unreliable in such cases. Following Verrall (1993), we propose a
generalization of WH smoothing that replaces the two-step procedure with
the direct maximization of a penalized log-likelihood. Instead of
smoothing pre-estimated rates, this method works directly with
aggregated event and exposure counts. The estimation is performed
iteratively using the PIRLS algorithm. We evaluate both methods on
simulated datasets reflecting typical life insurance portfolios. Results
show that, in smaller samples, the normal approximation in the
traditional method introduces notable bias. This supports the use of the
generalized approach---based on penalized log-likelihood---as a more
robust alternative when data are limited.

\subsubsection{How to select the smoothing
parameters?}\label{how-to-select-the-smoothing-parameters}

We now turn to the crucial choice of the smoothing parameter
\(\lambda\), which has long been left to actuarial judgment. Giesecke
and Center (1981) suggested choosing \(\lambda\) so that the variance of
the smoothed results matches the average variance of a Chi-square
statistic, but uses \(n - q\) as degrees of freedom, thus ignoring the
reduction in effective model dimension due to penalization. Brooks et
al. (1988) minimized the global cross-validation criterion introduced by
Wahba (1980), though this can result in severe under-smoothing as noted
by Wood (2011).

We instead propose to select \(\lambda\) by maximizing a marginal
likelihood function, a method first introduced by Patterson and Thompson
(1971) and later applied to smoothing parameter selection by Anderssen
and Bloomfield (1974). This approach is consistent with the Bayesian
framework discussed earlier and performs well in small samples, as shown
by Reiss and Todd Ogden (2009). This marginal likelihood function has a
closed-form expression and can be maximized numerically. For the
proposed generalization of WH smoothing, the marginal likelihood is no
longer available in closed form. Instead, we rely on the Laplace
Approximation of the Marginal Likelihood (LAML), which can be maximized
numerically. As both solving likelihood equations and selecting the
optimal smoothing parameter are iterative processes, we explore
different ways of nesting these iterations. We compare three nesting
strategies combined with three numerical optimization algorithms for
maximizing the marginal likelihood or LAML. Simulation results show that
all strategies have near-optimal accuracy, with the fastest performance
achieved using the outer iteration strategy combined with the Newton
algorithm

\subsubsection{How to improve smoothing computational
efficiency?}\label{how-to-improve-smoothing-computational-efficiency}

When the number of observations---and thus parameters---is large, the
computational cost of WH smoothing becomes a major challenge. This is
particularly relevant in actuarial contexts, such as smoothing
two-dimensional tables for disability or long-term care modelling.
Beyond actuarial applications, WH smoothing is also widely used in
economics for long time series, where it is known as the
Hodrick-Prescott filter (Hodrick and Prescott 1997). Although fast
algorithms have been developed to exploit the structure of the
penalization matrix (e.g., Weinert 2007; Cornea-Madeira 2017) they are
typically limited to the one-dimensional case and cannot be directly
extended to two dimensions.

After briefly outlining the main computational steps of (generalized) WH
smoothing-including smoothing parameter selection via marginal
likelihood or LAML-and their leading-order costs, we introduce two
complementary strategies to reduce the computational burden:

\begin{enumerate}
\def\labelenumi{\arabic{enumi}.}
\item
  Banded matrix exploitation: WH smoothing involves model and
  penalization matrices with banded structure. Taking advantage of this
  structure greatly accelerates key computations.
\item
  Reduced-rank basis via natural parametrization: Building on the work
  of Demmler and Reinsch (1975), we apply an eigendecomposition to the
  one-dimensional penalization matrices and drop components associated
  with the largest eigenvalues, which reduces the problem size. In two
  dimensions, we further improve efficiency using the Generalized Linear
  Array Model (GLAM) framework (Currie, Durban, and Eilers 2006) which
  leverages the rectangular shape of the data.
\end{enumerate}

In the two-dimensional case, we compare these strategies with a cubic
P-spline alternative using simulated datasets. Results show that the
banded implementation reduces computation time by up to a factor of 25.
The reduced-rank approach brings further gains---up to a factor of
250---at the cost of a slight reduction in accuracy. Its performance is
comparable to P-spline smoothing with a cubic basis of similar size.

\subsubsection{How to extrapolate smoothing
results?}\label{how-to-extrapolate-smoothing-results}

We conclude by addressing how to extrapolate smoothing results.
Semi-parametric models like WH and P-splines can extrapolate beyond the
observed data---similar to parametric models---but this feature is often
overlooked in actuarial practice. The existing literature is limited and
mostly focused on mortality forecasting.

Currie, Durban, and Eilers (2004) uses P-splines to fit and forecast
mortality rates by treating the extrapolated positions as zero-weight
observations (see also Delwarde, Denuit, and Eilers 2007; Currie 2013).
While this works well in one dimension, Carballo, Durban, and Lee (2021)
showed that it distorts the fit in two dimensions. To fix this, they
proposed adding constraints to preserve the values that would result
from fitting the observed data alone.

However, their approach to confidence intervals overlooks potential
innovation error beyond the observed data, effectively treating the
extrapolated process as perfectly smooth. In contrast, we propose an
approach that derives credible intervals for extrapolated values,
accounting for the underlying variability beyond the observed data
range.

\section{How to measure uncertainty in smoothing results?}\label{sec-CI}

The explicit solution given by Equation~\ref{eq-WH3} indicates that
\(\mathbb{E}(\hat{\mathbf{y}}) = (W + P_{\lambda})^{-1}W\mathbb{E}(\mathbf{y}) \ne \mathbb{E}(\mathbf{y})\)
when \(\lambda \ne 0\). This implies that penalization introduces a
smoothing bias, which prevents the construction of confidence intervals
for finite samples centred on \(\mathbb{E}(\mathbf{y})\). Therefore, in
this section, we turn to a Bayesian framework where smoothing can be
interpreted more naturally.

\subsection{Maximum a posteriori
estimate}\label{maximum-a-posteriori-estimate}

Suppose that
\(\mathbf{y} \mid \boldsymbol{\theta}\sim \mathcal{N}(\boldsymbol{\theta}, \sigma^2W^{-})\)
and \(\boldsymbol{\theta} \sim \mathcal{N}(0, \sigma^2P_{\lambda}^{-})\)
for some \(\sigma > 0\). The Bayes formula allows us to express the
posterior likelihood \(f(\boldsymbol{\theta} \mid \mathbf{y})\)
associated with these choices in the following form:

\[f(\boldsymbol{\theta} \mid \mathbf{y}) \propto f(\mathbf{y} \mid \boldsymbol{\theta}) f(\boldsymbol{\theta}) \propto \exp\left(- \frac{1}{2\sigma^2}\left[(\mathbf{y} - \boldsymbol{\theta})^{T}W(\mathbf{y} - \boldsymbol{\theta}) + \boldsymbol{\theta}^TP_{\lambda}\boldsymbol{\theta}\right]\right).\]

Hence the mode of the posterior distribution,
\(\hat{\boldsymbol{\theta}} = \text{argmax} [f(\boldsymbol{\theta} \mid \mathbf{y})]\),
also known as the \emph{maximum a posteriori} (MAP) estimate, coincides
with the solution \(\hat{\mathbf{y}}\) from Equation~\ref{eq-WH2}, whose
explicit form is given by Equation~\ref{eq-WH3}.

\subsection{\texorpdfstring{Posterior distribution of
\(\boldsymbol{\theta} \mid \mathbf{y}\)}{Posterior distribution of \textbackslash boldsymbol\{\textbackslash theta\} \textbackslash mid \textbackslash mathbf\{y\}}}\label{posterior-distribution-of-boldsymboltheta-mid-mathbfy}

A second-order Taylor expansion of the log-posterior likelihood around
\(\hat{\mathbf{y}} = \hat{\boldsymbol{\theta}}\) gives us:

\begin{equation}\phantomsection\label{eq-taylor1}{\ln f(\boldsymbol{\theta} \mid \mathbf{y}) = \ln f(\hat{\boldsymbol{\theta}} \mid \mathbf{y}) + \left.\frac{\partial \ln f(\boldsymbol{\theta} \mid \mathbf{y})}{\partial \boldsymbol{\theta}}\right|_{\boldsymbol{\theta} = \hat{\boldsymbol{\theta}}}^{T}(\boldsymbol{\theta} - \hat{\boldsymbol{\theta}}) + \frac{1}{2}(\boldsymbol{\theta} - \hat{\boldsymbol{\theta}})^{T} \left.\frac{\partial^2 \ln f(\boldsymbol{\theta} \mid \mathbf{y})}{\partial \boldsymbol{\theta} \partial \boldsymbol{\theta}^{T}}\right|_{\boldsymbol{\theta} = \hat{\boldsymbol{\theta}}}(\boldsymbol{\theta} - \hat{\boldsymbol{\theta}})}\end{equation}

\[\text{where} \quad \left.\frac{\partial \ln f(\boldsymbol{\theta} \mid \mathbf{y})}{\partial \boldsymbol{\theta}}\right|_{\boldsymbol{\theta} = \hat{\boldsymbol{\theta}}} = 0 \quad \text{and} \quad \left.\frac{\partial^2 \ln f(\boldsymbol{\theta} \mid \mathbf{y})}{\partial \boldsymbol{\theta} \partial \boldsymbol{\theta}^{T}}\right|_{\boldsymbol{\theta} = \hat{\boldsymbol{\theta}}} = - \frac{1}{\sigma^2}(W + P_{\lambda}).\]

As this last derivative no longer depends on \(\boldsymbol{\theta}\),
higher-order derivatives are all zero. The Taylor expansion allows for
an exact computation of \(\ln f(\boldsymbol{\theta} \mid \mathbf{y})\).
Substituting the result back into Equation~\ref{eq-taylor1} yields:

\[\begin{aligned}
f(\boldsymbol{\theta} \mid \mathbf{y}) &\propto \exp\left[\ln f(\hat{\boldsymbol{\theta}} \mid \mathbf{y}) - \frac{1}{2\sigma^2} (\boldsymbol{\theta} - \hat{\boldsymbol{\theta}})^{T}(W + P_{\lambda})(\boldsymbol{\theta} - \hat{\boldsymbol{\theta}})\right] \\
&\propto \exp\left[- \frac{1}{2\sigma^2} (\boldsymbol{\theta} - \hat{\boldsymbol{\theta}})^{T}(W + P_{\lambda})(\boldsymbol{\theta} - \hat{\boldsymbol{\theta}})\right]
\end{aligned}\]

which can immediately be recognized as the density of the
\(\mathcal{N}(\hat{\boldsymbol{\theta}},\sigma^2(W + P_{\lambda})^{- 1})\)
distribution.

\subsection{Consequence for the WH
smoothing}\label{consequence-for-the-wh-smoothing}

The prior
\(\boldsymbol{\theta} \sim \mathcal{N}(0, \sigma^2P_{\lambda}^{-})\)
provides a Bayesian interpretation of the smoothness penalty, expressing
an (improper) prior belief about the structure of \(\mathbf{y}\).

This Bayesian framework and the resulting credible intervals rely on the
assumption that
\(\mathbf{y} \mid \boldsymbol{\theta}\sim \mathcal{N}(\boldsymbol{\theta}, \sigma^2W^{-})\),
meaning that the components of \(\mathbf{y}\) are independent with known
variances (up to a constant \(\sigma^2\)). The weight vector
\(\mathbf{w}\) must then be proportional to the inverse variances, not
chosen empirically. If \(\sigma^2\) is known, \(100(1 - \alpha) \%\)
credible intervals take the form:

\begin{equation}\phantomsection\label{eq-CI}{\mathbb{E}(\mathbf{y}) \mid \mathbf{y} \in \left[\hat{\mathbf{y}} \pm \Phi^{- 1}\left(1 - \alpha / 2\right)\sqrt{\sigma^2\textbf{diag}\left\lbrace(W + P_{\lambda})^{-1}\right\rbrace}\right]}\end{equation}

where \(\hat{\mathbf{y}} = (W + P_{\lambda})^{- 1}W\mathbf{y}\) and
\(\Phi\) is the cumulative distribution function for the standard normal
distribution. According to Marra and Wood (2012), such intervals have
good Frequentist coverage.

If \(\sigma^2\) is unknown, it can be estimated as:

\[\hat{\sigma}^2 = \frac{(\mathbf{y} - \hat{\mathbf{y}})^TW(\mathbf{y} - \hat{\mathbf{y}})}{n - \text{tr}(H)}\quad \text{where}\quad H = (W + P_{\lambda})^{- 1}W.\]

In that case, \(\sigma^2\) is replaced by \(\hat{\sigma}^2\) and the
normal distribution in Equation~\ref{eq-CI} by the Student t -
distribution with \(n - \text{tr}(H)\) degrees of freedom.

\section{Which observation and weight vectors to
use?}\label{sec-vectors}

Section~\ref{sec-CI} highlighted that Whittaker-Henderson smoothing may
be interpreted in a robust statistical framework when applied to a
vector \(\mathbf{y}\) of independent, normally distributed observations
with known variances, and a weight vector \(\mathbf{w}\) proportional to
the inverses of those variances. In this section, we propose, within the
framework of duration models used for constructing experience tables for
life insurance risks, vectors \(\mathbf{y}\) and \(\mathbf{w}\) that
satisfy these conditions.

\subsection{Survival analysis
framework}\label{survival-analysis-framework}

We consider a longitudinal follow-up of \(m\) individuals, subject to
left truncation and non-informative right censoring, and aim to estimate
a distribution governed by a continuous explanatory variable \(x\)
(e.g., age). Let \(\mu\) denote the hazard function, also known as the
force of mortality in the study of the death risk. Under standard
survival analysis assumptions, the log-likelihood takes the following
continuous-time form:

\begin{equation}\phantomsection\label{eq-vraisemblance1}{\ell(\boldsymbol{\theta}) = \underset{i = 1}{\overset{m}{\sum}} \left[\delta_i \ln\mu(x_i + t_i,\boldsymbol{\theta}) - \underset{u = 0}{\overset{t_i}{\int}}\mu(x_i + u,\boldsymbol{\theta})\text{d}u\right].}\end{equation}

Here \(x_i\) is the age at the start of observation, \(t_i\) is the
follow-up duration for individual \(i\) and \(\delta_i\) is an event
indicator: 1 if the event is observed and 0 if censored.

Although model estimation can be based on direct maximization of
Equation~\ref{eq-vraisemblance1}, this approach scales poorly with large
\(m\) and generally requires numerical integration-except in simple
parametric cases. We instead adopt a discrete approximation by assuming
the hazard rate is piecewise constant over one-year intervals:

\[\mu(x + \epsilon) = \mu(x) \quad\text{for all}\quad x \in \mathbb{N}, \epsilon \in [0,1[.\]

Under this assumption, the log-likelihood simplifies to a sum over
discrete ages:

\begin{equation}\phantomsection\label{eq-vraisemblance2}{\ell(\boldsymbol{\theta}) = \underset{x = x_{\min}}{\overset{x_{\max}}{\sum}} \ln\mu(x,\boldsymbol{\theta}) d(x) - \mu(x,\boldsymbol{\theta}) e_c(x).}\end{equation}

Here \(d(x)\) is the number of observed events at age \(x\) and
\(e_c(x)\) is the central exposure to risk, i.e., the total duration
individuals are observed at age \(x\).

This discretization, first introduced by Hoem (1971), is widely used in
actuarial science. Its advantages are underlined for example in
Gschlössl, Schoenmaekers, and Denuit (2011). It extends naturally to the
two-dimensional case by assuming
\(\mu(x + \epsilon, z + \xi) = \mu(x, z)\) and summing over \((x, z)\)
pairs.

Details on the derivation of Equations \ref{eq-vraisemblance1} and
\ref{eq-vraisemblance2}, along with the computation of central exposures
and event counts, are provided in Section~\ref{sec-expo} of the
appendices.

\subsection{Likelihood equations}\label{sec-cons}

Assuming one parameter per observation and using the exponential link
\(\boldsymbol{\mu}(\boldsymbol{\theta}) = \boldsymbol{\exp}(\boldsymbol{\theta})\),
we recover the crude rates estimator, which models each age (or age
pair) independently. The exponential link ensures positive hazard rates.
The log-likelihood, in both one- and two-dimensional cases, takes the
vectorized form:

\begin{equation}\phantomsection\label{eq-vraisemblance3}{
\ell(\boldsymbol{\theta}) = \boldsymbol{\theta}^{T}\mathbf{d} - \boldsymbol{\exp}(\boldsymbol{\theta})^{T}\mathbf{e_c}
}\end{equation}

where \(\mathbf{d}\) and \(\mathbf{e}_c\) are the vectors of observed
deaths and central exposures.

The derivatives of this likelihood are:

\begin{equation}\phantomsection\label{eq-vraisemblance4}{\frac{\partial \ell}{\partial \boldsymbol{\theta}} = \mathbf{d} -\boldsymbol{\exp}(\boldsymbol{\theta}) \odot \mathbf{e_c} \quad \text{and} \quad \frac{\partial^2 \ell}{\partial\boldsymbol{\theta}\partial\boldsymbol{\theta}^T} = - \text{Diag}(\boldsymbol{\exp}(\boldsymbol{\theta}) \odot \mathbf{e_c}).}\end{equation}

These equations correspond to those of a Poisson GLM (Nelder and
Wedderburn 1972) with mean
\(\boldsymbol{\mu}(\boldsymbol{\theta}) \odot \mathbf{e}_c\), although
derived under different assumptions.

The model admits the closed-form solution
\(\hat{\boldsymbol{\theta}} = \ln(\mathbf{d} / \mathbf{e}_c)\). Under
standard regularity conditions, the maximum likelihood estimator
satisfies
\(\hat{\boldsymbol{\theta}} \sim \mathcal{N}(\boldsymbol{\theta}, W_{\hat{\boldsymbol{\theta}}}^{-1})\),
with \(W_{\hat{\boldsymbol{\theta}}} = \text{Diag}(\mathbf{d})\).

Notably, this asymptotic approximation depends on the number of
individuals \(m\) and not the dimension \(n\) of the aggregated vectors.

\subsection{Consequence for the WH
smoothing}\label{consequence-for-the-wh-smoothing-1}

We conclude that, under the duration model framework and using crude
rates, the log-estimate \(\ln(\mathbf{d} / \mathbf{e}_c)\) is
asymptotically normal:

\[\ln(\mathbf{d} / \mathbf{e}_c) \sim \mathcal{N}(\ln\boldsymbol{\mu}, W^{-1}) \quad\text{with}\quad W = \text{Diag}(\mathbf{d}).\]

This justifies applying Whittaker-Henderson smoothing to the observation
vector \(\mathbf{y} = \ln(\mathbf{d} / \mathbf{e}_c)\) with weight
vector \(\mathbf{w} = \mathbf{d}\). Using results from
Section~\ref{sec-CI}, and \(\sigma^2 = 1\), the credible intervals for
\(\ln\boldsymbol{\mu}\) are:

\[\ln\boldsymbol{\mu} \mid \mathbf{d}, \mathbf{e_c} \in \left[\hat{\boldsymbol{\theta}} \pm \Phi^{- 1}\left(1 -\alpha / 2\right)\sqrt{\textbf{diag}\left\lbrace(\text{Diag}(\mathbf{d}) + P_{\lambda})^{-1}\right\rbrace}\right]\]

with
\(\hat{\boldsymbol{\theta}} = (W + P_{\lambda})^{-1}W(\ln\mathbf{d} - \ln \mathbf{e}_c)\).
Credible intervals for \(\boldsymbol{\mu}\) itself are then obtained by
exponentiating the bounds.

\section{How to improve the accuracy of smoothing with limited data
volume?}\label{sec-generalisation}

\subsection{Generalized Whittaker-Henderson
smoothing}\label{generalized-whittaker-henderson-smoothing}

The approach described in Section~\ref{sec-cons} assumes that the crude
rates estimator is asymptotically normal, justifying the application of
WH smoothing to its logarithm. However, with limited data, this
approximation may introduce significant bias. We therefore propose an
alternative based directly on the exact likelihood in
Equation~\ref{eq-vraisemblance3}. Applying the Bayesian framework from
Section~\ref{sec-CI} and assuming
\(\boldsymbol{\theta} \sim \mathcal{N}(0, P_{\lambda}^{-})\), Bayes'
theorem gives:

\[f(\boldsymbol{\theta} \mid \mathbf{d}, \mathbf{e_c})
\propto f(\mathbf{d}, \mathbf{e_c} \mid \boldsymbol{\theta}) f(\boldsymbol{\theta})
\propto \exp\left[\ell(\boldsymbol{\theta}) - \frac{1}{2}\boldsymbol{\theta}^TP_{\lambda}\boldsymbol{\theta}\right].\]

We define the penalized log-likelihood as
\(\ell_P(\boldsymbol{\theta}) = \ell(\boldsymbol{\theta}) - \boldsymbol{\theta}^TP_{\lambda}\boldsymbol{\theta}/2\).
The maximum a posteriori estimate is the maximizer of \(\ell_P\).

Using a second-order Taylor expansion of the posterior log-likelihood
around \(\hat{\boldsymbol{\theta}}\) leads to the Laplace approximation:

\begin{equation}\phantomsection\label{eq-taylor2}{f(\boldsymbol{\theta} \mid \mathbf{d}, \mathbf{e_c}) \approx \mathcal{N}(\hat{\boldsymbol{\theta}},(W_{\hat{\boldsymbol{\theta}}} + P_{\lambda})^{- 1})}\end{equation}

where
\(W_{\hat{\boldsymbol{\theta}}} = \text{Diag}(\boldsymbol{\exp}(\hat{\boldsymbol{\theta}}) \odot \mathbf{e_c})\).
Unlike the normal case studied in Section~\ref{sec-CI}, the higher-order
derivatives of the posterior log-likelihood are not zero, and
Equation~\ref{eq-taylor2} only provides an approximation of the
posterior log-likelihood, which yields asymptotic credible intervals:

\[\ln \boldsymbol{\mu} \mid \mathbf{d}, \mathbf{e_c} \in \left[\hat{\boldsymbol{\theta}} \pm \Phi^{- 1}\left(1 -\alpha / 2\right)\sqrt{\textbf{diag}\left\lbrace(W_{\hat{\boldsymbol{\theta}}} + P_{\lambda})^{-1}\right\rbrace}\right].\]

Unlike the closed-form estimator in Equation~\ref{eq-vraisemblance4}, no
analytical solution for \(\hat{\boldsymbol{\theta}}\) exists here. We
solve numerically using Newton's algorithm, which iteratively updates:

\[\boldsymbol{\theta}_{k + 1} = 
\boldsymbol{\theta}_k + (W_k + P_{\lambda})^{- 1} (\mathbf{d} -\boldsymbol{\exp}(\boldsymbol{\theta}_k) \odot \mathbf{e_c} - P_{\lambda} \boldsymbol{\theta}_k)]\]

with
\(W_k = \text{Diag}(\boldsymbol{\exp}(\boldsymbol{\theta}_k) \odot \mathbf{e_c})\).
The update can be rewritten as:

\[\boldsymbol{\theta}_{k + 1} = 
(W_k + P_{\lambda})^{- 1}W_k\mathbf{z}_k\quad\text{where}\quad \mathbf{z}_k = \boldsymbol{\theta}_k + W_k^{-1}[\mathbf{d} -\boldsymbol{\exp}(\boldsymbol{\theta}_k) \odot \mathbf{e_c}].\]

Initializing with the crude rates estimator
\(\boldsymbol{\theta}_0 = \ln(\mathbf{d} / \mathbf{e_c})\) implies
\(W_0 = \text{Diag}(\mathbf{d})\) and
\(z_0 = \ln(\mathbf{d} / \mathbf{e_c})\), so the first iteration
recovers the classical WH smoothing result.

Subsequent iterations refine the observation and weight vectors. This
process can thus be interpreted as an iterative generalization of WH
smoothing, akin to how generalized linear models extend linear models.

We refer to this method as generalized Whittaker-Henderson smoothing.
The iterative estimation algorithm described above corresponds to the
Penalized Iteratively Reweighted Least Squares (PIRLS) algorithm, widely
used for fitting generalized additive models.

This framework naturally extends to other exponential family
distributions, such as the binomial case suggested in Verrall (1993), by
adapting the likelihood, link function, weight matrix, and working
vector. However, we advocate for the Poisson-like likelihood of
Equation~\ref{eq-vraisemblance3}, which offers several advantages: it
generalizes to competing risks, supports multiplicative covariate
effects via the log link, and allows the use of an external reference
table as a multiplicative offset.

\subsection{Impact of the normal approximation in the original
smoothing}\label{sec-impact-reg}

As discussed in Section~\ref{sec-vectors}, classical Whittaker-Henderson
smoothing can be viewed as an approximation to a penalized likelihood
maximization, relying on a crude rate estimator assumed to be
asymptotically normal. To assess the practical consequences of this
approximation, we conduct an empirical comparison based on six simulated
datasets reflecting the typical structure and volume of real insurance
portfolios:

\begin{itemize}
\item
  The first three datasets simulate annuity portfolios with 20,000,
  100,000, and 500,000 policyholders. The sole covariate is age, ranging
  from 50 to 95.
\item
  The next three mimic long-term care (LTC) portfolios of the same
  sizes. Modelling of LTC typically relies on the illness-death model
  (Fix and Neyman 1951; Clifford 1977). To get a two-dimensional
  illustration we focus on the transition between the disabled and dead
  states (the two other transitions would provide additional
  one-dimensional examples). Two covariates are used: age (70--100) and
  duration in LTC (0--15 years).
\end{itemize}

Each dataset consists of individual-level longitudinal data, from which
we derive event counts \(\mathbf{d}\) and exposures \(\mathbf{e_c}\),
aggregated by age \(x\) (for annuities) of by \((x, z)\) pairs (for
LTC). All datasets within each group share the same underlying structure
and differ only in size. Key dataset statistics are provided in
Table~\ref{tbl-datasets} and additional details about how those datasets
were generated are provided in Section~\ref{sec-datasets} of the
appendices.

\begin{table}

\caption{\label{tbl-datasets}Key figures associated with the 6 simulated
datasets}

\centering{

\centering
\begin{tabular}[t]{|>{}l|r|r|r|>{}r|}
\hline
\multicolumn{1}{|c|}{\cellcolor[HTML]{008851}{\textcolor{white}{\textbf{Portolio type}}}} & \multicolumn{1}{c|}{\cellcolor[HTML]{008851}{\textcolor{white}{\textbf{Dimensions}}}} & \multicolumn{1}{c|}{\cellcolor[HTML]{008851}{\textcolor{white}{\textbf{Head count}}}} & \multicolumn{1}{c|}{\cellcolor[HTML]{008851}{\textcolor{white}{\textbf{Exposure count}}}} & \multicolumn{1}{c|}{\cellcolor[HTML]{008851}{\textcolor{white}{\textbf{Death count}}}}\\
\hline
annuity & 45 & 20,000 & 136,524 & 1,722\\
\hline
annuity & 45 & 100,000 & 679,728 & 8,452\\
\hline
annuity & 45 & 500,000 & 3,405,892 & 42,499\\
\hline
LTC & 30 x 15 & 20,000 & 8,115 & 1,888\\
\hline
LTC & 30 x 15 & 100,000 & 40,004 & 9,281\\
\hline
LTC & 30 x 15 & 500,000 & 202,666 & 47,358\\
\hline
\end{tabular}

}

\end{table}%

We apply two methods:

\begin{enumerate}
\def\labelenumi{\arabic{enumi}.}
\item
  Original WH smoothing using
  \(\mathbf{y} = \ln(\mathbf{d} / \mathbf{e_c})\) and weights
  \(\mathbf{w} = \mathbf{d}\) as in Section~\ref{sec-vectors}.
\item
  Generalized WH smoothing, using the likelihood formulation of
  Section~\ref{sec-generalisation}.
\end{enumerate}

Both methods use the same smoothing parameter(s) \(\lambda\), to ensure
that prior assumptions on \(\boldsymbol{\theta} = \ln \boldsymbol{\mu}\)
are held constant. We fix the penalty order at \(q = 2\), corresponding
to second-order differences.

As both estimators target \(\boldsymbol{\theta}\), we compare them using
the following relative error metric:

\begin{equation}\phantomsection\label{eq-delta1}{\Delta(\boldsymbol{\theta}) = \frac{\ell_P(\hat{\boldsymbol{\theta}}_\text{ML}) - \ell_P(\boldsymbol{\theta})\;\;}{\ell_P(\hat{\boldsymbol{\theta}}_\text{ML}) - \ell_P(\hat{\boldsymbol{\theta}}_\infty)}.}\end{equation}

Here \(\hat{\boldsymbol{\theta}}_\infty\) maximizes the penalized
likelihood, while \(\hat{\boldsymbol{\theta}}_\infty\) corresponds to
the solution with \(\lambda \rightarrow \infty\), which we later show to
be the degree-\((q - 1)\) polynomial that maximizes the likelihood. By
construction:

\[\Delta(\hat{\boldsymbol{\theta}}_\text{ML}) = 0, \quad\Delta(\hat{\boldsymbol{\theta}}_\infty) = 1, \quad\text{and}\quad \Delta(\boldsymbol{\theta}) \ge 0.\]

A model with \(\Delta(\boldsymbol{\theta}) > 1\) performs worse than a
simple polynomial fit under the prior.

Table~\ref{tbl-approx-normale} presents the values of
\(\Delta(\hat{\boldsymbol{\theta}}_\text{norm})\) across the six
datasets. As expected, discrepancies decrease with portfolio size. For
annuities, the approximation performs reasonably well even at smaller
scales. In contrast, for LTC, it yields substantial errors, except for
the largest portfolio.

One explanation, supported by the Standardized Mortality Ratio (SMR)
also provided in Table~\ref{tbl-approx-normale}, is the positive
correlation between observed event counts and their use as weights. This
causes high crude rates to be overweighted, and low rates to be
underweighted---introducing systematic overestimation of mortality
rates. This bias is more severe in the LTC case where the observed
deaths by data point is lower. In contrast, generalized WH smoothing
preserves total event counts by construction, always yielding an SMR of
exactly 100\%. These results support adopting generalized WH smoothing
in most practical settings. It retains the advantages of the original
method while offering improved accuracy---even in small samples---and
remains straightforward to implement.

\begin{table}

\caption{\label{tbl-approx-normale}Impact of the approximation from the
original WH smoothing on the 6 simulated datasets}

\centering{

\centering
\begin{tabular}[t]{|>{}l|r|r|>{}l|}
\hline
\multicolumn{1}{|c|}{\cellcolor[HTML]{008851}{\textcolor{white}{\textbf{Portolio type}}}} & \multicolumn{1}{c|}{\cellcolor[HTML]{008851}{\textcolor{white}{\textbf{Head count}}}} & \multicolumn{1}{c|}{\cellcolor[HTML]{008851}{\textcolor{white}{\textbf{Relative Error}}}} & \multicolumn{1}{c|}{\cellcolor[HTML]{008851}{\textcolor{white}{\textbf{SMR}}}}\\
\hline
annuity & 20,000 & 1,91\% & 99,19\%\\
\hline
annuity & 100,000 & 0,02\% & 99,89\%\\
\hline
annuity & 500,000 & 0,00\% & 99,99\%\\
\hline
LTC & 20,000 & 93,27\% & 86,86\%\\
\hline
LTC & 100,000 & 5,12\% & 97,59\%\\
\hline
LTC & 500,000 & 0,24\% & 99,56\%\\
\hline
\end{tabular}

}

\end{table}%

\section{How to select the smoothing parameters?}\label{sec-selection}

\subsection{Impact of smoothing parameter
choice}\label{impact-of-smoothing-parameter-choice}

In the one-dimensional case, WH smoothing involves a single smoothing
parameter \(\lambda\); in two dimensions a pair
\(\lambda = (\lambda_x, \lambda_z)\). These parameters govern the
trade-off between fidelity to the data and smoothness of the estimate,
as defined in Equation~\ref{eq-WH1}.

\begin{figure}

\centering{

\includegraphics[width=1\linewidth,height=\textheight,keepaspectratio]{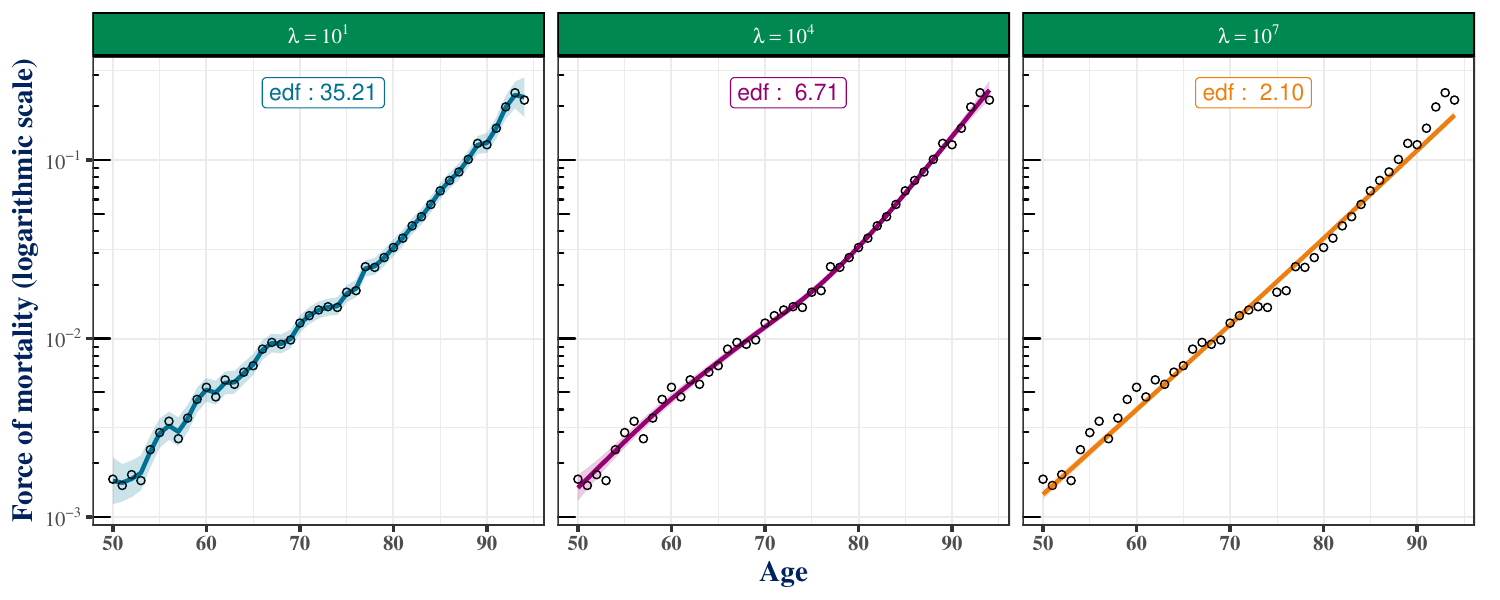}

}

\caption{\label{fig-illu}WH smoothing on a synthetic annuity portfolio
with 3 smoothing levels. Dots: crude rates; curves: smoothed estimates;
shaded areas: credibility intervals. \emph{edf}: effective degrees of
freedom.}

\end{figure}

Figure~\ref{fig-illu} illustrates this effect in a one-dimensional
annuity dataset (100,000 policyholders, see
Section~\ref{sec-impact-reg}), with three values of \(\lambda\). The
effective degrees of freedom (edf), computed as the trace of the hat
matrix \(H = (W + P_{\lambda})^{- 1}W\), are shown for each curve. This
quantity serves as a non-parametric analog of the number of free
parameters in classical models and can take fractional values.

As shown, a low value \(\lambda = 10^1\) yields an overfitted result
that mirrors sampling noise, while a high value \(\lambda = 10^7\)
oversmooths and obscures the underlying trend. A mid-range value
\(\lambda = 10^4\) appears visually balanced. However, selecting a
smoothing parameter by eye is unreliable: small-sample variability at
the extremes of the age range can easily be mistaken for meaningful
patterns.

\begin{figure}

\centering{

\includegraphics[width=1\linewidth,height=\textheight,keepaspectratio]{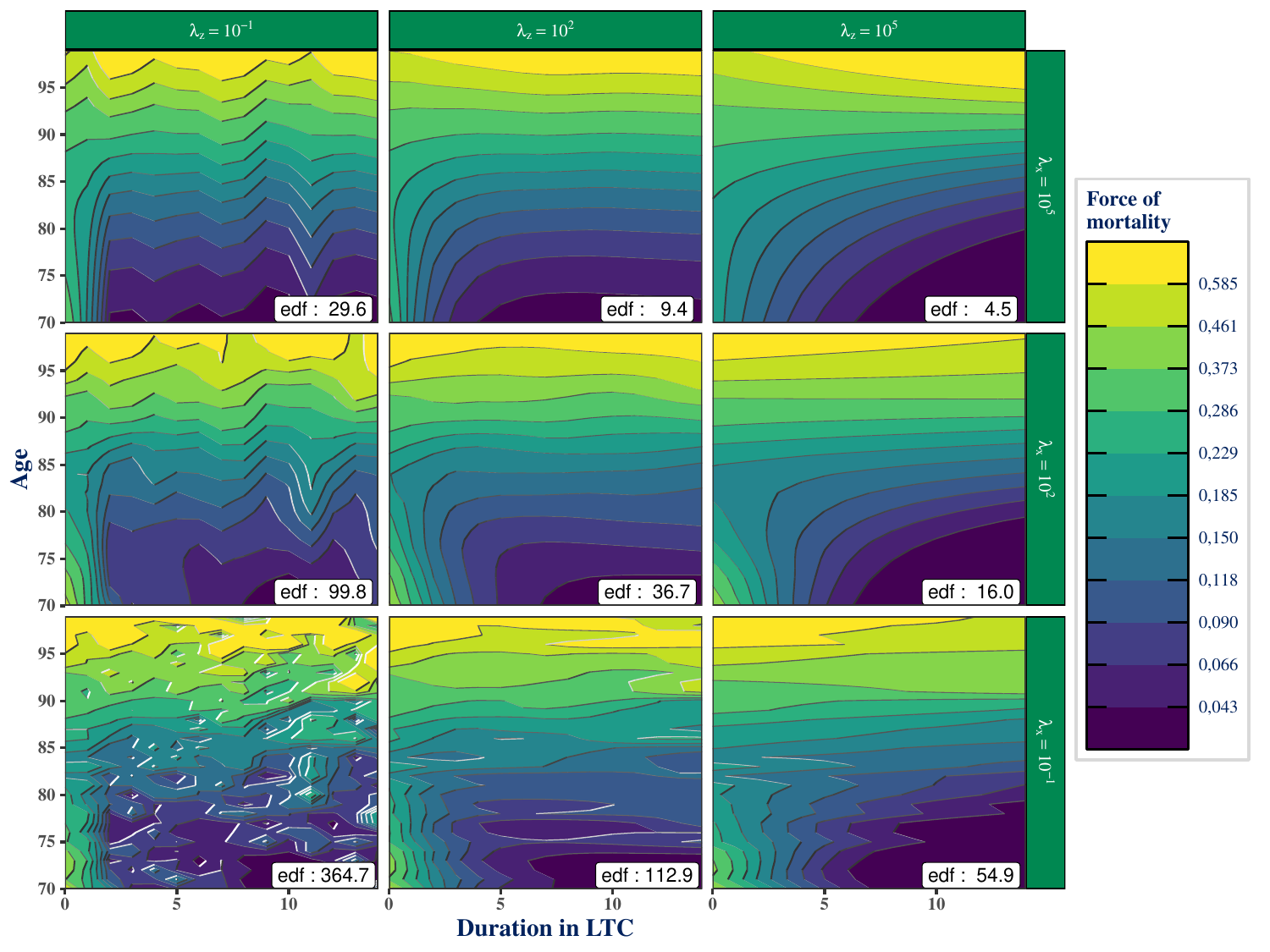}

}

\caption{\label{fig-illu2d}WH smoothing applied to disability-to-death
transitions in an LTC portfolio, using 9 combinations of smoothing
parameters. Contour lines and colours show the smoothed mortality
surface by age and LTC duration.}

\end{figure}

The two-dimensional case further illustrates this difficulty.
Figure~\ref{fig-illu2d} presents the smoothed transition rates from
disability to death in an LTC portfolio (100,000 policyholders), using 9
combinations of \((\lambda_x, \lambda_z)\). Choosing an appropriate
parameter pair visually becomes nearly impossible, reinforcing the need
for a data-driven statistical selection criterion.

\subsection{Statistical criteria for parameter
selection}\label{statistical-criteria-for-parameter-selection}

Smoothing parameter selection typically relies on two classes of
statistical criteria:

\begin{enumerate}
\def\labelenumi{\arabic{enumi}.}
\item
  Prediction-based criteria, which aim to minimize prediction error,
  such as the Akaike Information Criterion (AIC) (Akaike 1973) and
  Generalized Cross-Validation (GCV) (Wahba 1980);
\item
  Likelihood-based criteria, which maximize the marginal likelihood---an
  approach introduced by Patterson and Thompson (1971) (under the name
  REML in the Gaussian case) and adapted to smoothing by Anderssen and
  Bloomfield (1974).
\end{enumerate}

While prediction-based criteria have desirable asymptotic properties
(Wahba 1985; Kauermann 2005), their convergence toward optimal smoothing
parameters can be slow. In contrast, marginal likelihood criteria tend
to perform more robustly in finite samples (Reiss and Todd Ogden 2009;
Wood 2011).

To illustrate this, we apply AIC, GCV, and marginal likelihood to 100
replicates of the annuity portfolio with 100,000 policyholders (see
Section~\ref{sec-impact-reg}). For each replicate, we select the optimal
smoothing parameter and compute the corresponding effective degrees of
freedom.

\begin{figure}

\centering{

\includegraphics[width=1\linewidth,height=\textheight,keepaspectratio]{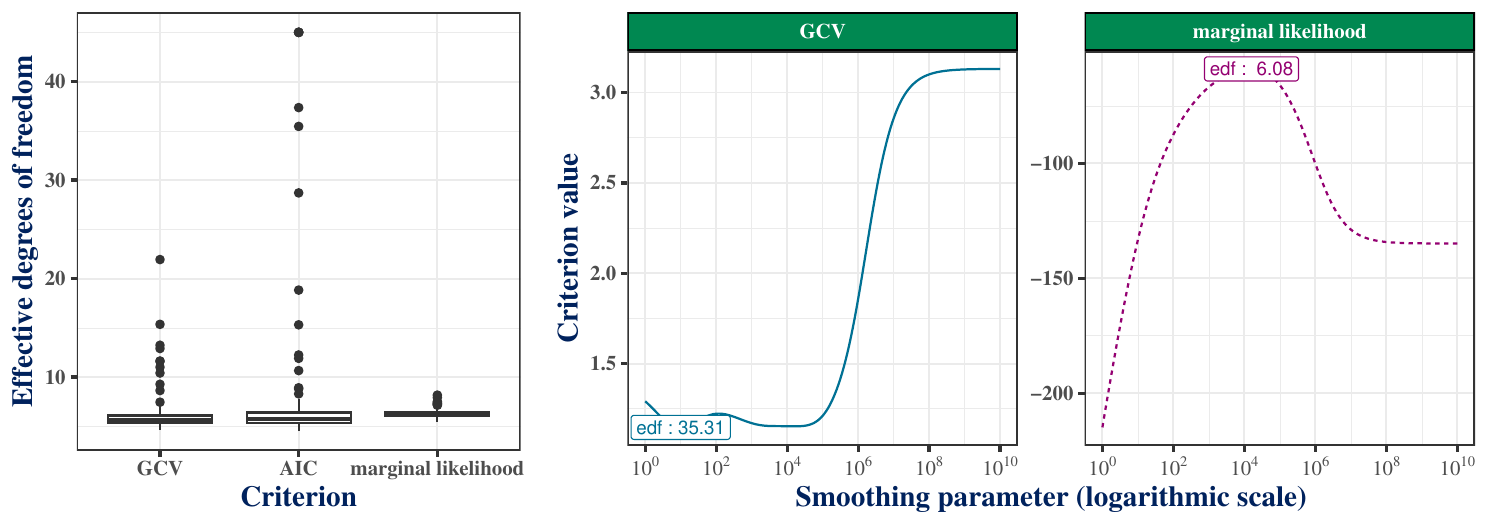}

}

\caption{\label{fig-crit}Comparison of criteria for selecting the
smoothing parameter in one-dimensional WH smoothing. Left: distribution
of effective degrees of freedom under AIC, GCV, and marginal likelihood
across 100 replicates. Right: GCV and marginal likelihood values for one
replicate as functions of the smoothing parameter.}

\end{figure}

As shown on the left side of Figure~\ref{fig-crit}, marginal likelihood
produces stable and coherent degrees of freedom across replicates,
whereas AIC and especially GCV often yield overly complex models. On the
right, we plot the GCV and marginal likelihood profiles for a single
replicate: marginal likelihood exhibits a well-defined maximum, while
GCV presents two local minima. One aligns with the marginal likelihood
optimum, but the global minimum corresponds to a model with
\textasciitilde35 degrees of freedom---an implausibly complex mortality
curve.

These observations support the use of marginal likelihood over
prediction-based criteria, especially in actuarial applications where
robustness is key. Moreover, this choice aligns naturally with the
Bayesian framework introduced in Sections \ref{sec-CI} to
\ref{sec-generalisation}.

We now detail its implementation---first for the original WH smoothing,
then for the generalized setting---introducing three optimization
strategies and three numerical algorithms and comparing their respective
performances.

\subsection{Selection in the original
smoothing}\label{sec-selection-originale}

We consider again the normal framework from Section~\ref{sec-CI}, where
\(\mathbf{y} \mid \boldsymbol{\theta} \sim \mathcal{N}(\boldsymbol{\theta}, \sigma^2W^{-})\)
and
\(\boldsymbol{\theta} \mid \lambda \sim \mathcal{N}(0, \sigma^2P_{\lambda}^{-})\).
In the empirical Bayes approach, the smoothing parameter \(\lambda\) is
estimated by maximizing the marginal likelihood:

\[\mathcal{L}^m_\text{norm}(\lambda) = f(\mathbf{y} \mid \lambda) = \int f(\mathbf{y}, \boldsymbol{\theta} \mid \lambda)\text{d}\boldsymbol{\theta} = \int f(\mathbf{y} \mid \boldsymbol{\theta}) f(\boldsymbol{\theta}  \mid \lambda)\text{d}\boldsymbol{\theta}.\]

This is simply the maximum likelihood method applied to the smoothing
parameter, treated as deterministic but unknown. A closed-form
expression for this integral can be derived using standard Gaussian
identities (see Section~\ref{sec-likelihood-computations} of the
appendices), yielding the marginal log-likelihood:

\[\ell^m_\text{norm}(\lambda) = - \frac{1}{2}\left[(\mathbf{y} -\hat{\boldsymbol{\theta}}_{\lambda})^{T}W(\mathbf{y} -\hat{\boldsymbol{\theta}}_{\lambda}) / \sigma^2 + \hat{\boldsymbol{\theta}}_{\lambda}^{T}P_{\lambda} \hat{\boldsymbol{\theta}}_{\lambda} / \sigma^2 + \ln|W + P_{\lambda}| - \ln |P_{\lambda}|_{+} + C\right].\]

where
\(\hat{\boldsymbol{\theta}}_\lambda = (W + P_{\lambda})^{-1}W\mathbf{y}\),
and \(C = - \ln|W|_{+} + (n_* - q)\ln(2\pi\sigma^2)\) is a constant
independent of \(\lambda\). This function is maximized numerically to
obtain \(\hat{\lambda}_\text{norm}\).

\subsection{Selection in the generalized
smoothing}\label{sec-selection-generalized}

The empirical Bayes approach introduced in the normal framework can be
extended to the generalized smoothing framework developed in
Section~\ref{sec-generalisation}. While no closed-form expression exists
for the marginal likelihood in this context, it can be approximated
using a second-order Taylor expansion of the log-posterior density
around its maximum \(\hat{\boldsymbol{\theta}}_\lambda\)---similarly to
what was done in the normal case. This yields the so-called Laplace
Approximation of the Marginal Likelihood (LAML), defined as:

\[\ell^m_\text{LAML}(\lambda) = \ell(\hat{\boldsymbol{\theta}}_{\lambda}) - \frac{1}{2}\left[\hat{\boldsymbol{\theta}}_{\lambda}^T P_{\lambda} \hat{\boldsymbol{\theta}}_{\lambda} + \ln|W_{\lambda} + P_{\lambda}| - \ln|P_{\lambda}|_{+} - q\ln(2\pi)\right]\]

where
\(W_\lambda = \text{Diag}(\exp(\hat{\boldsymbol{\theta}}_{\lambda}) \odot \mathbf{e_c})\)
and \(\ell(\hat{\boldsymbol{\theta}}_{\lambda})\) is the log-likelihood
evaluated at the penalized MLE. The detailed derivation of the Laplace
approximation in this setting is provided in
Section~\ref{sec-likelihood-computations} of the appendices. This
approximation plays a central role in the automatic selection of the
smoothing parameter \(\lambda\) in the generalized Whittaker-Henderson
smoothing framework. As in the normal case, the marginal likelihood
\(\ell^m_\text{LAML}(\lambda)\) must be maximized numerically. However,
a key distinction is that the penalized likelihood maximizer
\(\hat{\boldsymbol{\theta}}_\lambda\) now depends on \(\lambda\) and
must be recomputed at each iteration via the PIRLS algorithm. This leads
to a two-level optimization procedure:

\begin{itemize}
\item
  an inner loop estimating \(\hat{\boldsymbol{\theta}}_\lambda\) for
  fixed \(\lambda\) using PIRLS;
\item
  and an outer loop optimizing \(\ell^m_\text{LAML}(\lambda)\) with
  respect to \(\lambda\).
\end{itemize}

This outer iteration approach is the most principled method for
smoothing parameter selection in this setting.

Alternative strategies have been proposed to reduce computational
burden. The first one, known as performance-oriented iteration, was
introduced by Gu (1992) and relies on the observation that, at each
PIRLS step, the working response vector \(\mathbf{z}_k\) can be treated
as approximately normal:
\(\mathbf{z}_k \mid \boldsymbol{\theta} \sim \mathcal{N}(\boldsymbol{\theta}, W_k^{-1})\).
Assuming \(W_k\) independent of \(\lambda\), the marginal likelihood can
be maximized within each PIRLS step using the normal approximation
methodology of Section~\ref{sec-selection-originale}, with
\(\mathbf{y}\) replaced by \(\mathbf{z}_k\) and \(W\) by \(W_k\). This
effectively reverses the nesting structure, potentially saving
computational time when updating \(\lambda\) is less costly than
recomputing a PIRLS step. A formal justification of the method is
provided by Wood (2017, 149) which emphasizes that it does not actually
require \(\mathbf{z}_k\) to have a normal distribution to be
well-founded.

A third and even simpler strategy is the alternate iteration approach,
used for instance by Wood et al. (2017). It consists in alternating
updates of \(\boldsymbol{\theta}\) (via PIRLS) and \(\lambda\) (via
approximate marginal likelihood), without fully optimizing either at
each step. This relies on the empirical observation that a coarse update
of \(\lambda\) may suffice, as the marginal likelihood surface changes
between iterations.

Despite their efficiency, both performance-oriented and alternate
iteration approaches lack formal convergence guarantees. Unlike outer
iteration, they operate on different smoothing parameters at each step,
rendering penalized likelihood values non-comparable across iterations.
Moreover, they do not track the value of \(\ell^m_\text{LAML}(\lambda)\)
during the optimization, making it harder to assess convergence or apply
step-length controls.

Detailed algorithmic formulations of all three strategies in the
generalized WH smoothing framework are provided in
Section~\ref{sec-algorithms-appendix} of the appendices.

\subsection{Algorithms for the maximization of the marginal
likelihood}\label{sec-algorithms}

Several algorithms can be used to maximize the marginal likelihood or
its Laplace approximation (LAML). It is generally preferable to apply
these algorithms to the logarithm of the smoothing parameters, for three
main reasons:

\begin{enumerate}
\def\labelenumi{\arabic{enumi}.}
\item
  It ensures positivity of the smoothing parameters;
\item
  It simplifies the expressions of derivatives, when required;
\item
  It allows more uniform coverage of the range of interest (e.g., from
  \(\lambda = 10^1\) to \(10^7\), as in Figure~\ref{fig-illu},
  differences of comparable magnitude occur on a logarithmic scale).
\end{enumerate}

\subsubsection{Derivative-free
heuristics}\label{derivative-free-heuristics}

A first, operationally simple option is to use general-purpose
derivative-free optimization methods:

\begin{itemize}
\item
  Brent's method (Brent 1973) in the one-dimensional case;
\item
  The Nelder-Mead simplex algorithm (Nelder and Mead 1965) in higher
  dimensions.
\end{itemize}

These are readily available in base \(\mathtt{R}\) via the
\texttt{optimize} and \texttt{optim} functions. They only require
evaluating the marginal likelihood or LAML at each step, which is
computationally inexpensive. However, they typically require more
iterations to converge and cannot be combined with the alternate
iteration approach, as they do not guarantee systematic improvement of
the criterion at each step.

\subsubsection{Generalized Fellner-Schall
method}\label{generalized-fellner-schall-method}

A more specialized algorithm is the generalized Fellner-Schall method,
based on ideas from Fellner (1986) and Schall (1991), and adapted for
smoothing parameter selection in multidimensional generalized linear
models by Rodriguez-Alvarez et al. (2015). It may be summarised by the
update formula:

\begin{equation}\phantomsection\label{eq-fs}{\lambda_{j}^{\text{next}} = \frac{\text{tr}(P_\lambda^{-}P_j) - \text{tr}[(X^TWX + P_\lambda)^{- 1}P_j]}{\hat{\boldsymbol{\beta}}_\lambda^TP_j\hat{\boldsymbol{\beta}}_\lambda}\lambda_j^{\text{current}} \quad \text{for} \quad j\in\{x, z\} .}\end{equation}

where for WH smoothing \(P_j = D_{n_{+},q}^{T}D_{n_{+},q}\) in the
one-dimensional case and \(P_x\) (resp. \(P_z\)) is the marginal
penalization matrix \(I_{n_z} \otimes D_{n_x,q_x}^{T}D_{n_x,q_x}\)
(resp. \(D_{n_z,q_z}^{T}D_{n_z,q_z} \otimes I_{n_x}\)) in the
two-dimensional case.This update can be interpreted more intuitively as:

\[\hat{\boldsymbol{\beta}}_\lambda^T(\lambda_{j}^{\text{next}}P_j)\hat{\boldsymbol{\beta}}_\lambda = \text{tr}[P_\lambda^{-}\lambda_j^{\text{current}}P_j - (X^TWX + P_\lambda)^{- 1}\lambda_j^{\text{current}}P_j]\]

where the right-hand side corresponds to an effective degrees of freedom
associated with \(\lambda_j^{\text{current}}P_j\), and the left-hand
side to a squared error, normalized by the updated penalty precision.
This makes \(\lambda_j^{\text{next}}\) resemble a REML-based estimator
for the inverse variance. More details may be found in Rodríguez-Álvarez
et al. (2019). This method:

\begin{itemize}
\item
  May be combined with any of the three iteration nesting schemes
  (outer, performance, alternate);
\item
  Does not require explicit derivative computations;
\item
  Converges toward an approximate maximum of LAML in the generalized
  case, since it ignores the dependence of \(W\) on \(\lambda\);
\item
  Tends to take longer steps than EM-like algorithms (Dempster, Laird,
  and Rubin 1977), but shorter than Newton updates (see Wood and Fasiolo
  2017, which also provides a thorough justification for the method).
\end{itemize}

\subsubsection{Newton algorithm}\label{newton-algorithm}

A third option is the Newton method, which involves computing both the
first and second derivatives of the marginal likelihood (or LAML) with
respect to \(\ln\lambda\). Full derivations are provided in Wood (2011),
which covers a more general case. The method applies in both the normal
and generalized cases, but in the latter, derivative expressions are
more complex due to the dependence of \(W\) on \(\lambda\). The Newton
algorithm is fast and precise and applicable to all three nesting
strategies. The downside is the operational complexity associated with
this method, especially in the generalized case.

\subsection{Performance comparison}\label{sec-comp-outer-perf}

Sections \ref{sec-selection-generalized} and \ref{sec-algorithms}
introduced eight combinations of nesting strategies and optimization
algorithms applicable to the generalized WH smoothing. We now assess the
potential convergence issues and approximation errors associated with
each of them.

This analysis is based on 100 replicates of the simulated annuity and
LTC portfolios with 100,000 policyholders, as described in
Section~\ref{sec-impact-reg}. For each replicate and each method
combination, we compute the LAML at the selected smoothing parameters,
and compare this value to the (approximate) optimal value obtained
across all combinations, denoted \(\hat{\lambda}_\text{opt}\).

To quantify the discrepancy, we define the relative error:

\begin{equation}\phantomsection\label{eq-delta2}{\Delta(\lambda) = \frac{\ell^m_\text{LAML}(\hat{\lambda}_\text{opt}) - \ell^m_\text{LAML}(\lambda)}{\ell^m_\text{LAML}(\hat{\lambda}_\text{opt}) - \ell^m_\text{LAML}(\infty)}}\end{equation}

where \(\ell^m_\text{LAML}(\infty)\) corresponds to the LAML value when
using an infinite smoothing penalty, i.e., the overly smooth baseline.
By construction, \(\Delta(\lambda) \ge 0\) for all tested methods, with
\(\Delta(\hat{\lambda}_\text{opt}) = 0\) and \(\Delta(\infty) = 1\). In
the two-dimensional setting, we also compare average computation time
for each method across replicates.

\begin{figure}

\centering{

\includegraphics[width=1\linewidth,height=\textheight,keepaspectratio]{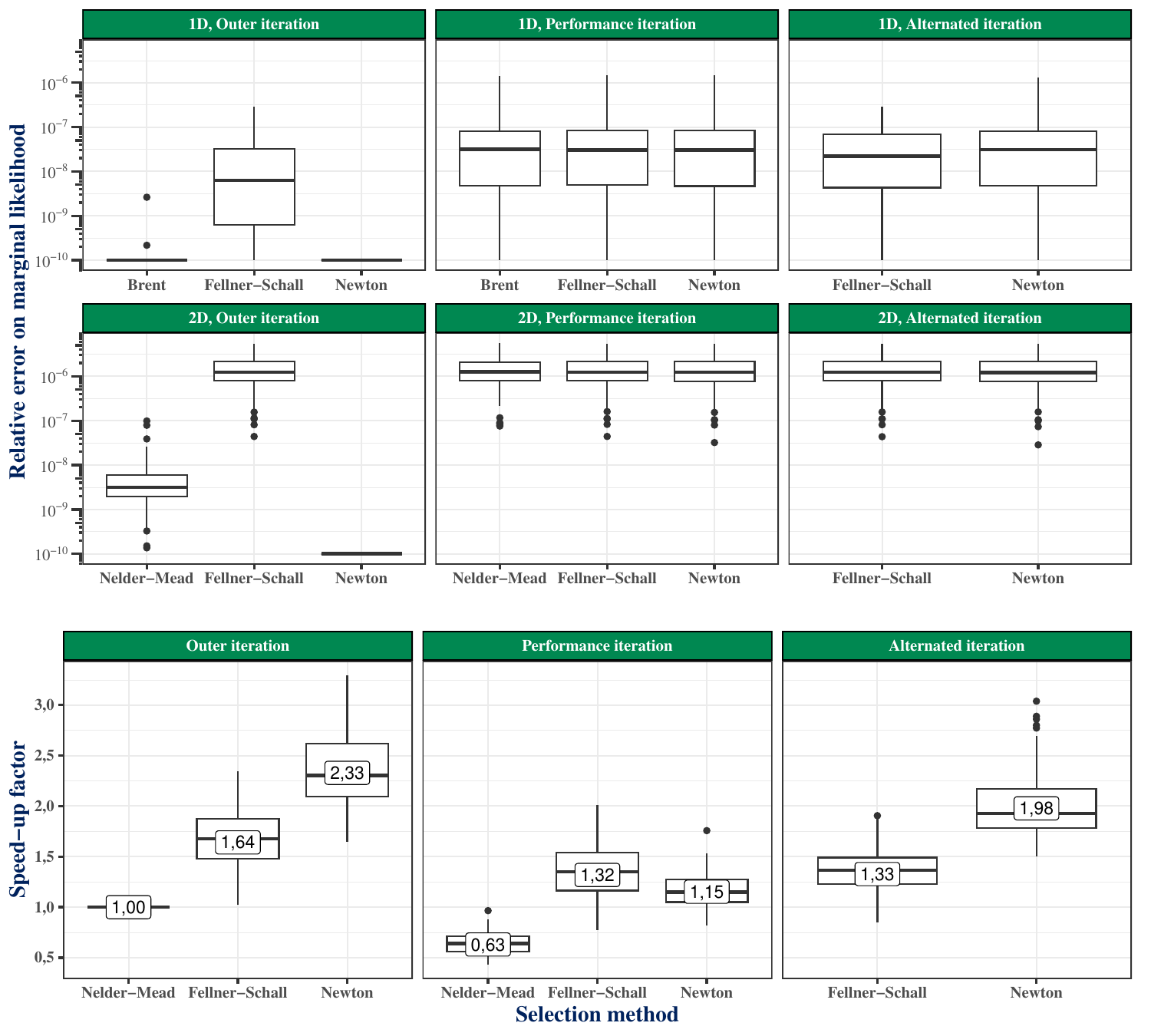}

}

\caption{\label{fig-comp-ecart}Comparison of the 8 nesting strategy and
algorithm combinations in the 1D and 2D simulated cases. Top: relative
error on the LAML (log scale). Bottom: improvement in average
computation time compared to the Nelder-Mead + outer iteration
reference.}

\end{figure}

Results are summarised in Figure~\ref{fig-comp-ecart}. The top panel
displays the relative error \(\Delta(\lambda)\) (capped below
\(10^{-10}\) for readability). In the outer iteration framework:

\begin{itemize}
\item
  The Newton method consistently achieves relative errors below
  \(10^{-10}\);
\item
  Brent and Nelder-Mead heuristics yield slightly higher errors but
  remain below \(10^{-7}\);
\item
  The generalized Fellner-Schall method produces higher errors, but
  still below \(10^{-5}\) and negligible in practice.
\end{itemize}

In the performance and alternate iteration frameworks, all methods yield
similar errors, consistently below \(10^{-5}\), with no convergence
issues observed in any replicate. These findings suggest that method
selection can be guided by practical considerations such as speed and
implementation ease.

The bottom panel of Figure~\ref{fig-comp-ecart} compares computation
times (relative to the Nelder-Mead + outer iteration baseline):

\begin{itemize}
\item
  In the outer iteration framework, the Newton method is the fastest,
  followed by the Fellner-Schall approach;
\item
  All outer iteration variants are faster than their performance or
  alternate counterparts.
\end{itemize}

This is unsurprising, as PIRLS steps are particularly lightweight in WH
smoothing (where the model matrix is the identity). However, alternate
strategies may remain useful for more general cases like those described
in Section~\ref{sec-rr}.

For reference, the average time required for a single iteration using
Nelder-Mead in the 2D outer iteration case is approximately 1.68 seconds
(versus 5 milliseconds in the 1D case).

\section{How to improve smoothing computational
efficiency?}\label{sec-reduction}

\subsection{Motivation}\label{sec-prob-red}

Whittaker-Henderson (WH) smoothing is a full-rank method, meaning that
it includes as many parameters as there are observation points. This
feature ensures a high degree of flexibility, allowing the estimator to
closely track the input signal when sufficient data is available.
Formally, WH smoothing is asymptotically unbiased since:

\[\mathbb{E}(\hat{\mathbf{y}}) = (W + P_{\lambda})^{- 1}W\mathbb{E}(\mathbf{y}) \overset {m \rightarrow \infty}{\rightarrow} \mathbb{E}(\mathbf{y}),\]

where \(m\) denotes the number of observed individuals, which influences
the matrix \(W\).

However, this flexibility comes at a computational cost. Some key
operations, such as (implicit) matrix inversions, scale cubically with
the number of parameters. As a result, WH smoothing may become
impractical with large number of combinations or when applied repeatedly
(e.g.~in simulations or bootstraps).

In one-dimensional settings, such as age-only models with annual
discretization, the number of points rarely exceeds 100, and computation
time is negligible. In contrast, two-dimensional use cases---common in
insurance---can lead to substantially larger datasets:

\begin{itemize}
\item
  Disability tables in France must cover entry ages from 18 to 61 and
  exit ages up to 62, resulting in
  \((62 - 18) \times (62 - 18 + 1) / 2 = 990\) combinations.
\item
  Transition tables from short-term incapacity to disability involve
  entry ages from 18 to 67 and monthly durations from 0 to 36 months,
  yielding \((67 - 18) \times 36 = 1,764\) combinations.
\item
  Long-term care (LTC) models require coverage over ages 50 to 100 and
  durations from 0 to 20 years, totalling
  \((100 - 50) \times (20 - 0) = 1,000\) combinations (in practice, this
  number may be lower due to data sparsity).
\end{itemize}

In such settings, computing WH smoothing---especially when paired with
smoothing parameter selection---can take several minutes per
application, limiting usability in iterative contexts.

To address this limitation, we now analyse the computational complexity
of the main steps in WH smoothing and smoothing parameter selection then
introduce two complementary strategies to reduce computation time:

\begin{itemize}
\item
  A structural optimization that exploits the specific form of WH
  penalization matrices;
\item
  A reduced-rank approximation that lowers the number of parameters
  while minimizing bias compared to the full-rank estimator.
\end{itemize}

Finally, we benchmark these strategies in terms of runtime and accuracy
using 100 replicates of the mid-size annuity and LTC portfolios from
Section~\ref{sec-impact-reg}. The structural optimization is compared to
the original WH method, while the reduced-rank approximation is
evaluated against the original method, the structural optimization, and
a reference P-spline smoothing approach.

\subsection{Practical computation for penalized
smoothers}\label{sec-practical}

Whittaker-Henderson (WH) smoothing belongs to a broader family of
penalized smoothing methods that produce estimates of the form:

\[\hat{\mathbf{y}}_\lambda = X\hat{\boldsymbol{\beta}}_{\lambda} \quad\text{where}\;\boldsymbol{\beta}_{\lambda}\;\text{solves}\: (X^TWX + P_{\lambda})\hat{\boldsymbol{\beta}}_\lambda = X^TW\mathbf{y}.\]

Here, \(X\) and \(P_\lambda\) denote the model and penalization matrices
of size \(n \times p\) and \(p \times p\) respectively, and \(W\) is a
diagonal matrix of positive weights of size \(n \times n\).

\subsubsection{Computational steps}\label{computational-steps}

The computation of \(\hat{\mathbf{y}}_\lambda\) for a given \(\lambda\)
typically involves the following steps:

\begin{enumerate}
\def\labelenumi{\arabic{enumi}.}
\item
  Absorb the weights in the model matrix and observation vector, forming
  \(W^{1/2}X\) and \(W^{1/2}\mathbf{y}\), which requires \(O(n^2)\) and
  \(O(n)\) operations respectively (multiplying each row of \(X\) and
  each element of \(\mathbf{y}\) by the corresponding element of
  \(\mathbf{w}\)).
\item
  Form the matrix \(P_\lambda\). The cost of this operation is typically
  \(O(p^2)\) in the general case.
\item
  Form the matrix \(X^TWX\) and the vector \(X^TW\mathbf{y}\), which
  requires up to \(O(np^2)\) and \(O(np)\) operations respectively.
\item
  Add together \(X^TWX\) and \(P_{\lambda}\) which requires \(O(p^2)\)
  operations in the general case.
\item
  Compute the Cholesky decomposition \(X^TWX + P_{\lambda} = R^TR\) at a
  cost of \(O(p^3)\).
\item
  Obtain \(\hat{\boldsymbol{\beta}}_\lambda\) by forward-backward
  substitution, first solving \(R^T\mathbf{u} = X^TW\mathbf{y}\) then
  \(R\hat{\boldsymbol{\beta}}_\lambda = \mathbf{u}\) with an associated
  cost of \(O(p^2)\) for each system.
\item
  Compute
  \(\hat{\mathbf{y}}_\lambda = X\hat{\boldsymbol{\beta}}_\lambda\) at a
  cost of \(O(np)\).
\end{enumerate}

As an alternative to Cholesky, QR decomposition may be used for greater
numerical stability (see Golub and Van Loan 2013). It applies to the
weighted design matrix stacked with a matrix \(B\) such that
\(B^T B = P_\lambda\).

\subsubsection{Simplifications for WH
smoothing}\label{simplifications-for-wh-smoothing}

In WH smoothing, \(X = I_n\), which simplifies computations:

\begin{itemize}
\item
  Step 7 is unnecessary, as well as the first part of step 3.
\item
  \(X^T W \mathbf{y} = W \mathbf{y}\) (step 3) is computed in \(O(n)\)
  by multiplying \(\mathbf{w}\) and \(\mathbf{y}\).
\item
  \(X^T W X + P_\lambda = W + P_\lambda\) (step 4) is also computed in
  \(O(n)\) by adding the vector \(\mathbf{w}\) to the leading diagonal
  of \(P_{\lambda}\).
\end{itemize}

\subsubsection{Generalized WH smoothing with outer
iteration}\label{generalized-wh-smoothing-with-outer-iteration}

When using the outer iteration approach (see
Section~\ref{sec-selection}), each candidate \(\lambda\) requires a full
PIRLS cycle to estimate \(\hat{\boldsymbol{\theta}}_\lambda\), with new
working vector \(\hat{\mathbf{z}}^k_\lambda\) and weight matrix
\(W^k_\lambda\). Steps 1--6 above are repeated until convergence of the
PIRLS algorithm, which may be assessed by monitoring the changes in
penalized deviance. The deviance may be computed at a \(O(n)\) cost. For
penalization based on differences matrices, computation of
\(\hat{\boldsymbol{\beta}}_\lambda{\,}^TP_\lambda\hat{\boldsymbol{\beta}}_\lambda\)
should be based on the expression of \(R_{\lambda,q}\) provided in
Section~\ref{sec-brief} for an associated cost of \(O(qp)\). In
addition, PIRLS iterations for each new \(\lambda\) can be initialized
using the previous estimate of \(\hat{\mathbf{y}}_\lambda\) for faster
convergence.

\subsubsection{LAML computation}\label{laml-computation}

Once the deviance is known, computing the marginal likelihood/LAML also
requires:

\begin{itemize}
\item
  \(\ln|X^TWX + P_{\lambda}|\), which may be computed at a cost of
  \(O(p)\) from the leading diagonal of the Cholesky/QR factor \(R\)
  computed at step 5 in the derivation of \(\hat{\mathbf{y}}_\lambda\).
\item
  \(\ln|P_{\lambda}|_{+}\), which may be obtained from the eigenvalues
  of the penalization matrix: Section~\ref{sec-vp} shows that in the
  two-dimensional case, it can be computed via eigendecomposition of
  \(D_{p_x,q_x}^T D_{p_x,q_x}\) and \(D_{p_z,q_z}^T D_{p_z,q_z}\),
  performed only once, at a \(O(p_x^3 + p_z^3)\) cost. Computation of
  \(\ln|P_{\lambda}|_{+}\) then only requires scaling the eigenvalues
  for a cost of \(O(p)\).
\end{itemize}

\subsubsection{Algorithm-specific
computations}\label{algorithm-specific-computations}

Brent and Nelder-Mead require only marginal likelihood/LAML evaluations.

The generalized Fellner-Schall algorithm relies on the update formula of
Equation~\ref{eq-fs}:

\[\lambda_{j}^{\text{next}} = \frac{\text{tr}(P_\lambda^{-}P_j) - \text{tr}[(X^TWX + P_\lambda)^{- 1}P_j]}{\hat{\boldsymbol{\beta}}_\lambda^TP_j\hat{\boldsymbol{\beta}}_\lambda}\lambda_j^{\text{current}} \quad \text{for} \quad j\in\{x, z\} .\]

Evaluation of \(\text{tr}(P_\lambda^{-}P_j)\) does not require any
matrix product. In the one-dimensional case it is simply
\((p - q)/\lambda\) while in the two-dimensional case it may be obtained
directly at a \(O(p)\) cost using the eigenvalues of the aforementioned
penalization matrices \(P_j\). Evaluation of
\(\text{tr}[(X^TWX + P_\lambda)^{- 1}P_j]\) may use the identity
\(\text{tr}(AB) = \sum_{i,j} A_{ij}B_{ji}\) and therefore be computed at
an \(O(p^2)\) cost if the matrix \((X^TWX + P_\lambda)^{- 1}\) and
\(P_j\) are available. Computation of \(V = (X^TWX + P_\lambda)^{- 1}\)
is done by first solving for the inverse \(K = R^{- 1}\) of the
Cholesky/QR factor and then forming \(V\) as \(KK^T\). Both operations
have a \(O(p^3)\) cost.

Newton method also requires computation of \(V\), as well as several
matrix products involving the penalization matrix \(P_j\). For example,
the second derivatives of marginal likelihood require
\(\text{tr}[VP_jVP_k]\) terms and the second derivatives of marginal
likelihood require
\(\text{tr}[V(X(\partial W/\partial \rho_j)X + P_j)V(X(\partial W/\partial \rho_k)X + P_k)]\)
terms where \(\rho_j = \ln(\lambda_j)\), \(j = k = x\) in the
one-dimensional case and \(\{j,k\} \in \{x, z\}\) in the two-dimensional
case. The identity \(\text{tr}(AB) = \sum_{i,j} A_{ij}B_{ji}\) can also
be used in this case but matrix products \(VP_j\) or
\(V[X(\partial W/\partial \rho_j)X + P_j]\) still need to be explicitly
computed, for a respective cost of \(O(p^3)\) and \(O(n p^2)\) each.

These additional computations make Newton updates more expensive than
generalized Fellner-Schall updates, but they generally yield faster
convergence and higher precision (see
Section~\ref{sec-comp-outer-perf}).

\subsection{Banded optimization for WH
smoothing}\label{banded-optimization-for-wh-smoothing}

We now consider how to exploit the banded structure of the penalization
matrix in Whittaker-Henderson (WH) smoothing. This structure enables
significant computational gains, especially when dealing with large
number of observations. Throughout this section, we assume \(X = I_n\)
and \(p = n\), which holds for both the original and generalized WH
smoothing.

\subsubsection{One-Dimensional Case}\label{one-dimensional-case}

In one dimension, the penalization matrix takes the form:
\(P_\lambda = \lambda D_{n,q}^TD_{n,q}\). This matrix is symmetric and
banded with bandwidth \(q\). As a consequence:

\begin{enumerate}
\def\labelenumi{\arabic{enumi}.}
\item
  Compact storage: \(P_\lambda\) can be stored in a compact form with
  dimensions \(n \times (q + 1)\), and updated for new \(\lambda\) at a
  cost of \(O(qn)\). The matrix \(W + P_\lambda\) shares this structure.
\item
  Efficient Cholesky decomposition: the Cholesky factor \(R\) of
  \(W + P_\lambda\) can be computed in \(O(q^2 n)\) instead of
  \(O(n^3)\), and \(R\) is also banded with the same bandwidth.
\item
  Efficient back-substitution: computing
  \(\hat{\mathbf{y}}_\lambda = \hat{\boldsymbol{\beta}}_\lambda\) using
  \(R\) is now \(O(qn)\) instead of \(O(n^2)\).
\item
  Efficient inversion of \(R\): the inverse \(K = R^{-1}\) costs
  \(O(qn^2)\), an improvement over the \(O(n^3)\) cost for dense
  matrices.
\end{enumerate}

However, \(K\) is a dense triangular matrix, meaning the computation of
\(V = K K^T\) remains a \(O(n^3)\) operation. Fortunately, the
generalized Fellner-Schall algorithm only requires the diagonal of
\(V\), which can be obtained from \(K\) in \(O(n^2)\), since:

\[\lambda_j[\text{tr}(P_\lambda^{-}P_j) - \text{tr}(VP_j)] = (n - q) - (n - \text{tr}[VW]) = \textbf{diag}(V)^T\mathbf{w} - q.\]

Furthermore, the Newton algorithm benefits as well: the trace terms
involving \(V P_\lambda\) or
\(V (\partial W / \partial \ln\lambda + P_\lambda)\) are based on banded
matrices, making those products computable in \(O(qn^2)\) instead of
\(O(n^3)\).

\subsubsection{Two-dimensional Case}\label{two-dimensional-case}

In two dimensions, the penalization matrix is
\(P_\lambda = \lambda_xP_x + \lambda_zP_z\) where :

\begin{itemize}
\item
  \(P_x = I_{n_z} \otimes D_{n_x,q_x}^{T}D_{n_x,q_x}\)
\item
  \(P_z = D_{n_z,q_z}^{T}D_{n_z,q_z} \otimes I_{n_x}\).
\end{itemize}

This structure has the following key properties:

\begin{itemize}
\item
  Both matrices are made of \(n_z \times n_z\) square blocks of
  dimensions \(n_x \times n_x\) each.
\item
  \(P_x\) is block-diagonal with \(n_z\) identical \(n_x \times n_x\)
  banded blocks (bandwidth \(q_x\)).
\item
  \(P_z\) is block-banded with bandwidth \(q_z\). Each block is a scaled
  identity matrix.
\item
  As a whole, \(P_z\) and \(P_\lambda\) may be viewed as banded matrices
  with bandwidth \(q = q_z \times n_x\).
\end{itemize}

This implies that all statements made in the one-dimensional case carry
over to the two-dimensional case with this value of \(q\). It also
suggests that, if \((q_x + 1) / n_x < (q_z + 1) / n_z\), dimensions
\(x\) and \(z\) should be permuted before applying WH smoothing for
maximal efficiency.

As in the one-dimensional case, the generalized Fellner-Schall update
formula does not require the full computation of \(V\). Indeed, to
compute \(\text{tr}[VP_x]\) and \(\text{tr}[VP_z]\), we only need access
to elements of \(V\) for which either \(P_x\) or \(P_z\) is non-zero.
From what precedes, \(P_x\) has bandwidth \(q_x\) while \(P_z\) only
contains \(q_z\) non-zero diagonals on each side of the leading
diagonal. As \(V\) is symmetric, we only need to compute
\(q_x + q_z + 1\) diagonals of \(V\) for an associated cost of
\(O([q_x + q_z]n^2)\) instead of \(O(n^3)\).

With the Newton method, while computing \(V = K K^T\) still incurs a
\(O(n^3)\) cost, matrix multiplications like \(V P_j\) or
\(V (\partial W / \partial \rho_j + P_j)\) can be performed block-wise.
It may easily be checked for example that the products \(VP_x\) and
\(V(\partial W/\partial \rho_x + P_x)\) have a cost of \(O(q_xn^2)\)
while the products \(VP_z\) and \(V(\partial W/\partial \rho_z + P_z)\)
have a cost of \(O(q_zn^2)\).

\subsubsection{Summary of complexity
gains}\label{summary-of-complexity-gains}

Thanks to the banded structure, most computations involved in WH
smoothing can be accelerated by a factor of \(n / (q + 1)\) in the 1D
case and \(\max(n_x / (q_z + 1), n_z / (q_x + 1))\) in the 2D case.
There are 3 notable exceptions:

\begin{itemize}
\item
  Cholesky decomposition is improved from \(O(n^3)\) to \(O(q^2 n)\)---a
  quadratic speed-up.
\item
  Computation of \(V = K K^T\) remains \(O(n^3)\).
\item
  Some matrix products required by Newton method get a full
  \(n / (q_z + 1)\) or \(n / (q_z + 1)\) speed-up in the 2D case.
\end{itemize}

Table~\ref{tbl-comp-speed} summarises theoretical complexities across
different frameworks, including a typical generalized additive model
framework for which the penalization matrix is diagonal. This last
framework is used by the rank-reduced WH smoothing approach introduced
next, as well as the P-spline alternative used for comparison.

\begin{table}

\caption{\label{tbl-comp-speed}Compared theoretical leading-order costs
associated with the key steps in smoothing computations for several
frameworks. All cells should be read as O(\ldots).}

\centering{

\centering
\begin{tabular}[t]{|>{}l|c|c|>{}c|}
\hline
\multicolumn{1}{|c|}{\cellcolor[HTML]{008851}{\textcolor{white}{\textbf{Computation}}}} & \multicolumn{1}{c|}{\cellcolor[HTML]{008851}{\textcolor{white}{\textbf{Dense}}}} & \multicolumn{1}{c|}{\cellcolor[HTML]{008851}{\textcolor{white}{\textbf{Banded}}}} & \multicolumn{1}{c|}{\cellcolor[HTML]{008851}{\textcolor{white}{\textbf{Rank-reduced}}}}\\
\hline
$X^T W X$ & $\emptyset$ & $\emptyset$ & $np^2$\\
\hline
$X^T W z$ & $n$ & $n$ & $np$\\
\hline
$P_\lambda$ & $n^2$ & $qn$ & $p$\\
\hline
$X^TWX + P_\lambda$ & $n$ & $n$ & $p$\\
\hline
$R$ & $n^3$ & $q^2 n$ & $p^3$\\
\hline
$\hat{\boldsymbol{\beta}}_\lambda$ & $n^2$ & $qn$ & $p^2$\\
\hline
$\hat{\mathbf{y}}_\lambda = X\hat{\boldsymbol{\beta}}_\lambda$ & $\emptyset$ & $\emptyset$ & $np$\\
\hline
ML/LAML & $qn$ & $qn$ & $n$\\
\hline
$K = R^{-1}$ & $n^3$ & $qn^2$ & $p^3$\\
\hline
$V = K K^T$ & $n^3$ & $n^3$ & $p^3$\\
\hline
Brent/Nelder-Mead & $n^3$ & $q^2 n$ & $p^3$\\
\hline
Fellner-Schall & $n^3$ & $qn^2$ & $p^3$\\
\hline
Newton & $n^3$ & $n^3$ & $p^3$\\
\hline
\end{tabular}

}

\end{table}%

\subsubsection{Empirical gains}\label{empirical-gains}

\begin{figure}

\centering{

\includegraphics[width=1\linewidth,height=\textheight,keepaspectratio]{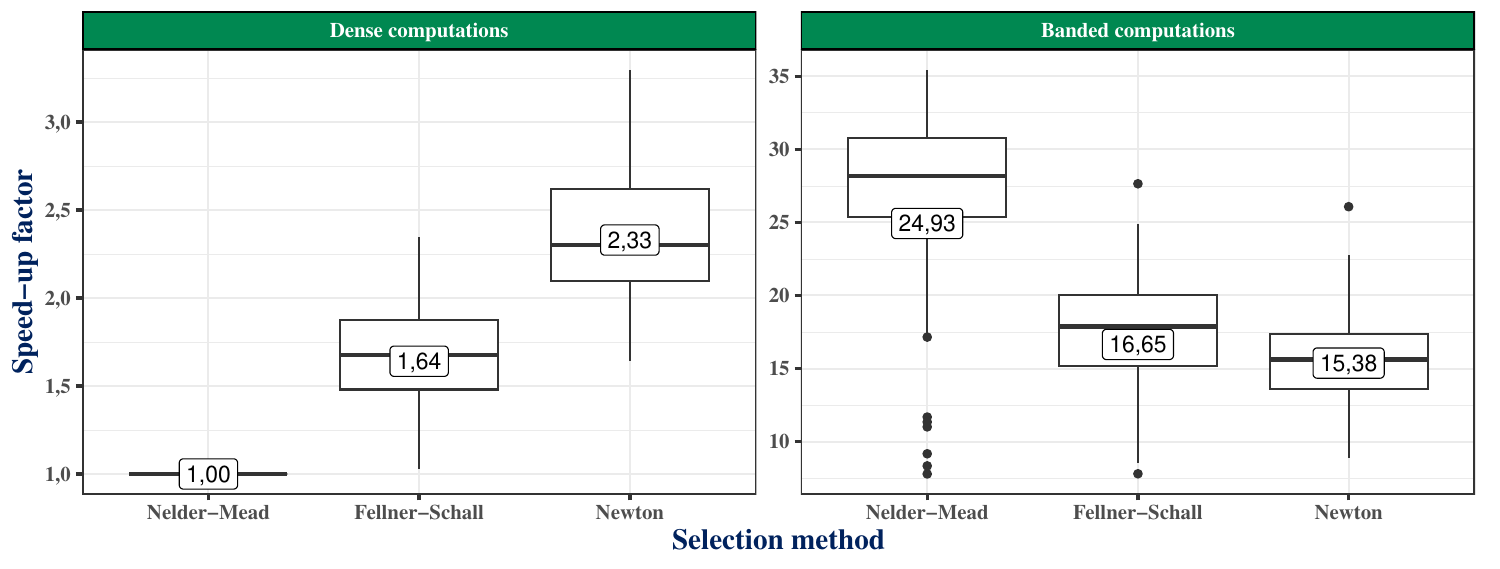}

}

\caption{\label{fig-comp-banded-dense}Computation time comparison for 2D
generalized WH smoothing with outer iteration. The speed-up factor is
computed relative to the original dense method using the Nelder-Mead
algorithm.}

\end{figure}

Figure~\ref{fig-comp-banded-dense} compares actual computation times of
WH smoothing (two-dimensional, outer iteration), showing that adapting
the implementation to exploit banded structures results in large speed
gains:

\begin{itemize}
\item
  The Nelder-Mead method benefits the most, with a 25 \(\times\) speedup
  compared to dense computation.
\item
  Newton and Fellner-Schall methods see 6.6 \(\times\) and 10 \(\times\)
  improvements, respectively, making them fall behind the Nelder-Mead
  method.
\end{itemize}

As a final advantage, Brent and Nelder-Mead heuristic methods rely
solely on banded matrices that can be stored as compact matrices of
dimensions \((q + 1)\times n\), adding further efficiency.

\subsection{Natural parameterization and rank reduction of WH
smoothing}\label{sec-vp}

Demmler and Reinsch (1975) proposed a natural parameterization for
penalized smoothers using the eigendecomposition of the penalization
matrix. This provides both an intuitive interpretation of the smoothing
mechanism and a foundation for dimension reduction via rank-restricted
estimation.

\subsubsection{One-dimensional Case}\label{one-dimensional-case-1}

In one dimension, let \(D_{n,q}^T D_{n,q} = U \Sigma U^T\) be the
eigendecomposition of the penalty matrix, where \(U\) is orthogonal and
\(\Sigma\) diagonal with non-negative eigenvalues. A change of variable
\(\boldsymbol{\theta} = U\boldsymbol{\beta}\) transforms the WH
optimization into:

\[
\hat{\mathbf{y}} = U\hat{\boldsymbol{\beta}} \quad\text{where}\quad \hat{\boldsymbol{\beta}} = \underset{\boldsymbol{\beta}}{\text{argmin}}\left\lbrace (\mathbf{y} - U\boldsymbol{\beta})^{T}W(\mathbf{y} - U\boldsymbol{\beta}) + \lambda\boldsymbol{\beta}^{T}\Sigma\boldsymbol{\beta}\right\rbrace
\]

yielding the solution:

\[
\hat{\mathbf{y}} = U(U^TWU + S_{\lambda})^{-1}U^TW\mathbf{y} \quad \text{where} \quad S_{\lambda} = \lambda\Sigma.
\] This formulation shows that WH smoothing decomposes the signal into
eigenvector components and attenuates each according to the associated
eigenvalue---the higher the eigenvalue, the stronger the shrinkage.

We refer to Section~\ref{sec-natural-parametrization} of the appendices
for graphical illustrations of:

\begin{itemize}
\item
  the basis eigenvectors of \(D_{n,q}^T D_{n,q}\);
\item
  the evolution of their effective degrees of freedom under smoothing.
\end{itemize}

These figures show that only the first few components retain substantial
degrees of freedom under moderate smoothing, motivating dimensionality
reduction.

\subsubsection{Two-dimensional case}\label{two-dimensional-case-1}

In two dimensions, the penalization matrix takes the form:

\[P_{\lambda} = \lambda_x I_{n_z} \otimes D_{n_x,q_x}^{T}D_{n_x,q_x} + \lambda_z D_{n_z,q_z}^{T}D_{n_z,q_z} \otimes I_{n_x},\]

with eigendecompositions
\(D_{n_x,q_x}^T D_{n_x,q_x} = U_x \Sigma_x U_x^T\) and
\(D_{n_z,q_z}^T D_{n_z,q_z} = U_z \Sigma_z U_z^T\).

Define \(U = U_z \otimes U_x\), and
\(\boldsymbol{\theta} = U \boldsymbol{\beta}\). Then the WH estimate
becomes:

\[
\hat{\mathbf{y}} = U(U^TWU + S_{\lambda})^{- 1}U^TW\mathbf{y} \quad \text{where} \quad S_{\lambda} = \lambda_x I_{n_z} \otimes \Sigma_x + \lambda_z \Sigma_z \otimes I_{n_x}.
\]

As in the one-dimensional case, this representation reveals how
smoothing operates via coordinate-wise shrinkage in the eigenbasis.
Section~\ref{sec-natural-parametrization} of the appendices displays the
corresponding per-parameter effective degrees of freedom.

\subsubsection{Rank reduction strategy}\label{sec-rr}

Inspection of the effective degrees of freedom reveals that many
components are heavily shrunk, especially those associated with high
eigenvalues. This suggests reducing the dimension by keeping only the
\(p < n\) components with the lowest eigenvalues.

In the one-dimensional case, the reduced-rank approximation is:

\[\hat{\mathbf{y}}_p = U_p (U_p^TWU_p + \lambda\Sigma_p)^{- 1}U_p^TW\mathbf{y}\]

where \(U_p\) and \(\Sigma_p\) consist of the first \(p\) eigenvectors
and their corresponding eigenvalues respectively.

In the two-dimensional case, we retain \(p_x\) and \(p_z\) eigenvectors
in each dimension and use:

\[\hat{\mathbf{y}}_{p_x,p_z} = U_{p_x,p_z} (U_{p_x,p_z}^TWU_{p_x,p_z} + \lambda_x I_{p_z} \otimes \Sigma_{x,p_x} + \lambda_z \Sigma_{z,p_z} \otimes I_{p_x})^{- 1}U_{p_x,p_z}^TW\mathbf{y}\]

with \(U_{p_x,p_z} = U_{z,p_z} \otimes U_{x,p_x}\). In that case, given
a target number of parameters \(p_\text{max}\), we propose selecting
\((p_x, p_z)\) such that \(p_x p_z \le p_{\text{max}}\) and
\(p_x / n_x \approx p_z / n_z\) using the rule:

\[\kappa = \sqrt{p_\text{max} /n_x n_z},\quad p_x = \lfloor \text{min}(\kappa,1)n_x\rfloor,\quad p_z = \lfloor \text{min}(\kappa,1)n_z\rfloor.\]
Adaptations for generalized WH smoothing follow by replacing
\((\mathbf{y}, W)\) with \((\mathbf{z}_k, W_k)\) in the above
expressions.

\subsubsection{Efficient computation via
GLAM}\label{efficient-computation-via-glam}

Currie, Durban, and Eilers (2006) propose a general framework,
Generalized Linear Array Models (GLAM), that exploits Kronecker
structure for efficient computations. In our context, the model matrix
\(U_{p_x,p_z}\) inherits a Kronecker product form, allowing operations
that rely on this matrix to be executed dimension-wise without explicit
construction of the full matrix. This significantly reduces memory use
and computation time in the two-dimensional rank-reduced WH framework.

\subsubsection{Impact of using the rank-reduced
basis}\label{impact-of-using-the-rank-reduced-basis}

We now evaluate the impact of the rank-reduced WH basis introduced in
Section~\ref{sec-rr} in terms of both smoothing accuracy and
computational speed. For context, results are compared against those
obtained using P-spline smoothing with the same number of basis
functions.

To ensure a fair comparison, both approaches were implemented in the
same computational framework, including the use of GLAM in the
two-dimensional case---only the structure of the basis (and hence the
model matrix) differs. The penalty structure, as well as the unpenalized
fixed effects (polynomials of degree \(q - 1\)), are identical.

In addition to the full basis of size 450 (\(30 \times 15\)), three
reduced basis of respective size 288 (\(24 \times 12\)), 128
(\(16 \times 8\)) and 32 (\(8 \times 4\)) were considered.

As in Section~\ref{sec-comp-outer-perf}, accuracy is assessed using the
relative LAML error defined in Equation~\ref{eq-delta2}. Note, however,
that since both reduced-rank and P-spline smoothers rely on different
bases and penalization matrices, their LAML expressions are different
from the one used for full-rank WH smoothing. Hence, a reduced model can
exhibit a higher LAML than the full-rank version at its selected
smoothing parameter.

\begin{figure}

\centering{

\includegraphics[width=1\linewidth,height=\textheight,keepaspectratio]{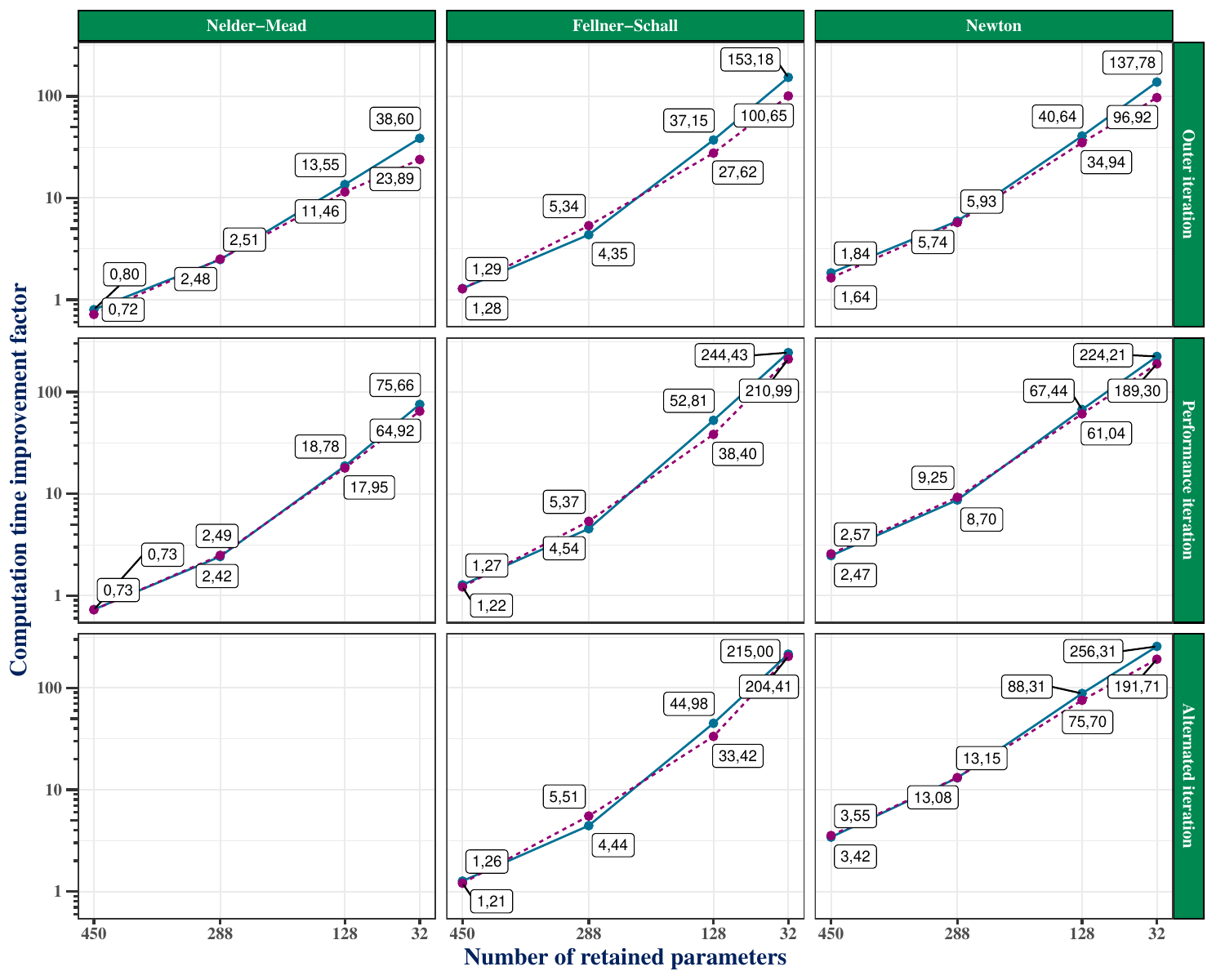}

}

\caption{\label{fig-red-speed}Computation speed improvement from WH
smoothing with a reduced-rank basis (solid lines) or P-spline basis
(dashed lines), relative to unoptimized full-rank WH smoothing, as a
function of basis size.}

\end{figure}

Figure~\ref{fig-red-speed} summarises the average speed-up achieved by
both the reduced-rank and P-spline smoothers compared to the full-rank
WH smoothing. As the number of retained parameters decreases,
computation time drops substantially. Compared to the full-rank WH
smoothing (unoptimized):

\begin{itemize}
\item
  the 128-parameter basis achieves an 88 \(\times\) speed-up;
\item
  the 32-parameter basis achieves up to 256 \(\times\) faster
  computation.
\end{itemize}

The alternate iteration and performance iteration strategies outperform
the outer iteration in the reduced setting, primarily because model
matrix construction becomes the new computational bottleneck---even with
the use of the GLAM framework. In this context, the Newton algorithm
combined with alternate iteration proves to be the most efficient, with
the generalized Fellner-Schall update being nearly as competitive for
smaller bases.

\begin{figure}

\centering{

\includegraphics[width=0.8\linewidth,height=\textheight,keepaspectratio]{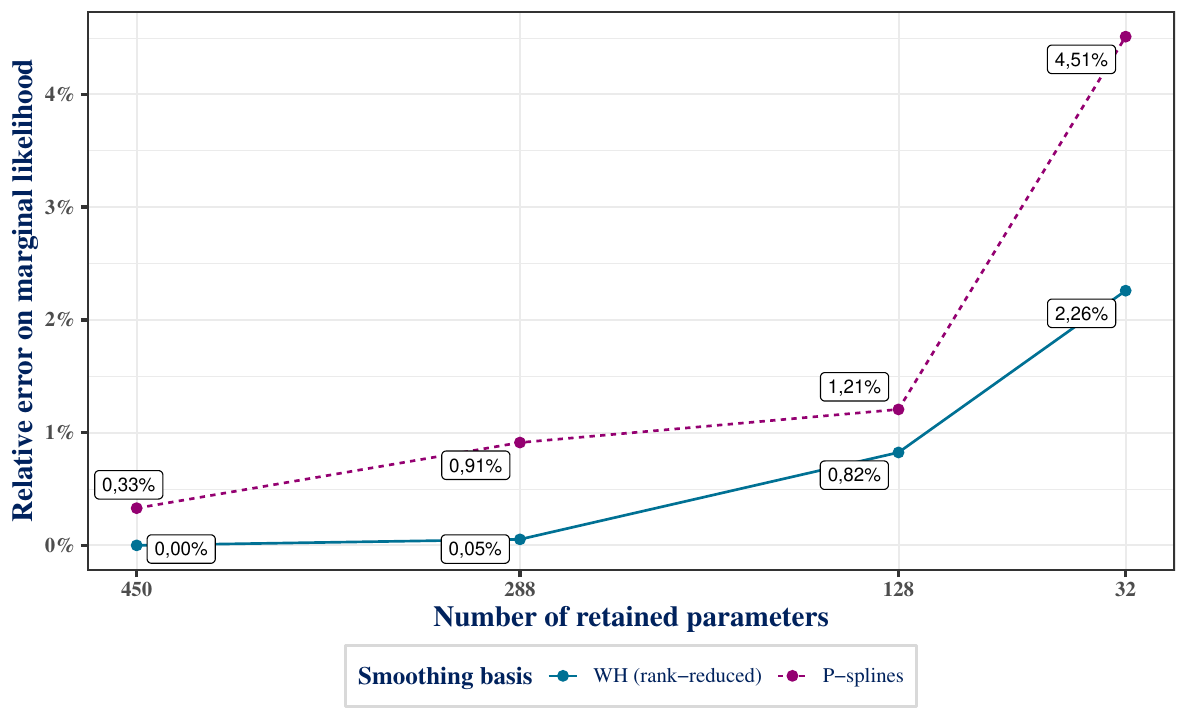}

}

\caption{\label{fig-red-ecart}Relative LAML error of WH smoothing with a
reduced-rank basis (solid lines) or a P-spline basis (dashed lines),
with respect to unoptimized full-rank WH smoothing, as a function of
basis size.}

\end{figure}

The gains in computational speed come with a moderate tradeoff in
estimation accuracy. As shown in Figure~\ref{fig-red-ecart}, the
relative LAML errors remain small:

\begin{itemize}
\item
  For the 128-parameter basis, the average error is just 0.82\%.
\item
  For the 32-parameter basis, it rises to 2.26\%.
\end{itemize}

Across all sizes, the reduced-rank WH smoother slightly outperforms the
P-spline smoother in terms of LAML error, confirming its effectiveness
as a principled dimension reduction strategy.

\section{How to extrapolate the smoothing?}\label{sec-extra}

Semi-parametric methods such as P-splines and Whittaker-Henderson (WH)
smoothing naturally allow for extrapolation---that is, predicting values
outside the range of the original data. Extrapolation is handled by
solving an extended smoothing problem where extrapolated positions are
associated with zero-weight observations.

However, in the two-dimensional case, extrapolation must be performed
carefully: constraints are needed to ensure that the extrapolated
solution remains consistent with the original smoothing result over the
observed data. Following the approach introduced by Carballo, Durban,
and Lee (2021) for P-splines, we now extend WH smoothing to support
extrapolation while also enabling the construction of credibility
intervals that capture uncertainty both inside and outside the original
observation domain.

\subsection{Defining the extrapolation of the
smoothing}\label{defining-the-extrapolation-of-the-smoothing}

Let \(\hat{\mathbf{y}}\) be the WH smoothing result obtained from an
observation vector \(\mathbf{y}\) defined over positions \(\mathbf{x}\)
(in 1D) or \((\mathbf{x}, \mathbf{z})\) (in 2D). We wish to extend
predictions to a larger domain \(\mathbf{x}+\) (or
\((\mathbf{x}+, \mathbf{z}+)\)), with \(\mathbf{x} \subset \mathbf{x}+\)
and similarly for \(\mathbf{z}\).

To preserve WH smoothing's requirement for evenly spaced points, we
assume that \(\mathbf{x}+\) and \(\mathbf{z}+\) are sequences of
consecutive integers. Let \(n_+\) be the length of \(\mathbf{x}_+\) in
the on-dimensional case. In the two-dimensional case, let \(n_{x+}\) and
\(n_{z+}\) be the lengths of \(\mathbf{x}+\) and \(\mathbf{z}+\) and
note \(n_+ = n_{x+} \times n_{z+}\).

We define matrices \(C_x\) and \(C_z\) such that each extracts the
indices of the original data from the larger domain. Specifically:
\(C_j = (O \mid I_{n_j} \mid O)\), where \(j \in \{x, z\}\) and
\(I_{n_j}\) is an identity matrix aligned with the observed positions.
Define the matrix \(C\) as:

\[C = \begin{cases}C_x & \text{in the one-dimensional case}, \\ C_z \otimes C_x & \text{in the two-dimensional case.}\end{cases}\]

Then \(C\) has the following useful properties:

\begin{itemize}
\item
  For any full-domain vector \(\mathbf{y}+\), \(C\mathbf{y}+\) returns
  the observed values only.
\item
  \(C^T \mathbf{y}\) embeds the observed values into a larger
  zero-padded vector.
\item
  \(CC^T = I_n\) and \(C^TC\) is a \(2 \times 2\) block matrix with an
  identity matrix block and zeros everywhere else.
\end{itemize}

The extrapolated WH smoothing is defined as the solution to the
following extended problem:

\begin{equation}\phantomsection\label{eq-extra1}{\hat{\mathbf{y}}_+ = \underset{\boldsymbol{\theta}_+}{\text{argmin}}\left\lbrace (\mathbf{y}_+ - \boldsymbol{\theta}_+)^TW_+(\mathbf{y}_+ - \boldsymbol{\theta}_+) + \boldsymbol{\theta}_+^TP_+\boldsymbol{\theta}_+\right\rbrace}\end{equation}

where:

\begin{itemize}
\item
  \(\mathbf{y}_+ = C^{T}\mathbf{y}\) is the extended data vector (zeros
  for unobserved points),
\item
  \(W_+ = C^{T}WC\) is the extended weight matrix (zeros for unobserved
  points),
\item
  \(P_+\) is the penalization matrix over the extended grid, defined as:
\end{itemize}

\[P_+ = \begin{cases}\lambda D_{n_{+},q}^{T}D_{n_{+},q} & \text{in the one-dimensional case,} \\ \lambda_x I_{z+} \otimes D_{n_{x+},q_x}^{T}D_{n_{x+},q_x} + \lambda_z D_{n_{z+},q_z}^{T}D_{n_{z+},q_z} \otimes I_{x+} & \text{in the two-dimensional case.}\end{cases}\]

Importantly, the smoothing parameters \(\lambda\), \(\lambda_x\), and
\(\lambda_z\) must remain fixed during extrapolation---they are
inherited from the original fit and no new information is introduced.

The fidelity term in Equation~\ref{eq-extra1} simplifies to:

\[(\mathbf{y}_+ - \boldsymbol{\theta}_+)^TW_+(\mathbf{y}_+ - \boldsymbol{\theta}_+) = (C^{T}\mathbf{y} - \boldsymbol{\theta}_+)^TC^{T}WC(C^{T}\mathbf{y} - \boldsymbol{\theta}_+) = (\mathbf{y} - \boldsymbol{\theta})^TW(\mathbf{y} - \boldsymbol{\theta})\]

where \(\boldsymbol{\theta} = C \boldsymbol{\theta}_+\). This is the
fidelity term from the original fit.

The smoothness criterion, on the other hand, now applies to the entire
extended domain, constraining the extrapolated parts of
\(\hat{\mathbf{y}}_+\) to remain smooth and consistent with the trend
learned from the data.

The same extrapolation approach applies directly to generalized WH
smoothing, simply by replacing \(\mathbf{y}\) by \(\mathbf{z}_k\) and
\(W\) by \(W_k\), obtained at convergence of the PIRLS algorithm and
setting \(\sigma^2 = 1\) in the derived credible intervals.

\subsection{Unconstrained solution for the 1D
case}\label{unconstrained-solution-for-the-1d-case}

The solution to the extrapolation problem in Equation~\ref{eq-extra1}
can be obtained directly, as in Section~\ref{sec-explicit}, by taking
derivatives with respect to \(\boldsymbol{\theta}_+\) and setting them
to zero. This yields the closed-form solution:

\[
\hat{\mathbf{y}}_+ = (W_+ + P_+)^{- 1}W_+\mathbf{y}_+ \quad\text{where}\quad \mathbf{y}_+ = C^{T}\mathbf{y} \quad\text{and}\quad W_+ = C^{T}WC. 
\]

Assuming a Bayesian model where
\(\mathbf{y}+ \mid \boldsymbol{\theta}+ \sim \mathcal{N}(\boldsymbol{\theta}+, \sigma^2 W+^{-1})\)
and \(\boldsymbol{\theta}+ \sim \mathcal{N}(0, \sigma^2 P+^{-1})\), we
obtain, as in Section~\ref{sec-CI}, the following credible interval:

\[\mathbb{E}(\mathbf{y}_+) \mid \mathbf{y}_+ \in \left[(W_+ + P_+)^{- 1}W_+\mathbf{y}_+ \pm \Phi^{- 1}\left(1 -\alpha / 2\right)\sqrt{\sigma^2\textbf{diag}\left\lbrace(W_+ + P_+)^{-1}\right\rbrace}\right].\]

To get a better understanding about how the variance-covariance matrix
\(V_+ = (W_+ + P_+)^{- 1}\) for the unconstrained extrapolation problem
of Equation~\ref{eq-extra1} is related to the variance-covariance matrix
\(V = (W + P_\lambda)^{- 1}\) of the original smoothing problem,
introduce matrices \(\overline{C}_j\) (for \(j \in {x, z}\)) which
selects the rows in the extrapolated domain that are not part of the
original data and define:

\[\quad \overline{C} = \begin{cases}\overline{C}_x & \text{in the one-dimensional case}, \\ \overline{C}_z \otimes \overline{C}_x & \text{in the two-dimensional case},\end{cases} \quad\text{and}\quad Q = \begin{bmatrix} C \\ \overline{C} \end{bmatrix}.\]

With this definition, \(Q\) is a permutation matrix moving observed
positions to the top.

In the unidimensional case, the extended difference matrix \(D_{n_+,q}\)
takes the block-wise form:

\[D_{n_+,q} = \begin{bmatrix}D_{2-} & D_{1-} & 0 \\ 0 & D_{n,q} & 0 \\ 0 &D_{1+} & D_{2+}\\ \end{bmatrix} = Q^T\begin{bmatrix}D_{n,q} & 0 \\ D_1 & D_2 \\ \end{bmatrix}Q \quad\text{where} \; D_1 = \begin{bmatrix}D_{1-} \\ D_{1+} \\ \end{bmatrix} \;\text{and}\; D_2 = \begin{bmatrix}D_{2-} & 0 \\ 0 & D_{2+} \\ \end{bmatrix}.\]

The extended weight and penalization matrices may be rewritten:

\[W_+ = Q^T\begin{bmatrix}W & 0 \\ 0 & 0 \\ \end{bmatrix}Q \quad\text{and}\quad P_+ = D_{n_+,q}^TD_{n_+,q} = \lambda Q^T\begin{bmatrix}P_\lambda + P_+^{11} & P_+^{12} \\ P_+^{21} & P_+^{22} \\ \end{bmatrix}Q\]

where \(P_+^{ij} = \lambda D_i^TD_j\), for \(i,j\in\{1,2\}\).

This block structure allows us to apply standard results for partitioned
matrix inverses to derive:

\[V_+ = Q^T\begin{bmatrix}V_+^{11} & V_+^{12}\\ V_+^{21} & V_+^{22} \\ \end{bmatrix}Q = Q^T\begin{bmatrix}V_+^{11} & - V_+^{11}P_+^{12}(P_+^{22})^{- 1} \\ - (P_+^{22})^{- 1}P_+^{21}V_+^{11} & (P_+^{22})^{- 1}P_+^{21}V_+^{11}P_+^{12}(P_+^{22})^{- 1} + (P_+^{22})^{- 1}\end{bmatrix}Q\]

with
\(V_+^{11} = [W + P_\lambda + P_+^{11} - P_+^{12}(P_+^{22})^{- 1}P_+^{21}]^{- 1}\).

From the above, we retrieve:

\[C\hat{\mathbf{y}}_+ = CV_+W_+y_+ = CQ^TV_+QC^TWy = V_+^{11}Wy.\]

This coincides with the original fit \(\hat{\mathbf{y}}\) only if
\(V_+^{11} = V\). In general, this equality does not hold, since the
extrapolation solution minimizes the total smoothness of the extended
vector, not just of the observed part.

In \(V_+^{22}\), we identify:

\begin{itemize}
\item
  a propagation term:
  \((P_+^{22})^{-1} P_+^{21} V_+^{11} P_+^{12} (P_+^{22})^{-1}\),
  capturing the uncertainty transferred from the known part to the
  extrapolated part;
\item
  an innovation error term: \((P_+^{22})^{- 1}\) associated with the
  prior on the extrapolated coefficients
  \(\overline{C} \hat{\mathbf{y}}_+\).
\end{itemize}

In the one-dimensional case, \(D_2\) is block-diagonal with invertible
triangular blocks, so:

\[P_+^{11} - P_+^{12}(P_+^{22})^{- 1}P_+^{21} = D_1^TD_1 - D_1^TD_2(D_2^TD_2)^{- 1}D2^TD_1 = 0\]

which means that \(V_+^{11} = (W + P_\lambda)^{- 1} = V\). This confirms
the result from Carballo et al. (2021), namely that with a
difference-based penalty, a perfectly smooth extrapolation that leaves
the original fit unchanged can always be constructed in the
one-dimensional case.

\begin{figure}

\centering{

\includegraphics[width=0.8\linewidth,height=\textheight,keepaspectratio]{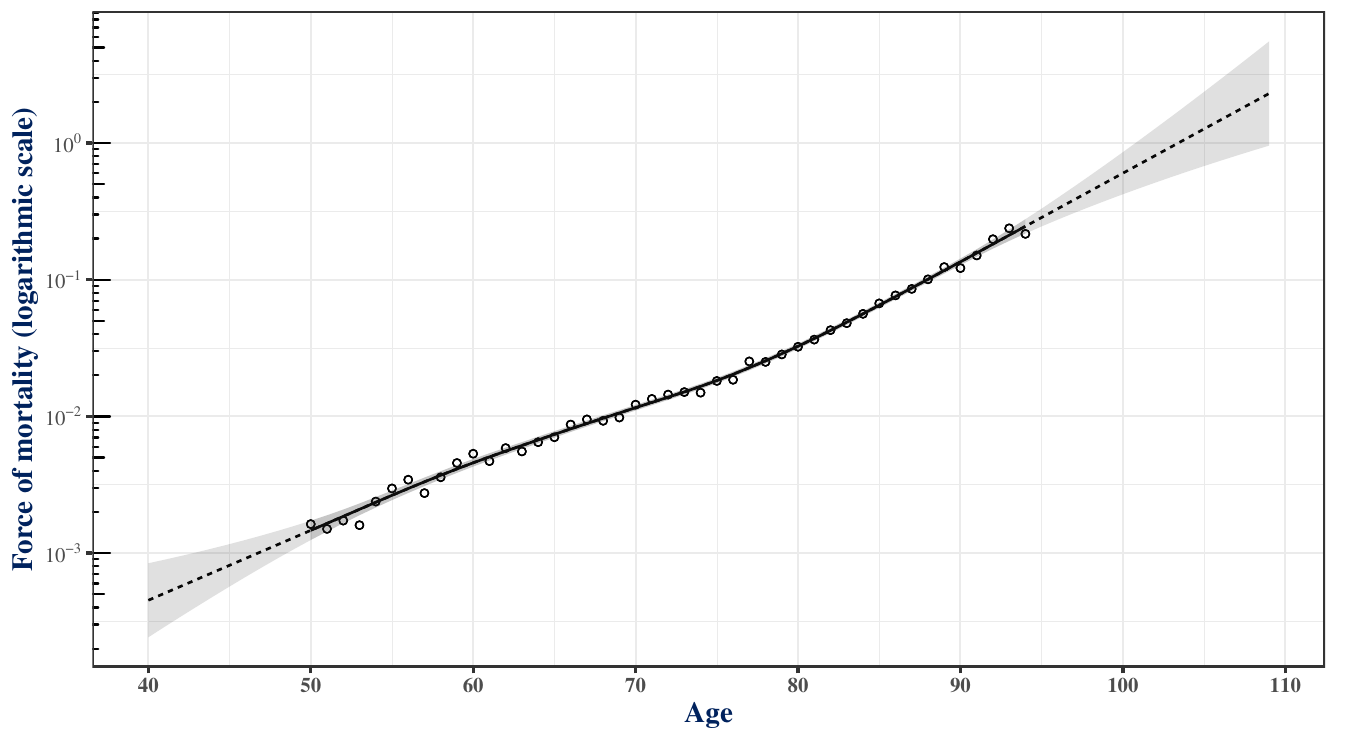}

}

\caption{\label{fig-extra1}Extrapolation of one-dimensional WH
smoothing. The smoother is extrapolated on both sides of the initial
observation range following a polynomial of degree q - 1 (in this case a
straight line as q = 2).}

\end{figure}

This behaviour is illustrated in Figure~\ref{fig-extra1}, which shows
the extrapolated fit (with \(q = 2\)) obtained from generalized WH
smoothing applied to the annuity portfolio used previously. The
extrapolation follows a straight line---the polynomial of degree
\(q - 1 = 1\)---and joins smoothly with the original curve.

\subsection{Constrained solution for the 2D
case}\label{constrained-solution-for-the-2d-case}

In the two-dimensional case, while the extended penalization matrix
\(P_+\) still takes the same structure as previously described, the
expressions of its block components \(P_+^{11}\), \(P_+^{12}\),
\(P_+^{21}\) and \(P_+^{22}\) are more complex. In particular, we no
longer have the simplification
\(P_+^{11} - P_+^{12}(P_+^{22})^{-1}P_+^{21} = 0\), therefore
\(V_+^{11} \ne V\) and \(C\hat{\mathbf{y}}_+ \ne \hat{\mathbf{y}}\).
Solving the unconstrained extrapolation problem thus leads to a
modification of the estimated coefficients for the observed data
positions, as demonstrated by Carballo, Durban, and Lee (2021).

This difference arises because, unlike the one-dimensional case, the
smoothness criterion in two dimensions penalizes both rows and columns
simultaneously, making it impossible to extrapolate without increasing
the penalization. Since no new data is introduced in the extrapolated
region, the smoothness criterion weighs more heavily in the
optimization, prompting adjustments to the originally fitted values in
order to produce a globally smoother estimate.

To address this, we follow the approach proposed by Carballo, Durban,
and Lee (2021) and formulate a constrained optimization problem that
enforces preservation of the original fitted values in the smoothing
region. This is done by introducing a Lagrange multiplier
\(\boldsymbol{\omega}\) and solving the following constrained problem:

\[(\hat{\mathbf{y}}_+^\ast, \hat{\boldsymbol{\omega}}) = \underset{\boldsymbol{\theta}_+^\ast, \boldsymbol{\omega}}{\text{argmin}}\left\lbrace (\mathbf{y}_+ - \boldsymbol{\theta}_+^\ast)^TW_+(\mathbf{y}_+ - \boldsymbol{\theta}_+^\ast) + \boldsymbol{\theta}_+^{\ast T}P_{+}\boldsymbol{\theta}_+^\ast + 2 \boldsymbol{\omega}^T(C\boldsymbol{\theta}_+^\ast - \hat{\mathbf{y}})\right\rbrace.\]

This optimization admits a closed-form solution for the constrained
extrapolated estimator \(\hat{\mathbf{y}}_+^\ast\) as a linear
transformation of \(\hat{\mathbf{y}}\). The derivation details are
provided in Section~\ref{sec-extrapolation-computation} of the
appendices. The final form is:

\[\hat{\mathbf{y}}^\ast_+ = Q^T\begin{bmatrix}I \\ - (P_+^{22})^{- 1}P_+^{21}\end{bmatrix}\hat{\mathbf{y}}\]

and the associated variance-covariance matrix is:

\[V_+^{\ast} = Q^T\begin{bmatrix}V & - V P_+^{12}(P_+^{22})^{- 1} \\ - (P_+^{22})^{- 1}P_+^{21}V & (P_+^{22})^{- 1}P_+^{21}V P_+^{12}(P_+^{22})^{- 1} + (P_+^{22})^{- 1} \end{bmatrix}Q.\]

This formulation differs from the variance matrix of the unconstrained
solution. Indeed, it enforces the constraint that the initial
coefficients remain unchanged, as reflected by the presence of \(V\)
(the original variance matrix) instead of \(V_+^{11}\). The
corresponding credible intervals are:

\[\mathbb{E}(\mathbf{y}_+) \mid \mathbf{y}_+ \in \left[\hat{\mathbf{y}}^\ast_+ \pm \Phi^{- 1}\left(1 -\alpha / 2\right)\sqrt{\sigma^2\textbf{diag}(V_+^{\ast})}\right].\]

The following figures illustrate the impact of the constrained
extrapolation procedure discussed above, using the LTC portfolio of
100,000 policyholders as a case study.

\begin{figure}

\centering{

\includegraphics[width=1\linewidth,height=\textheight,keepaspectratio]{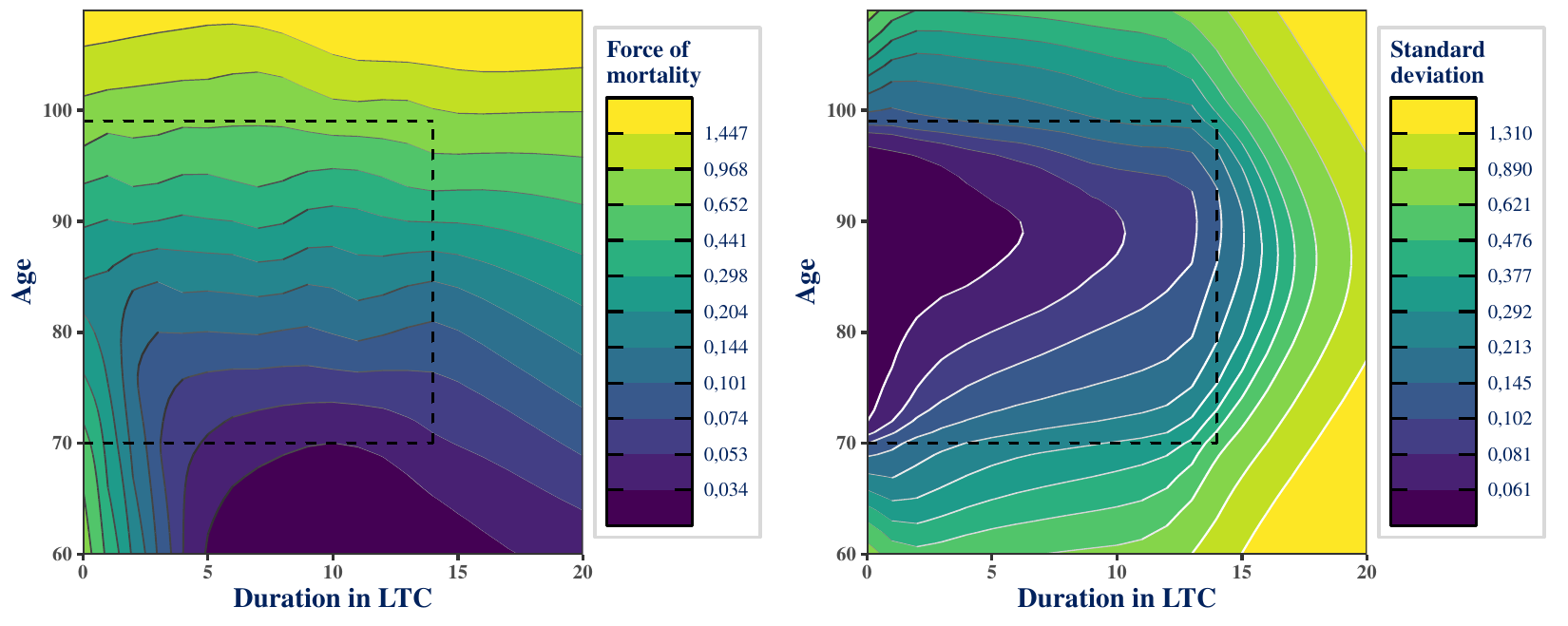}

}

\caption{\label{fig-extra2}Constrained extrapolation of 2D WH smoothing.
The contour lines of mortality rates and the associated standard
deviation are depicted. The dotted lines delimit the boundaries of the
initial smoothing region.}

\end{figure}

\begin{itemize}
\item
  Figure~\ref{fig-extra2}, left (mortality rates): this panel shows the
  estimated mortality rates obtained after applying the constrained
  extrapolation procedure to the two-dimensional WH smoothing model. The
  dotted lines indicate the boundaries of the original smoothing region.
  Visually, the transition from the smoothing region to the extrapolated
  area is seamless---the extrapolated surface naturally extends the
  smoothed mortality rates while respecting the original fitted values
  within the data range.
\item
  Figure~\ref{fig-extra2}, right (standard deviation): this panel
  displays the posterior standard deviation (or credible interval width)
  associated with the extrapolated estimates. It reflects both the
  uncertainty from the original smoothing and the innovation error
  introduced in the extrapolated region. As expected, the standard
  deviation increases as we move away from the observed region,
  illustrating growing uncertainty about farther values.
\end{itemize}

\begin{figure}

\centering{

\includegraphics[width=0.7\linewidth,height=\textheight,keepaspectratio]{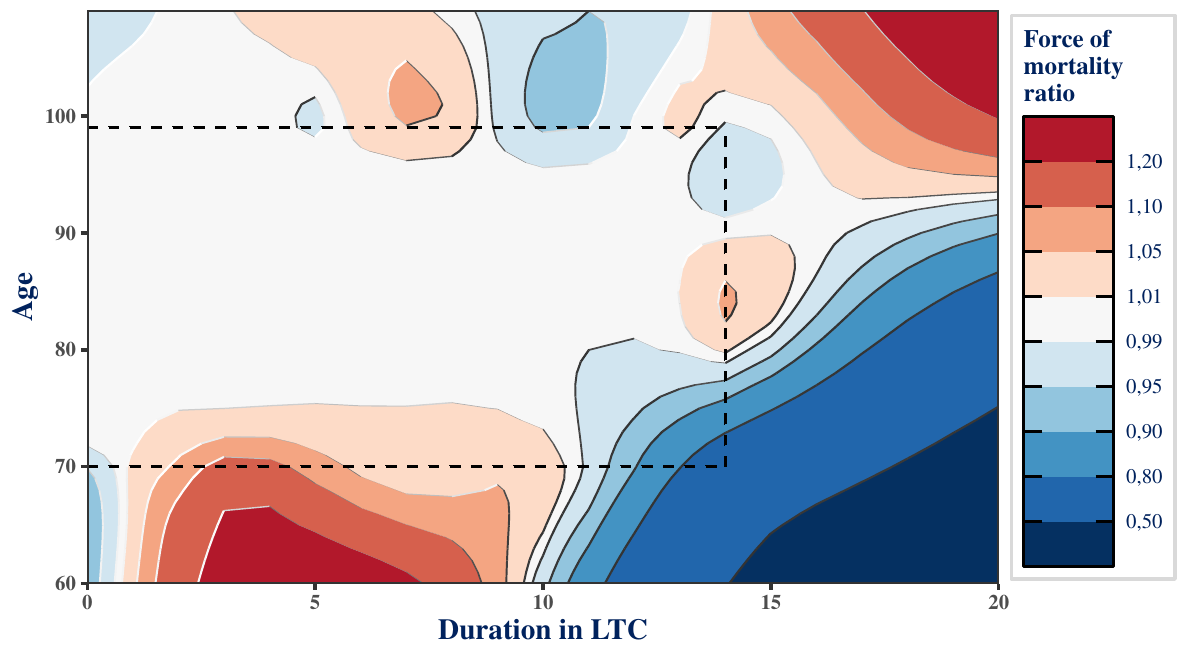}

}

\caption{\label{fig-extra2ratio}Ratio of mortality rates resulting from
the extrapolation of 2D WH smoothing. The numerator corresponds to the
unconstrained extrapolation and the denominator to the constrained
extrapolation presented in Figure~\ref{fig-extra2}.}

\end{figure}

\begin{itemize}
\tightlist
\item
  Figure~\ref{fig-extra2ratio} (ratio of mortality rates): this heatmap
  shows the pointwise ratio between the unconstrained and constrained
  extrapolation of the mortality rates. A value above 1 indicates that
  the unconstrained version overshoots the constrained one at that
  location, while values below 1 indicate underestimation. We observe
  that discrepancies exist not only in the extrapolated region but also
  within the original data region---confirming that the unconstrained
  approach distorts the original estimates in order to achieve overall
  smoothness.
\end{itemize}

\begin{figure}

\centering{

\includegraphics[width=1\linewidth,height=\textheight,keepaspectratio]{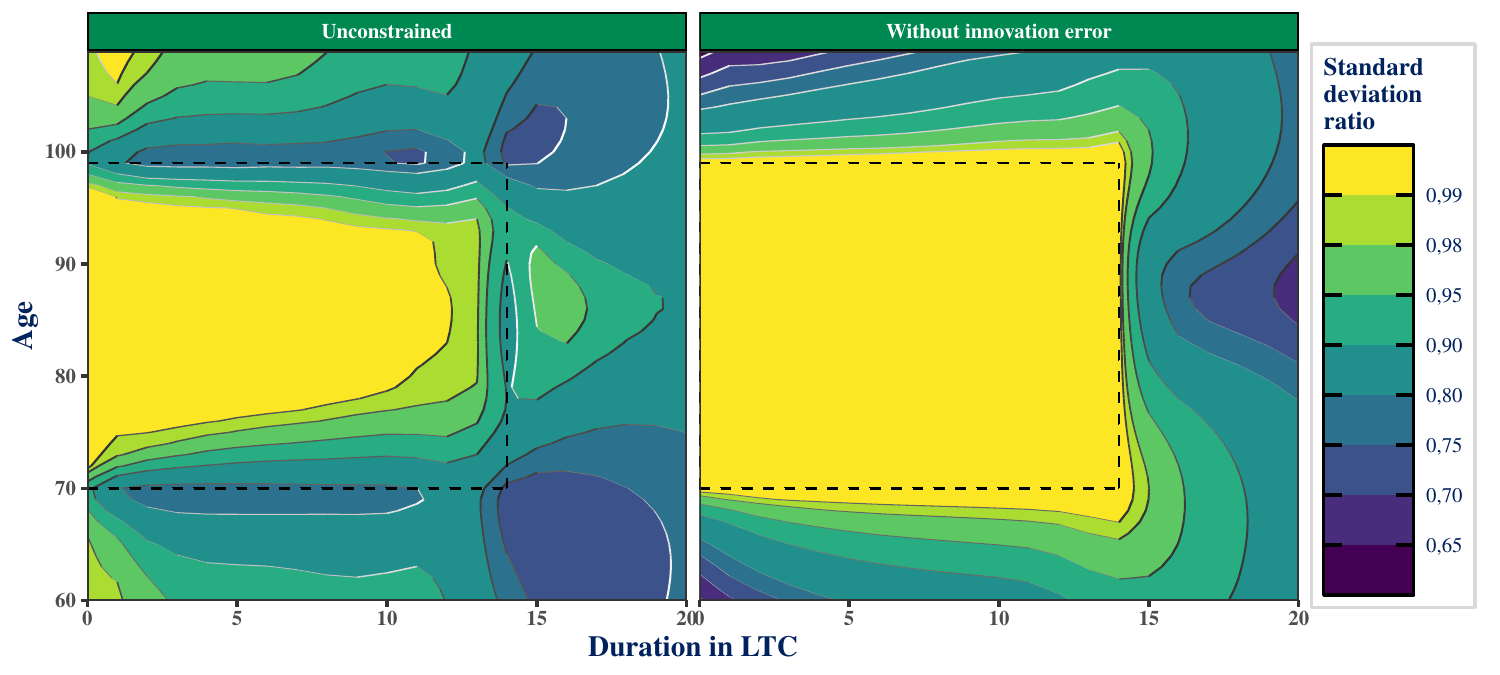}

}

\caption{\label{fig-extra2ratiostd}Ratio of standard deviation of
log-mortality rates from the three extrapolation methods. Left:
unconstrained vs constrained with innovation error. Right: constrained
without vs with innovation error. In both, the denominator is the fully
constrained method of Figure~\ref{fig-extra2}.}

\end{figure}

\begin{itemize}
\item
  Figure~\ref{fig-extra2ratiostd} (ratio of standard deviations): this
  final figure includes two panels comparing uncertainty estimates.

  \begin{itemize}
  \item
    Left panel: ratio of standard deviation from the unconstrained
    extrapolation over that from the constrained extrapolation
    (including innovation error). The unconstrained version
    underestimate the actual uncertainty not only in the extrapolated
    region but also within the original data region, again reflecting
    the adjustments made to the original estimates in order to achieve
    overall smoothness.
  \item
    Right panel: ratio of standard deviation from the constrained
    extrapolation without innovation error over the fully constrained
    version with innovation error. This illustrates the contribution of
    the innovation error to the total uncertainty---it is substantial
    and should not be neglected.
  \end{itemize}
\end{itemize}

\section{Discussion}\label{sec-discussion}

\subsubsection{Choosing the order of the
penalization}\label{choosing-the-order-of-the-penalization}

Throughout this work, we have assumed second-order difference matrices
for penalization. This choice is both standard and meaningful: from a
Bayesian perspective, it corresponds to a prior belief that the
log-transformed quantity of interest evolves linearly, which implies
exponential behaviour on the original scale---consistent with actuarial
models such as Gompertz.

The difference order directly shapes both the estimated trend and its
extrapolation: higher-order penalties allow for more flexibility, but
may induce unstable or erratic behaviour outside the data range. While
Whittaker originally used third-order differences and higher orders can
marginally improve model fit according to information criteria such as
AIC, second-order penalties typically offer a robust compromise between
smoothness, interpretability, and extrapolation stability. A detailed
evaluation is provided in Section~\ref{sec-difference-order} of the
appendices.

\subsubsection{Summary of contributions}\label{summary-of-contributions}

This paper revisits the classical Whittaker-Henderson (WH) smoothing
approach through the lens of modern statistical modelling. Each section
brought forward a key practical insight:

\begin{itemize}
\item
  Section~\ref{sec-CI} established that WH smoothing is more than an
  empirical method. It has a firm Bayesian foundation. Under Gaussian
  assumptions, credibility intervals may be derived and used as
  practical substitutes to confidence intervals.
\item
  Section~\ref{sec-vectors} clarified how to construct observation and
  weight vectors in survival analysis models: using log-crude rates as
  observations and event counts as weights yields a sound statistical
  formulation.
\item
  Section~\ref{sec-generalisation} introduced generalized WH smoothing,
  in which the penalization is applied directly to the likelihood rather
  than a normal approximation. This refined method yields more accurate
  results, especially in situations where the available data volume is
  limited but the number of combinations is high, such as in the
  two-dimensional case.
\item
  Section~\ref{sec-selection} advocated for smoothing parameter
  selection via marginal likelihood (or its Laplace approximation,
  LAML), offering a principled and robust alternative to heuristic
  criteria like AIC or GCV.
\item
  Section~\ref{sec-reduction} presented two computational improvements:
  one exploits the banded structure of WH matrices to reduce runtime by
  up to a factor of 25; the other relies on reduced-rank smoothing
  basis, leading to even faster estimation (up to 250 \(\times\)
  speed-up) with limited loss in accuracy, slightly outperforming
  P-splines.
\item
  Section~\ref{sec-extra} addressed extrapolation: while WH smoothing
  naturally extends beyond the data range, constraints are needed in two
  dimensions to preserve the original fit. We proposed a method to
  extrapolate while accounting for both structural uncertainty and
  innovation error, and provided credible intervals accordingly.
\end{itemize}

All these techniques are available in the \texttt{WH} package for the
statistical software \(\mathtt{R}\) (R Core Team 2025), including
automated smoothing parameter selection and constrained extrapolation
with uncertainty quantification.

\subsubsection{Limitations and outlook}\label{limitations-and-outlook}

Despite its strong practical appeal, WH smoothing has limitations that
suggest several avenues for future work:

\begin{itemize}
\item
  Regular spacing requirement: WH smoothing assumes evenly spaced
  observations, which aligns well with standard life insurance grids
  (age and/or duration). However, this is less suitable when events are
  concentrated in a short period, such as in disability or long-term
  care claims. One solution is to combine finer discretization in early
  durations with methods like P-splines that accommodate irregular
  grids. Alternatively, and adaptive WH smoothing procedure (based on
  the ideas in Ruppert and Carroll 2000; Krivobokova, Crainiceanu, and
  Kauermann 2008) could offer a way to retain regular spacing while
  varying the smoothness locally.
\item
  Limited covariate handling: The basic WH framework does not
  accommodate additional explanatory variables (e.g., gender or policy
  features). However, WH smoothing can be extended using ideas from
  smoothing spline ANOVA and hierarchical models (Lee and Durban 2011;
  Gu 2013), allowing for structured random effects and flexible
  interactions. This opens the door to richer, more personalized
  experience modelling while preserving interpretability.
\end{itemize}

In sum, revisiting WH smoothing through a modern lens reinforces its
theoretical foundations and offers practitioners fast, transparent, and
adaptable tools for experience modelling. It remains a compelling
alternative to more recent---yet often more opaque---techniques when
working with evenly spaced discrete data.

\section*{References}\label{references}
\addcontentsline{toc}{section}{References}

\phantomsection\label{refs}

\section*{Appendices}\label{appendices}
\addcontentsline{toc}{section}{Appendices}

\appendix
\renewcommand{\thesubsection}{\Alph{subsection}}
\setcounter{secnumdepth}{2}
\setcounter{subsection}{0}

\subsection{Exposure computation in the survival analysis
framework}\label{sec-expo}

This appendix outlines the methodology used to compute central exposure
to risk in survival analysis, for both univariate and bivariate
settings.

The derivation relies on standard assumptions of left truncation and
non-informative right censoring, and models the hazard rate as piecewise
constant over one-year intervals. This discretization allows
reformulating the continuous-time log-likelihood in terms of aggregated
death counts and exposure durations.

In the one-dimensional case, exposure corresponds to the total time
under observation within each integer age interval. The same principle
extends naturally to the two-dimensional case, such as mortality
modelling by age and duration in LTC, where exposure is computed over
the age-duration grid using the same piecewise constant assumption.

These quantities---death counts and central exposures---form the inputs
to the generalized Whittaker-Henderson smoothing approach used
throughout the paper.

\subsubsection{One-dimensional case}\label{one-dimensional-case-2}

Consider the observation of \(m\) individuals in a longitudinal study
subject to left truncation and non-informative right censoring. Suppose
we aim to estimate a distribution that depends on only one continuous
explanatory variable, denoted by \(x\). One may for example think of a
mortality distribution with the explanatory variable of interest \(x\)
representing age. Such a distribution is fully characterized by either
of the following quantities:

\begin{itemize}
\tightlist
\item
  the cumulative distribution function \(F(x)\) or its complement, the
  survival function \(S(x) = 1 - F(x)\),
\item
  the associated probability density function
  \(f(x) = - \frac{\text{d}}{\text{d}x}S(x)\),
\item
  the instantaneous hazard function
  \(\mu(x) = - \frac{\text{d}}{\text{d}x}\ln S(x)\).
\end{itemize}

Those 3 quantities are related by the following relationships:

\begin{equation}\phantomsection\label{eq-relationships}{
S(x) = \exp\left(\underset{u = 0}{\overset{x}{\int}}\mu(u)\text{d}u\right)\quad \text{and} \quad f(x) = \mu(x)S(x).
}\end{equation}

Suppose that the considered distribution depends on a vector of
parameters \(\boldsymbol{\theta}\) estimated using maximum likelihood.
The likelihood associated with the observation of the individuals takes
the form:

\begin{equation}\phantomsection\label{eq-vraisemblance1-appendix}{
\mathcal{L}(\boldsymbol{\theta}) = \underset{i = 1}{\overset{m}{\prod}} \left[\frac{f(x_i + t_i,\boldsymbol{\theta})}{S(x_i,\boldsymbol{\theta})}\right]^{\delta_i}\left[\frac{S(x_i + t_i,\boldsymbol{\theta})}{S(x_i,\boldsymbol{\theta})}\right]^{1 - \delta_i}
}\end{equation}

where \(x_i\) represents the age at the start of observation, \(t_i\)
represents the duration of observation for individual \(i\) and
\(\delta_i\) is the indicator of event observation, which takes the
value 1 if the event of interest is observed and 0 if the observation is
instead censored. We will not go into the details of how these three
quantities are derived, however they should take into account
individual-specific information such as the subscription date, lapse
date if applicable, as well as the global characteristics of the product
such as the presence of a waiting period or medical selection
phenomenon, and the choice of a restricted observation period due to
delays in the reporting of event of interests. These factors typically
result in a narrower effective observation window than the actual time
individuals spend in the portfolio.

Using Equation~\ref{eq-relationships}, the log-likelihood associated
with Equation~\ref{eq-vraisemblance1-appendix} can be rewritten using
only the instantaneous hazard function (also known as force of mortality
in the case of the death risk):

\begin{equation}\phantomsection\label{eq-vraisemblance2-appendix}{
\ell(\boldsymbol{\theta}) = \underset{i = 1}{\overset{m}{\sum}} \left[\delta_i \ln\mu(x_i + t_i,\boldsymbol{\theta}) - \underset{u = 0}{\overset{t_i}{\int}}\mu(x_i + u,\boldsymbol{\theta})\text{d}u\right]
}\end{equation}

We discretize the problem by assuming that the mortality rate is
piecewise constant over one-year intervals between two integer ages or
more formally \(\mu(x + \epsilon) = \mu(x)\) for all
\(x \in \mathbb{N}\) and \(\epsilon \in [0,1[\). Further note that, if
\(\mathbf{1}\) denotes the indicator function, then for any
\(x_{\min} \leq a < x_{\max}\), we have
\(\sum_{x = x_{\min}}^{x_{\max}} \mathbf{1}(x \leq a < x + 1) = 1\),
where \(x_{\min} = \min(\mathbf{x})\) and
\(x_{\max} = \max(\mathbf{x})\).
Equation~\ref{eq-vraisemblance2-appendix} may therefore be rewritten as:

\[
\begin{aligned}
\ell(\boldsymbol{\theta}) = \underset{i = 1}{\overset{m}{\sum}} &\left[\underset{x = x_{\min}}{\overset{x_{\max}}{\sum}} \delta_i\mathbf{1}(x \le x_i + t_i < x + 1)\ln\mu(x_i + t_i,\boldsymbol{\theta})\right. \\ 
&- \left.\underset{u = 0}{\overset{t_i}{\int}}\underset{x = x_{\min}}{\overset{x_{\max}}{\sum}} \mathbf{1}(x \le x_i + u < x + 1)\mu(x_i + u,\boldsymbol{\theta})\text{d}u\right].
\end{aligned}
\]

The assumption of piecewise constant mortality rates implies that:

\[
\begin{aligned}
\mathbf{1}(x \le x_i + t_i < x + 1) \ln\mu(x_i + t_i,\boldsymbol{\theta}) &= \mathbf{1}(x \le x_i + t_i < x + 1) \ln\mu(x,\boldsymbol{\theta})\quad\text{and}\\ \mathbf{1}(x \le x_i + u < x + 1)\mu(x_i + u,\boldsymbol{\theta}) &= \mathbf{1}(x \le x_i + u < x + 1) \mu(x,\boldsymbol{\theta}).
\end{aligned}
\]

It is then possible to interchange the two summations to obtain the
following expressions:

\[
\begin{aligned}
\ell(\boldsymbol{\theta}) &= \underset{x = x_{\min}}{\overset{x_{\max}}{\sum}} \left[\ln\mu(x,\boldsymbol{\theta}) d(x) - \mu(x,\boldsymbol{\theta}) e_c(x)\right] \quad \text{where} \\
d(x) & = \underset{i = 1}{\overset{m}{\sum}} \delta_i \mathbf{1}(x \le x_i + t_i < x + 1) \quad \text{and} \\
e_c(x) & = \underset{i = 1}{\overset{m}{\sum}}\underset{u = 0}{\overset{t_i}{\int}}\mathbf{1}(x \le x_i + u < x + 1)\text{d}u = \underset{i = 1}{\overset{m}{\sum}} \left[\min(t_i, x - x_i + 1) - \max(0, x - x_i)\right]^+
\end{aligned}
\]

by denoting \(a^+ = \max(a, 0)\), where \(d(x)\) and \(e_c(x)\)
correspond to the number of observed deaths between ages \(x\) and
\(x + 1\) and the sum of observation durations of individuals between
these ages, respectively. The latter quantity is also known as central
exposure to risk.

\subsubsection{Two-dimensional case}\label{two-dimensional-case-2}

The extension of the proposed approach to the two-dimensional framework
requires only minor adjustments to the previous reasoning. Let
\(z_{\min} = \min(\mathbf{z})\) and \(z_{\max} = \max(\mathbf{z})\). The
piecewise constant assumption for the mortality rate needs to be
extended to the second dimension. Formally, assume that
\(\mu(x + \epsilon, z + \xi) = \mu(x, z)\) for all pairs
\(x, z \in \mathbb{N}\) and \(\epsilon, \xi \in [0,1[\). The sums
involving the variable \(x\) are then replaced by double sums
considering all combinations of \(x\) and \(z\). The log-likelihood
becomes:

\[
\begin{aligned}
\ell(\boldsymbol{\theta}) &= \underset{x = x_{\min}}{\overset{x_{\max}}{\sum}} \underset{z = z_{\min}}{\overset{z_{\max}}{\sum}}\left[\ln\mu(x,z,\boldsymbol{\theta}) d(x,z) - \mu(x,z,\boldsymbol{\theta}) \mathbf{e_c}(x,z)\right] \quad \text{where} \\
d(x,z) & = \underset{i = 1}{\overset{m}{\sum}} \delta_i \mathbf{1}(x \le x_i + t_i < x + 1) \mathbf{1}(z \le z_i + t_i < z + 1) \quad \text{and}\\
\mathbf{e_c}(x,z) & = \underset{i = 1}{\overset{m}{\sum}}\underset{u = 0}{\overset{t_i}{\int}}\mathbf{1}(x \le x_i + u < x + 1)\mathbf{1}(z \le z_i + u < z + 1)\text{d}u \\
& = \underset{i = 1}{\overset{m}{\sum}} \left[\min(t_i, x + 1 - x_i, z + 1 - z_i) - \max(0, x - x_i, z - z_i)\right]^+.
\end{aligned}
\]

\subsection{Simulated datasets}\label{sec-datasets}

This appendix details the simulation process used to generate the
datasets on which the comparative analysis of accuracy and computational
efficiency is based.

\subsubsection*{General approach}\label{general-approach}

The simulated datasets were constructed following a five-step
methodology:

\begin{enumerate}
\def\labelenumi{\arabic{enumi}.}
\item
  Define hypothetical underlying laws to serve as the ground truth.
\item
  Generate an initial population of insured individuals.
\item
  Simulate life outcomes for each individual.
\item
  Extract samples of a predetermined size from the simulated population.
\item
  Compute aggregated event counts and central exposures for each sample.
\end{enumerate}

\subsubsection*{Defining the underlying
laws}\label{defining-the-underlying-laws}

The synthetic laws used in the simulations are not meant to be accurate
representations of real-world phenomena. However, they incorporate
features commonly observed in mortality and long-term care (LTC)
experience studies to ensure plausible dynamics for testing purposes.

\subsubsection*{General population
mortality}\label{general-population-mortality}

Mortality rates are derived from the Human Mortality Database for France
in 2019. The dataset includes death counts and central exposure to risk
by age (0 to 109) and gender. A smoothing algorithm is applied to reduce
sampling noise, particularly at younger ages. The resulting rates define
the general population mortality.

\subsubsection*{Insured population
mortality}\label{insured-population-mortality}

To reflect the well-documented observation that insured individuals
generally exhibit lower mortality than the general
population---especially at younger ages---we apply a smooth logistic
adjustment factor to the general population mortality. This factor
transitions from 40\% at age 30 to 100\% at age 90, with a midpoint of
70\% at age 60. The adjusted rates define the insured population
mortality, which is used for all annuity simulations.

\subsubsection*{Long-Term Care (LTC) transition and mortality
laws}\label{long-term-care-ltc-transition-and-mortality-laws}

For LTC simulations, assumptions are needed for:

\begin{itemize}
\item
  Autonomous mortality (i.e., mortality of individuals not in LTC),
\item
  Incidence of entry into LTC,
\item
  Mortality within LTC.
\end{itemize}

We draw upon assumptions from a technical note published by the
reinsurer SCOR in 1995, adapted to use the previously defined insured
population mortality \(q_\text{ref}(x, g)\) for age \(x\) and gender
\(g\):

\begin{itemize}
\item
  \(q_a(x,g) = 0.8 \times q_\text{ref}(x,g)\) (autonomous mortality),
\item
  \(i(x) = 5.535 \times 10^{-3} \exp[(x - 52) / 8]\) (LTC incidence),
\item
  \(q_i(x,g) = 2 \times q_\text{ref}(x,g) + 0.035\) (initial LTC
  mortality).
\end{itemize}

We refine the disabled mortality law to include a shock at LTC onset,
reflecting heightened mortality during the first years in LTC, caused by
cancer-related admissions (see for example Biessy 2015 for details about
this phenomenon and impacts on curves for mortality in LTC). The refined
mortality is:

\[q_i(x,t,g) = 2 \times q_\text{ref}(x,g) + 0.035 + K(g)f(x -t)h(t).\]

where:

\begin{itemize}
\item
  \(t\) is the time since onset of LTC,
\item
  \(K(g)\) encodes a gender-specific intensity (0.5 for females, 0.75
  for males),
\item
  \(f\) and \(h\) are logistic-shaped modifiers to reflect attenuation
  with age and time since entry, respectively.
\end{itemize}

This formulation captures the elevated initial mortality due to severe
conditions like terminal cancers, which tapers off within two years or
by age 90. For further discussion, see Biessy (2016).

\subsubsection*{Simulating a population of insured
lives}\label{simulating-a-population-of-insured-lives}

We simulate 1,000,000 insured individuals over a subscription period
from 1990 to 2009. Eligibility spans:

\begin{itemize}
\item
  ages 20 to 65 for annuity policies;
\item
  ages 50 to 75 for LTC policies.
\end{itemize}

Subscription age, year, and gender are drawn with replacement from
weighted distributions based on French population demographics. Birth
dates and subscription dates are assigned uniformly at random within
valid ranges.

\subsubsection*{Simulating life
trajectories}\label{simulating-life-trajectories}

For each individual, we simulate life outcomes from their subscription
date up to December 31, 2024. The process involves:

\begin{enumerate}
\def\labelenumi{\arabic{enumi}.}
\item
  Dividing the time axis into intervals delimited by birthdays.
\item
  For each interval, computing event probabilities based on applicable
  mortality or incidence rates.
\item
  Drawing events using uniform random variables:
\end{enumerate}

\begin{itemize}
\item
  In the LTC case, distinguishing between autonomous death and LTC
  incidence.
\item
  Recording the event date accordingly.
\end{itemize}

\begin{enumerate}
\def\labelenumi{\arabic{enumi}.}
\setcounter{enumi}{3}
\tightlist
\item
  If no terminating event occurs, proceeding to the next interval.
\end{enumerate}

For individuals who enter LTC, a second simulation phase begins,
spanning from LTC entry to the end of observation. This time, the
timeline is segmented by both age and duration-in-LTC anniversaries.
Within each subperiod, we compute and simulate death-in-LTC events.

\subsubsection*{Sampling subsets from the simulated
population}\label{sampling-subsets-from-the-simulated-population}

\subsubsection*{Sample from the simulated population to get a subset of
desired
size}\label{sample-from-the-simulated-population-to-get-a-subset-of-desired-size}

To match study requirements, we extract 100 independent samples of
100,000 individuals from the simulated population without replacement.
While not fully independent, overlap is limited (\textasciitilde10\% on
average), which is deemed acceptable for the study's objectives.

\subsubsection*{Aggregating events and
exposures}\label{aggregating-events-and-exposures}

Each sample is aggregated to produce death counts and central exposures
:

\begin{itemize}
\item
  by age in the annuity case;
\item
  by age and duration in the LTC case.
\end{itemize}

Aggregation is performed using the methodology described in Appendix A.

\subsection{Derivation of marginal likelihood and
LAML}\label{sec-likelihood-computations}

This appendix presents the derivations of the marginal likelihood and
its Laplace approximation (LAML) used for the automatic selection of
smoothing parameters.

In the Gaussian case, where the model assumes normal conditional and
prior distributions, the marginal likelihood can be computed in closed
form by integrating out the latent parameters. This yields an explicit
expression involving the penalty and weight matrices, and forms the
basis of the outer iteration strategy.

For more general models in the exponential family, no closed-form
solution is available. Instead, we apply a Laplace approximation to the
marginal likelihood, based on a second-order expansion of the penalized
log-likelihood around its maximum. The resulting LAML criterion is used
in the performance and alternated iteration approaches for efficient and
principled smoothing parameter selection.

\subsubsection{Marginal likelihood}\label{marginal-likelihood}

Assume that
\(\mathbf{y} \mid \boldsymbol{\theta} \sim \mathcal{N}(\boldsymbol{\theta}, \sigma^2W^{-})\)
and
\(\boldsymbol{\theta} \mid \lambda \sim \mathcal{N}(0, \sigma^2P_{\lambda}^{-})\).
In the empirical Bayes approach, the smoothing parameter \(\lambda\) is
estimated by maximizing the marginal likelihood:

\[\mathcal{L}^m_\text{norm}(\lambda) = f(\mathbf{y} \mid \lambda) = \int f(\mathbf{y}, \boldsymbol{\theta} \mid \lambda)\text{d}\boldsymbol{\theta} = \int f(\mathbf{y} \mid \boldsymbol{\theta}) f(\boldsymbol{\theta} \mid \lambda)\text{d}\boldsymbol{\theta}.\]

The conditional and prior densities involved in this integral are:

\[\begin{aligned}
f(\mathbf{y} \mid \boldsymbol{\theta}) &= \sqrt{\frac{|W|_{+}}{(2\pi\sigma^2)^{n_*}}}\exp\left(- \frac{1}{2\sigma^2}(\mathbf{y} - \boldsymbol{\theta})^{T}W(\mathbf{y} - \boldsymbol{\theta})\right) \\
f(\boldsymbol{\theta} \mid \lambda) &= \sqrt{\frac{|P_{\lambda}|_{+}}{(2\pi\sigma^2)^{p - q}}} \exp\left(- \frac{1}{2\sigma^2}\boldsymbol{\theta}^{T}P_{\lambda} \boldsymbol{\theta}\right)
\end{aligned}\]

where \(n_*\) is the number of non-zero weights and \(q\) is the order
of the penalization (corresponding to the rank deficiency of
\(P_\lambda\)).

We then apply a second-order Taylor expansion of the joint log-density
\(\ln f(\mathbf{y}, \boldsymbol{\theta} \mid \lambda)\) around its mode
\(\hat{\boldsymbol{\theta}}_{\lambda}\) to approximate the integral:

\[
\ln f(\mathbf{y}, \boldsymbol{\theta} \mid \lambda) = \ln f(\mathbf{y}, \hat{\boldsymbol{\theta}}_{\lambda} \mid \lambda) + \frac{1}{2}(\boldsymbol{\theta} - \hat{\boldsymbol{\theta}}_{\lambda})^{T} (W + P_{\lambda})(\boldsymbol{\theta} - \hat{\boldsymbol{\theta}}_{\lambda})
\] This gives the following expression for the marginal likelihood:

\[\mathcal{L}^m_\text{norm}(\lambda) = f(\mathbf{y}, \hat{\boldsymbol{\theta}}_{\lambda} \mid \lambda) \int \exp\left[- \frac{1}{2\sigma^2}(\boldsymbol{\theta} - \hat{\boldsymbol{\theta}}_{\lambda})^{T}(W + P_{\lambda})(\boldsymbol{\theta} - \hat{\boldsymbol{\theta}}_{\lambda}) \right]\text{d}\boldsymbol{\theta}\]

which evaluates to:

\[\mathcal{L}^m_\text{norm}(\lambda) = \sqrt{\frac{|W|_{+}|P_{\lambda}|_{+}}{(2\pi\sigma^2)^{n_* - q}|W + P_{\lambda}|}} \exp\left(- \frac{1}{2\sigma^2}\left[(\mathbf{y} -\hat{\boldsymbol{\theta}}_{\lambda})^{T}W(\mathbf{y} -\hat{\boldsymbol{\theta}}_{\lambda}) + \hat{\boldsymbol{\theta}}_{\lambda}^T P_{\lambda} \hat{\boldsymbol{\theta}}_{\lambda}\right]\right)\]

where
\(\hat{\boldsymbol{\theta}}_{\lambda} = (W + P_{\lambda})^{-1} W \mathbf{y}\).

Taking the logarithm yields the marginal log-likelihood:

\[\ell^m_\text{norm}(\lambda) = - \frac{1}{2}\left[(\mathbf{y} -\hat{\boldsymbol{\theta}}_{\lambda})^{T}W(\mathbf{y} -\hat{\boldsymbol{\theta}}_{\lambda}) / \sigma^2 + \hat{\boldsymbol{\theta}}_{\lambda}^{T}P_{\lambda} \hat{\boldsymbol{\theta}}_{\lambda} / \sigma^2 + \ln|W + P_{\lambda}| - \ln |P_{\lambda}|_{+} + C\right].\]

where
\(\hat{\boldsymbol{\theta}}_\lambda = (W + P_{\lambda})^{-1}W\mathbf{y}\),
and \(C = - \ln|W|_{+} + (n_* - q)\ln(2\pi\sigma^2)\) is a constant
independent of \(\lambda\).

\subsubsection{Laplace approximation of the marginal
likelihood}\label{laplace-approximation-of-the-marginal-likelihood}

Assume that the log-likelihood \(\ell(\boldsymbol{\theta})\) is combined
with a Gaussian prior on the parameter vector:
\(\boldsymbol{\theta} \sim \mathcal{N}(0, P_\lambda^{-1})\). The
marginal likelihood of the data \((\mathbf{d}, \mathbf{e}_c)\) given
\(\lambda\) is:

\[\mathcal{L}^m_\text{ML}(\lambda) = f(\mathbf{d}, \mathbf{e}_c \mid \lambda) = \int f(\mathbf{d}, \mathbf{e}_c, \boldsymbol{\theta} \mid \lambda) f(\boldsymbol{\theta} \mid \lambda)\text{d}\boldsymbol{\theta}.\]

Since no closed-form expression exists for this integral in the general
exponential family case, we apply a second-order Taylor expansion of the
log-posterior around its mode
\(\hat{\boldsymbol{\theta}}_\lambda = \arg\max \ell_P(\boldsymbol{\theta})\),
where
\(\ell_P(\boldsymbol{\theta}) = \ell(\boldsymbol{\theta}) - \tfrac{1}{2} \boldsymbol{\theta}^T P_\lambda \boldsymbol{\theta}\)
is the penalized log-likelihood.

The resulting Laplace approximation of the marginal likelihood is:

\[\mathcal{L}^m_\text{ML}(\lambda) \approx \exp\left(\ell_P(\hat{\boldsymbol{\theta}}_{\lambda})\right) \sqrt{\frac{(2\pi)^p}{|W_{\lambda} + P_{\lambda}|}}, \]

where
\(W_\lambda = \text{Diag}(\exp(\hat{\boldsymbol{\theta}}_\lambda) \odot \mathbf{e}_c)\)
is the observed Fisher information. Taking the logarithm leads to the
LAML criterion:

\[\ell^m_\text{ML}(\lambda) \approx \ell(\hat{\boldsymbol{\theta}}_{\lambda}) - \frac{1}{2}\left[\hat{\boldsymbol{\theta}}_{\lambda}^T P_{\lambda} \hat{\boldsymbol{\theta}}_{\lambda} + \ln|W_{\lambda} + P_{\lambda}| - \ln|P_{\lambda}|_{+} - q\ln(2\pi)\right] \overset{\text{def}}{=} \ell^m_\text{LAML}(\lambda).\]

\subsection{Algorithms}\label{sec-algorithms-appendix}

This appendix presents the computational procedures used to implement
the generalized Whittaker-Henderson (WH) smoothing framework and the
various automatic selection methods for the associated smoothing
parameters.

\subsubsection*{Generalized WH
smoothing}\label{generalized-wh-smoothing}

Algorithm \ref{algo1} implements the core iterative procedure for
generalized WH smoothing, as introduced in Section 4. It details the
iterative computation of the estimated log-rates
\(\hat{\boldsymbol{\theta}}\) given fixed smoothing parameters and a
chosen differencing order. The algorithm iteratively solves a penalized
weighted least-squares problem until convergence is achieved, based on a
predefined deviance threshold.

\begin{algorithm}
\caption{Iterative solution of generalized Whittaker-Henderson smoothing\label{algo1}}
\SetKwInOut{Input}{inputs}
\SetKwInOut{Output}{outputs}
\SetKwInOut{Param}{parameters}
\DontPrintSemicolon

\Input{$\mathbf{d}$ and $\mathbf{e_c}$}
\Output{$\hat{\boldsymbol{\theta}}$}
\Param{$\lambda$, $q$, $\epsilon_d = 10^{-8}$}

\Begin{
Construct the penalty matrix $P_{\lambda}$ based on the difference matrices of order $q$.\;

$k \leftarrow 0$\;
$\boldsymbol{\theta}_0 \leftarrow \ln(\mathbf{d}^* / \mathbf{e_c})$, where $\mathbf{d}^* = \max(\mathbf{d}, \epsilon)$\;
$\text{dev}_0 \leftarrow \infty,\quad \text{cond} \leftarrow \text{true}$\;

\While{$\text{cond}$}{

 $\mathbf{w}_k \leftarrow \text{Diag}(\exp(\boldsymbol{\theta}_k) \odot \mathbf{e_c})$\;
 $\mathbf{z}_k \leftarrow \boldsymbol{\theta}_k + \mathbf{d} / \mathbf{w}_k - 1$\;
 Form $W_k + P_\lambda$ by adding $\mathbf{w}_k$ to the diagonal of $P_\lambda$.\;
 Find the Cholesky factor $R$ of $W_k + P_{\lambda}$.\;
 Find $\mathbf{u}$ such that $R^T \mathbf{u} = \mathbf{w}_k \odot \mathbf{z}_k$ by forward substitution.\;
 Find $\boldsymbol{\theta}_{k+1}$ such that $R \boldsymbol{\theta}_{k+1} = \mathbf{u}$ by backward substitution.\;
 $\text{dev}_{k+1} \leftarrow \text{dev}_P(\boldsymbol{\theta}_{k+1})$,\quad $\text{cond} \leftarrow \text{dev}_{k+1} \le (1 - \epsilon_d)\text{dev}_k$\;
 $k \leftarrow k + 1$\;
}
$\hat{\boldsymbol{\theta}} \leftarrow \boldsymbol{\theta}_k$\;
}
\end{algorithm}

\subsubsection*{Smoothing parameter selection
approaches}\label{smoothing-parameter-selection-approaches}

In Section 5, we introduced three alternative strategies to
automatically calibrate the smoothing parameter \(\lambda\), based on
marginal likelihood maximization. The following three algorithms
formalize their respective procedures.

Together, these algorithms offer a modular and flexible framework for
implementing WH smoothing and its data-driven calibration in practical
applications.

\paragraph*{Outer iteration}\label{outer-iteration}

Algorithm \ref{algo2} corresponds to the outer iteration approach. In
this strategy, a series of candidate values for \(\lambda\) are tested
sequentially. For each candidate, the generalized WH smoother of
\ref{algo1} is applied until convergence, and the corresponding marginal
likelihood is evaluated. The process continues until no further
improvement is observed. This approach separates the parameter selection
and smoothing steps into nested loops.

\begin{algorithm}
\caption{Smoothing parameter selection for generalized Whittaker-Henderson smoothing - outer iteration approach.\label{algo2}}
\SetKwInOut{Input}{inputs}\SetKwInOut{Output}{outputs}\SetKwInOut{Param}{parameters}
\DontPrintSemicolon
\Input{$\mathbf{d}$ and $\mathbf{e_c}$}
\Output{$\hat{\lambda}$}
\Param{$q$, $\epsilon_d = 10^{- 8}$, $\epsilon_\text{laml} = 10^{- 8}$}
\Begin{
$k \leftarrow 0$\;
$\text{laml}_0 \leftarrow \infty, \quad \text{cond}_\text{laml} \leftarrow \text{true}$\;
\While{$\text{cond}_\text{laml}$}{
If $k = 0$, choose an arbitrary initial value $\lambda_1$ for the smoothing parameter(s); otherwise, choose the next value $\lambda_{k + 1}$ using the desired heuristic.\;
$k \leftarrow k + 1$\;
Use Algorithm 1 to determine the vector $\hat{\boldsymbol{\theta}}_{\lambda_k}$ associated with the choice of $\lambda_k$, using a convergence treshold of $\epsilon_d$.\;
Calculate the marginal likelihood $\ell^m_\text{LAML}(\lambda_k)$ associated with the choice of $\lambda_k$ using the intermediate quantities calculated during the estimation of $\hat{\boldsymbol{\theta}}_{\lambda_k}.$\;
$\text{laml}_k \leftarrow \ell^m_\text{LAML}(\lambda_k), \quad \text{cond}_\text{laml} \leftarrow \text{laml}_k \ge (1 + \epsilon_\text{laml})\text{laml}_{k - 1}$\;
}
$\hat{\lambda} \leftarrow \lambda_k$\;
}
\end{algorithm}

\paragraph*{Performance iteration}\label{performance-iteration}

Algorithm \ref{algo3} implements the performance iteration strategy.
Here, the smoothing parameter is optimized at each step based on updated
pseudo-response and weight vectors. The smoother and the parameter
estimation are intertwined, with \(\lambda\) being re-optimized after
each update of the linear predictor. This often leads to faster
convergence toward the maximum marginal likelihood compared to the outer
iteration approach.

Let us note that in the algorithm,
\(\ell^m_\text{LAML}(\lambda_{k + 1} \mid \boldsymbol{\theta}_{k + 1})\)
denotes an approximate LAML, in which
\(\boldsymbol{\theta}_{k + 1}\)---the maximizer of the approximate
normal marginal likelihood constructed from \(\mathbf{w}_k\) and
\(\mathbf{z}_k\)---replaces the true penalized likelihood maximizer
\(\hat{\boldsymbol{\theta}}_{\lambda_{k + 1}}\), which is not available
in the performance iteration approach.

\begin{algorithm}
\caption{Parameter selection for generalized Whittaker-Henderson smoothing - performance iteration approach\label{algo3}}
\SetKwInOut{Input}{inputs}\SetKwInOut{Output}{outputs}\SetKwInOut{Param}{parameters}
\DontPrintSemicolon
\Input{$\mathbf{d}$ and $\mathbf{e_c}$}
\Output{$\hat{\lambda}$}
\Param{$q$, $\epsilon_\text{ml} = 10^{- 8}$, $\epsilon_\text{laml} = 10^{- 8}$}
\Begin{
$k \leftarrow 0$\;
$\boldsymbol{\theta}_0 \leftarrow \ln(\mathbf{d} / \mathbf{e_c})$\;
$\text{laml}_0 \leftarrow \infty, \quad \text{cond}_\text{laml} \leftarrow \text{true}$\;
\While{$\text{cond}_\text{laml}$}{
$\mathbf{w}_k \leftarrow \text{Diag}(\exp(\boldsymbol{\theta}_k) \odot \mathbf{e_c})$\;

 $\mathbf{z}_k \leftarrow \boldsymbol{\theta}_k + \mathbf{d} / \mathbf{w}_k - 1$\;
Find the parameter $\lambda_{k + 1}$ maximizing the marginal likelihood $\ell^m_\text{norm}$ associated with the observation vector $\mathbf{z}_k$ and the weight vector $\mathbf{w}_k$, using the desired heuristic, using a convergence treshold of $\epsilon_\text{ml}$.\;
Form $W_k + P_{\lambda_{k + 1}}$ by adding $\mathbf{w}_k$ to the diagonal of $P_{\lambda_{k + 1}}$.\;
Find the Cholesky factor $R$ of $W_k + P_{\lambda_{k + 1}}$.\;
Find $\mathbf{u}$ such that $R^T \mathbf{u} = \mathbf{w}_k \odot \mathbf{z}_k$ by forward substitution.\;
Find $\boldsymbol{\theta}_{k+1}$ such that $R \boldsymbol{\theta}_{k+1} = \mathbf{u}$ by backward substitution.\;
$\text{laml}_{k + 1} \leftarrow \ell^m_\text{LAML}(\lambda_{k + 1} \mid \boldsymbol{\theta}_{k + 1}), \quad \text{cond}_\text{laml} \leftarrow \text{laml}_{k + 1} \ge (1 + \epsilon_\text{laml})\text{laml}_k$\;
$k \leftarrow k + 1$\;
}
$\hat{\lambda} \leftarrow \lambda_k$;
}
\end{algorithm}

\paragraph*{Alternated iteration}\label{alternated-iteration}

Finally, Algorithm \ref{algo4} presents the alternated iteration
approach. This hybrid method alternates between updating the smoothing
parameter \(\lambda\) on one hand, and and an updated pseudo-response
and weight vector on the other hand until convergence of an approximate
marginal likelihood.

As in the performance iteration approach,
\(\ell^m_\text{LAML}(\lambda_{k + 1} \mid \boldsymbol{\theta}_{k + 1})\)
denotes an approximate LAML, based on the maximizer of an approximate
normal marginal likelihood rather than the true penalized likelihood
maximizer.

\begin{algorithm}
\caption{Parameter selection for generalized Whittaker-Henderson smoothing - alternated iteration approach\label{algo4}}
\SetKwInOut{Input}{inputs}\SetKwInOut{Output}{outputs}\SetKwInOut{Param}{parameters}
\DontPrintSemicolon
\Input{$\mathbf{d}$ and $\mathbf{e_c}$}
\Output{$\hat{\lambda}$}
\Param{$q$, $\epsilon_\text{laml} = 10^{- 8}$}
\Begin{
$k \leftarrow 0$\;
$\boldsymbol{\theta}_0 \leftarrow \ln(\mathbf{d} / \mathbf{e_c})$\;
$\text{laml}_0 \leftarrow \infty, \quad \text{cond}_\text{laml} \leftarrow \text{true}$\;
\While{$\text{cond}_\text{laml}$}{
$\mathbf{w}_k \leftarrow \text{Diag}(\exp(\boldsymbol{\theta}_k) \odot \mathbf{e_c})$\;

 $\mathbf{z}_k \leftarrow \boldsymbol{\theta}_k + \mathbf{d} / \mathbf{w}_k - 1$\;
 If $k = 0$, choose an arbitrary value $\lambda_1$ for the smoothing parameter(s); otherwise, choose the next value $\lambda_{k + 1}$ to improve the marginal likelihood $\ell^m_\text{norm}$ using the desired heuristic.\;
Form $W_k + P_{\lambda_{k + 1}}$ by adding $\mathbf{w}_k$ to the diagonal of $P_{\lambda_{k + 1}}.$\;
Find the Cholesky factor $R$ of $W_k + P_{\lambda_{k + 1}}.$\;
Find $\mathbf{u}$ such that $R^T \mathbf{u} = \mathbf{w}_k \odot \mathbf{z}_k$ by forward substitution.\;
Find $\boldsymbol{\theta}_{k + 1}$ such that $R \boldsymbol{\theta}_{k + 1} = \mathbf{u}$ by backward substitution.\;
$\text{laml}_{k + 1} \leftarrow \ell^m_\text{LAML}(\lambda_{k + 1} \mid \boldsymbol{\theta}_{k + 1}), \quad \text{cond}_\text{laml} \leftarrow \text{laml}_{k + 1} \ge (1 + \epsilon_\text{laml})\text{laml}_k$\;
$k \leftarrow k + 1$\;
}
$\hat{\lambda} \leftarrow \lambda_k$\;
}
\end{algorithm}

\subsection{Illustrations for the natural parameterization of WH
smoothing}\label{sec-natural-parametrization}

This appendix provides an interpretation of Whittaker-Henderson (WH)
smoothing in the framework of natural parameterization, based on the
eigendecomposition of the penalty matrix. This approach reveals how
smoothing acts as a spectral filter that progressively attenuates
components of the signal associated with rougher variations.

In the one-dimensional case, the smoothing problem is re-expressed in a
rotated basis formed by the eigenvectors of the penalty operator,
leading to a clear decomposition of the signal into smoother and rougher
components. The effect of smoothing is visualized via effective degrees
of freedom associated with each component.

The two-dimensional extension leverages Kronecker product identities to
generalize the spectral interpretation to tensor-product smoothing
penalties. Illustrations are provided to highlight how smoothing
parameters affect the influence of each spectral component in both
dimensions.

\subsubsection{One-dimensional case}\label{one-dimensional-case-3}

\begin{figure}

\centering{

\includegraphics[width=1\linewidth,height=\textheight,keepaspectratio]{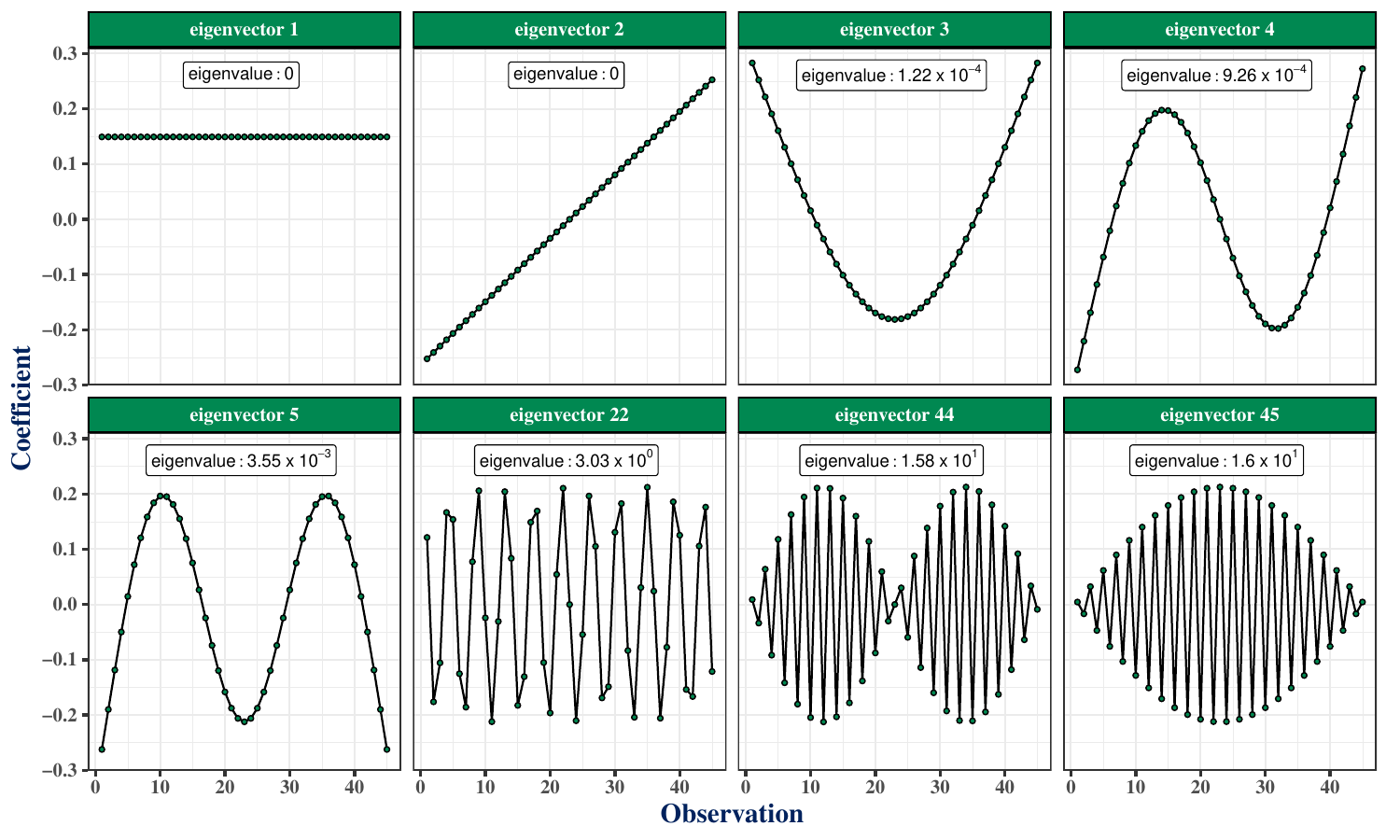}

}

\caption{\label{fig-eigen}A subset of the eigenvectors for the
penalization matrix \(D_{n,q}^TD_{n,q}\) with \(n = 45\) and \(q = 2\).}

\end{figure}

Building on the natural parameterization introduced by Demmler and
Reinsch (1975), the WH estimator can be reformulated as the solution to
the following optimization problem:

\[
\hat{\mathbf{y}} = U\hat{\boldsymbol{\beta}},\quad \hat{\boldsymbol{\beta}} = \underset{\boldsymbol{\beta}}{\text{argmin}}\left\lbrace (\mathbf{y} - U\boldsymbol{\beta})^{T}W(\mathbf{y} - U\boldsymbol{\beta}) + \lambda\boldsymbol{\beta}^{T}\Sigma\boldsymbol{\beta}\right\rbrace.
\]

\(\hat{\boldsymbol{\beta}}\) can be interpreted as a vector of
coordinates in the basis of eigenvectors of \(P_{\lambda}\), yielding a
spectral decomposition of the signal into components with varying
degrees of smoothness, as determined by the associated eigenvalues.
Figure~\ref{fig-eigen} represents 8 of the eigenvectors associated with
\(q = 2\) for a basis of size \(n = 45\). The first \(q\) eigenvalues
are zero as \(D_{n,q}\) is of rank \(n - q\).

By using the fact that \(U^{-1} = U^T\) and by linking this expression
back to the original smoothing formulation, we obtain the explicit
solution:

\begin{equation}\phantomsection\label{eq-interpretation}{
\hat{\mathbf{y}} = U(U^TWU + S_{\lambda})^{-1}U^TW\mathbf{y} \quad \text{where} \quad S_{\lambda} = \lambda\Sigma.
}\end{equation}

In order to interpret Equation~\ref{eq-interpretation}, consider the
special case where all weights are equal to 1 and therefore:

\[
\hat{\mathbf{y}} = U(U^TU + \lambda\Sigma)^{-1}U^T\mathbf{y} = U(I_n + \lambda\Sigma)^{-1}U^T\mathbf{y}.
\]

The transformation from \(\mathbf{y}\) to \(\hat{\mathbf{y}}\) can then
be seen as a 3-step process, reading the equation from right to left:

\begin{enumerate}
\def\labelenumi{\arabic{enumi}.}
\item
  Decomposition of the signal \(\mathbf{y}\) in the basis of
  eigenvectors through the left multiplication by \(U^T\).
\item
  Attenuation of the signal components based on the eigenvalues
  associated with these components. If we denote
  \(s = \text{diag}(\Sigma)\), then
  \((I_n + \lambda\Sigma)^{-1} = \text{Diag}[1 / (1 + \lambda s)]\).
  After the left multiplication by \((I_n + \lambda\Sigma)^{-1}\), each
  component is hence divided by a factor \(1 + \lambda s \geq 1\). This
  coefficient increases linearly with \(\lambda\), but the rate of
  increase varies with the magnitude of the corresponding eigenvalue.
\item
  Recomposition of the attenuated signal in the canonical basis through
  the left multiplication by \(U\).
\end{enumerate}

When weights are not uniform, the structure becomes more complex since
\(U^TWU\) is no longer a diagonal matrix. However, it is still possible
to interpret the effect of smoothing from the diagonal of the matrix
\(F = (U^TWU + S_{\lambda})^{-1}U^TWU\). Indeed:

\[
U^T\hat{\mathbf{y}} = U^TU\hat{\boldsymbol{\theta}} = \hat{\boldsymbol{\theta}} = (U^TWU + S_{\lambda})^{-1}U^TW\mathbf{y} = (U^TWU + S_{\lambda})^{-1}U^TWU U^T\mathbf{y} = FU^T\mathbf{y}.
\]

Since the vectors \(U^T\mathbf{y}\) and
\(\hat{\boldsymbol{\beta}} = U^T\hat{\mathbf{y}}\) represent the
coordinates of \(\mathbf{y}\) and \(\hat{\mathbf{y}}\) respectively in
the basis of eigenvectors of \(D_{n,q}^TD_{n,q}\), \(F\) thus acts as a
transformation matrix on the spectral coordinates, analogous to the role
played by the hat matrix \(H = U(U^TWU + S_{\lambda})^{-1}U^TW\) for the
observations. The diagonal values of \(F\) may be interpreted as the
effective degrees of freedom associated with each eigenvector after
smoothing. It can be verified that:

\[
\text{tr}(F) = \text{tr}[(U^TWU + S_{\lambda})^{-1}U^TWU] = \text{tr}[U(U^TWU + S_{\lambda})^{-1}U^TW] = \text{tr}(H)
\]

which means that the sum of the effective degrees of freedom remains the
same whether it is counted per observation or per parameter.

\begin{figure}

\centering{

\includegraphics[width=1\linewidth,height=\textheight,keepaspectratio]{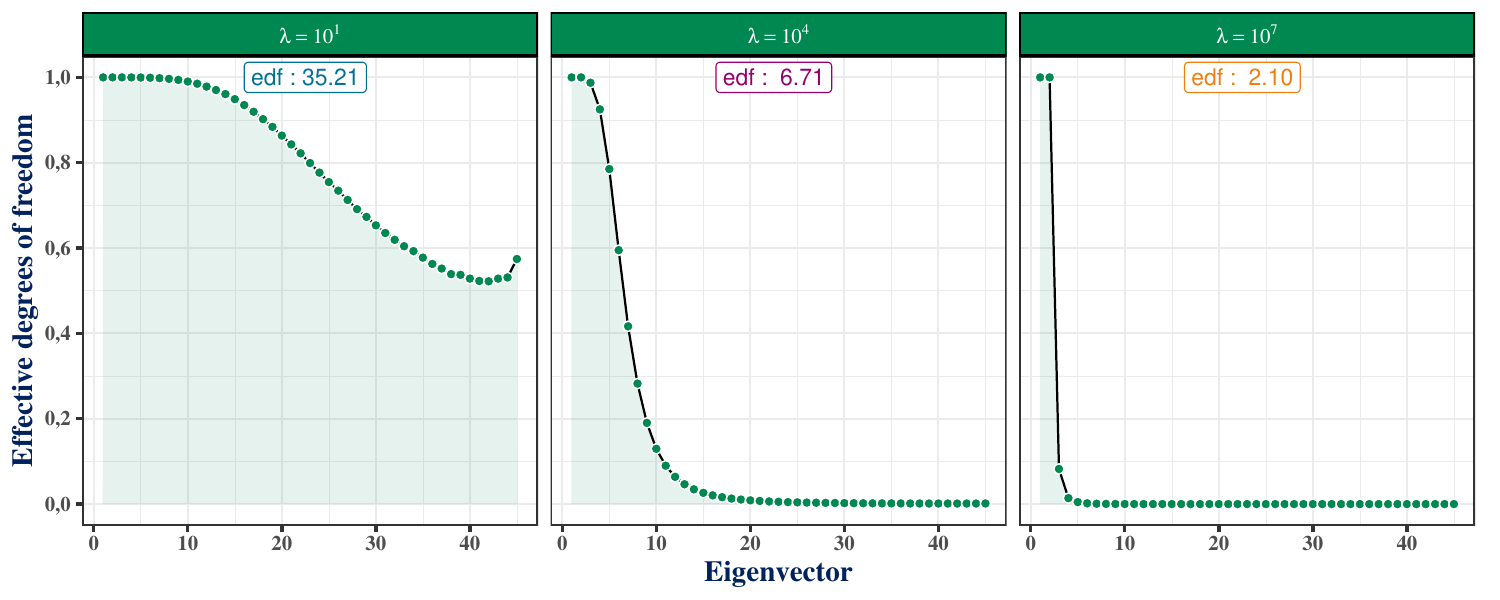}

}

\caption{\label{fig-edf}Effective degrees of freedom per eigenvector
after applying 1D WH smoothing, for different combinations of smoothing
parameter.}

\end{figure}

Figure~\ref{fig-edf} represents the effective degrees of freedom per
parameter in the previous illustration of smoothing. The first \(q\)
eigenvectors are never penalized, so their effective degrees of freedom
are always equal to 1, regardless of the smoothing parameter used. The
other eigenvectors have strictly decreasing effective degrees of freedom
with \(\lambda\). These degrees of freedom are mostly decreasing with
increasing eigenvalues of \(D_{n,q}^TD_{n,q}\), although in the presence
of non-unit weights and for small values of \(\lambda\), this is not
always the case.

\subsubsection{Two-dimensional case}\label{two-dimensional-case-3}

In the two-dimensional case, we have
\(P_{\lambda} = \lambda_x I_{n_z} \otimes D_{n_x,q_x}^{T}D_{n_x,q_x} + \lambda_z D_{n_z,q_z}^{T}D_{n_z,q_z} \otimes I_{n_x}\).
Similar to the one-dimensional case, we can perform the
eigendecomposition of the matrices \(D_{n_x,q_x}^{T}D_{n_x,q_x}\) and
\(D_{n_z,q_z}^{T}D_{n_z,q_z}\), yielding
\(D_{n_x,q_x}^{T}D_{n_x,q_x} = U_x\Sigma_x U_x^{T}\) and
\(D_{n_z,q_z}^{T}D_{n_z,q_z} = U_z\Sigma_z U_z^{T}\). Define
\(U = U_z \otimes U_x\) and perform the reparametrization
\(\boldsymbol{\beta} = U^T\boldsymbol{\theta} \Leftrightarrow \boldsymbol{\theta} = U\boldsymbol{\beta}\).
By leveraging the properties of the Kronecker product, we can rewrite
the smoothness criterion in a simplified form:

\[\boldsymbol{\theta}^TP_{\lambda}\boldsymbol{\theta} = (U\boldsymbol{\beta})^{T}P_{\lambda}(U\boldsymbol{\beta})= \boldsymbol{\beta}^{T}(\lambda_x I_{n_z} \otimes \Sigma_x + \lambda_z \Sigma_z \otimes I_{n_x})\boldsymbol{\beta}.\]

This leads to an alternative formulation of the optimization problem:

\[
\hat{\mathbf{y}} = U\hat{\boldsymbol{\beta}},\quad\hat{\boldsymbol{\beta}} = \underset{\boldsymbol{\beta}}{\text{argmin}} \left\lbrace (\mathbf{y} - U\boldsymbol{\beta})^{T}W(\mathbf{y} - U\boldsymbol{\beta}) + \lambda\boldsymbol{\beta}^{T}(\lambda_x I_{n_z} \otimes \Sigma_x + \lambda_z \Sigma_z \otimes I_{n_x})\boldsymbol{\beta} \right\rbrace.
\]

The solution to the smoothing problem, as in the one-dimensional case,
is given by:

\[
\hat{\mathbf{y}} = U(U^TWU + S_{\lambda})^{- 1}U^TW\mathbf{y} \quad \text{where} \quad S_{\lambda} = \lambda_x I_{n_z} \otimes \Sigma_x + \lambda_z \Sigma_z \otimes I_{n_x}.
\]

\begin{figure}

\centering{

\includegraphics[width=1\linewidth,height=\textheight,keepaspectratio]{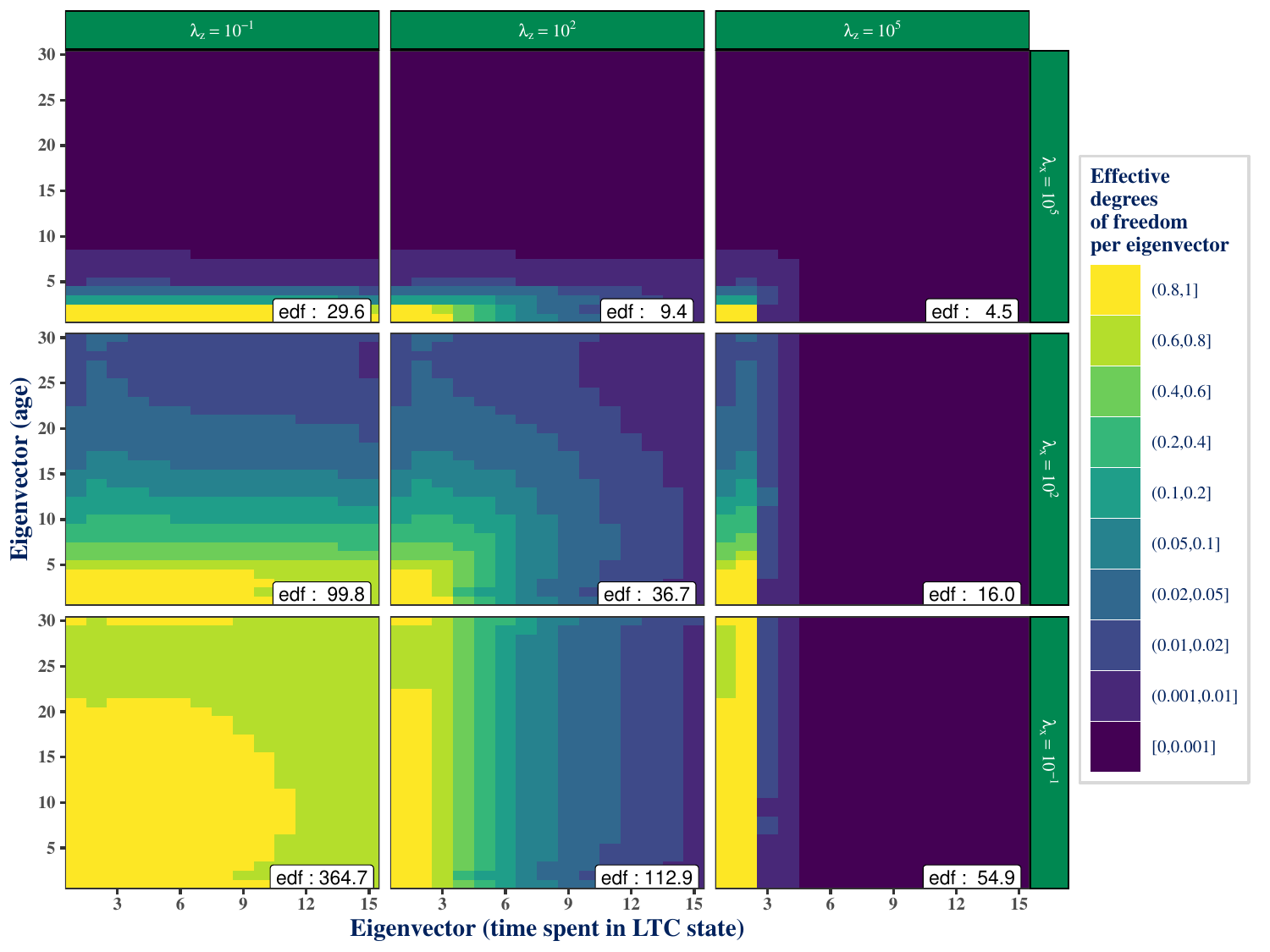}

}

\caption{\label{fig-edf2d}Residual degrees of freedom per eigenvector
after applying 2D WH smoothing, for different combinations of smoothing
parameters.}

\end{figure}

Figure~\ref{fig-edf2d} represents the residual degrees of freedom
associated with each parameter after applying the smoothing, in the
two-dimensional case, for different combinations of the smoothing
parameters. Similar to the one-dimensional case, these degrees of
freedom decrease as the smoothing parameters increase and are
particularly small for higher eigenvalues. The eigenvectors are sorted
in ascending order of eigenvalues for each one-dimensional penalty
matrix \(D_{n_x,q_x}^{T}D_{n_x,q_x}\) and
\(D_{n_z,q_z}^{T}D_{n_z,q_z}\).

\subsection{Derivation of constrained extrapolation in the 2D
case}\label{sec-extrapolation-computation}

This appendix provides a closed-form expression for the constrained
extrapolated estimator, which extends the smoothed surface beyond the
observed domain while preserving the values estimated during the initial
fit. It also derives the associated variance-covariance matrix,
accounting for both propagated uncertainty and additional variability in
the extrapolated region.

To obtain an estimator \(\hat{\mathbf{y}}^\ast_+\) that minimizes the
extended optimization problem under the constraint of preserving the
initial coefficients,
i.e.~\(C\hat{\mathbf{y}}^\ast_+ = \hat{\mathbf{y}}\), we follow the
approach proposed by Carballo, Durban, and Lee (2021) and introduce a
Lagrange multiplier \(\boldsymbol{\omega}\). The associated constrained
extended optimization problem is now written as:

\begin{equation}\phantomsection\label{eq-optim-lagrange}{(\hat{\mathbf{y}}_+^\ast, \hat{\boldsymbol{\omega}}) = \underset{\boldsymbol{\theta}_+^\ast, \boldsymbol{\omega}}{\text{argmin}}\left\lbrace (\mathbf{y}_+ - \boldsymbol{\theta}_+^\ast)^TW_+(\mathbf{y}_+ - \boldsymbol{\theta}_+^\ast) + \boldsymbol{\theta}_+^{\ast T}P_{+}\boldsymbol{\theta}_+^\ast + 2 \boldsymbol{\omega}^T(C\boldsymbol{\theta}_+^\ast - \hat{\mathbf{y}})\right\rbrace.}\end{equation}

Taking the partial derivatives of Equation~\ref{eq-optim-lagrange} with
respect to \(\boldsymbol{\theta}_+^\ast\) and \(\boldsymbol{\omega}\)
gives:

\[\begin{aligned}
\frac{\partial}{\partial \boldsymbol{\theta}_+^\ast}\left\lbrace(\mathbf{y}_+ - \boldsymbol{\theta}_+^\ast)^TW_+(\mathbf{y}_+ - \boldsymbol{\theta}_+^\ast) + \boldsymbol{\theta}_+^{\ast T}P_{+}\boldsymbol{\theta}_+^\ast + 2 \boldsymbol{\omega}^T(C\boldsymbol{\theta}_+^\ast - \hat{\mathbf{y}})\right\rbrace &= - 2 W_+(\mathbf{y}_+ - \boldsymbol{\theta}_+^\ast) +2P_+\boldsymbol{\theta}_+^\ast + 2 \boldsymbol{\omega}^TC\\
\frac{\partial}{\partial \boldsymbol{\omega}}\left\lbrace(\mathbf{y}_+ - \boldsymbol{\theta}_+^\ast)^TW_+(\mathbf{y}_+ - \boldsymbol{\theta}_+^\ast) + \boldsymbol{\theta}_+^{\ast T}P_{+}\boldsymbol{\theta}_+^\ast + 2 \boldsymbol{\omega}^T(C\boldsymbol{\theta}_+^\ast - \hat{\mathbf{y}})\right\rbrace &= 2(C\boldsymbol{\theta}_+^\ast - \hat{\mathbf{y}})
\end{aligned}\]

Setting these derivatives to zero yields the linear system:

\[\begin{bmatrix} W_+ + P_+& C^T \\ C & 0\end{bmatrix}\begin{bmatrix}\hat{\mathbf{y}}^\ast_+ \\ \hat{\boldsymbol{\omega}}\end{bmatrix} = \begin{bmatrix}W_+\mathbf{y}_+ \\ \hat{\mathbf{y}}\end{bmatrix}\]

The solution for \(\hat{\mathbf{y}}^\ast_+\) can be derived using
formulas for the inversion of a symmetric matrix partitioned with
\(2 \times 2\) blocks, setting \(V_+ = (W_+ + P_+)^{- 1}\):

\[\hat{\mathbf{y}}^\ast_+ = V_+\left\lbrace I - C^T[CV_+C^T]^{- 1}CV_+\right\rbrace W_+\mathbf{y}_+ + V_+ C^T[CV_+C^T]^{- 1}\hat{\mathbf{y}}.\]

Since \(W_+ = C^TWC\), the first term is actually zero, and this
expression simplifies to:

\[\hat{\mathbf{y}}^\ast_+ = V_+C^{T}[CV_+C^{T}]^{- 1}\hat{\mathbf{y}} = V_+C^{T}(V_+^{11})^{- 1}\hat{\mathbf{y}} = Q^T\begin{bmatrix}I \\ - (P_+^{22})^{- 1}P_+^{21}\end{bmatrix}\hat{\mathbf{y}}\]

which is a linear transformation of \(\hat{\mathbf{y}}\).

Defining
\(A_+^{*} = Q^T\begin{bmatrix}I \\ - (P_+^{22})^{- 1}P_+^{21}\end{bmatrix}\),
a variance-covariance of \(\mathbf{y}^*_+ \mid \boldsymbol{\theta}_+\)
based on this expression is given by:

\begin{equation}\phantomsection\label{eq-Psi-extra2}{A_+^{*}V A_+^{*T} = Q^T\begin{bmatrix}V & - V P_+^{12}(P_+^{22})^{- 1} \\ - (P_+^{22})^{- 1}P_+^{21}V & (P_+^{22})^{- 1}P_+^{21}V P_+^{12}(P_+^{22})^{- 1} \end{bmatrix}Q.}\end{equation}

Equation~\ref{eq-Psi-extra2} is very similar to the equivalent
formulation in the one-dimensional case with however two main
differences. First, every occurrence of \(V_+^{11}\) is replaced by
\(V\). This is consistent with the constraint that the coefficients over
the initial domain are held fixed at their estimated values. Second, as
the solution to the constrained extended optimization problem of
Equation~\ref{eq-optim-lagrange} was expressed as a linear
transformation of \(\hat{\mathbf{y}}\), Equation~\ref{eq-Psi-extra2} is
missing the innovation error term \((P_+^{22})^{- 1}\) associated with
the prior on the extrapolated coefficients. Not including this term
would be tantamount to considering that \(\boldsymbol{\theta}_+\) has
some degree of variability in the region of the initial data but is
perfectly smooth beyond this range. Adding this innovation error, we
obtain the following variance-covariance matrix for the constrained
optimization problem:

\[V_+^{\ast} = Q^T\begin{bmatrix}V & - V P_+^{12}(P_+^{22})^{- 1} \\ - (P_+^{22})^{- 1}P_+^{21}V &  (P_+^{22})^{- 1}P_+^{21}V P_+^{12}(P_+^{22})^{- 1} + (P_+^{22})^{- 1} \end{bmatrix}Q.\]

which still verifies

\[V_+^{\ast}W_+\mathbf{y}_+ = V_+^{\ast}CW\mathbf{y} = Q^T\begin{bmatrix}I \\ - (P_+^{22})^{- 1}P_+^{21}\end{bmatrix}V W\mathbf{y} = \hat{\mathbf{y}}^\ast_+.\]

\subsection{Choosing the order of the difference
matrices}\label{sec-difference-order}

This appendix investigates how the choice of the difference order \(q\)
in WH smoothing affects model performance. It provides empirical
guidance for selecting \(q\) based on AIC values computed across
simulation replicates, and confirms that second-order differences offer
a good balance between fit quality and extrapolation stability.

In Whittaker-Henderson (WH) smoothing, the order of the difference
matrix used in the penalization term, typically denoted \(q\) (or
\((q_x, q_z)\) in the two-dimensional case), governs the smoothness
prior imposed on the underlying signal. While Whittaker originally
proposed using third-order differences, second-order differences have
since become the standard choice. This aligns with the common
statistical definition of smoothness as the integrated squared second
derivative, a justification also used for cubic spline smoothing
(Reinsch 1967) and P-splines (Eilers and Marx 1996).

From a Bayesian perspective, penalizing second-order differences is
equivalent to assuming that the log-transformed target function follows
a locally affine (i.e., linear) trajectory. For mortality data, this
implies a prior belief in exponential growth of the mortality rate---a
widely accepted assumption consistent with the Gompertz model. For other
biometric risks, such as disability or long-term care, this prior is
also appropriate for the age dimension if included.

Penalizations of order \(q > 3\) are harder to justify in actuarial
practice. They may produce undesirable artefacts in the extrapolated
regions, where the extrapolation from the fit tends to follow a
polynomial of degree \(q - 1\).

\begin{figure}

\centering{

\includegraphics[width=0.8\linewidth,height=\textheight,keepaspectratio]{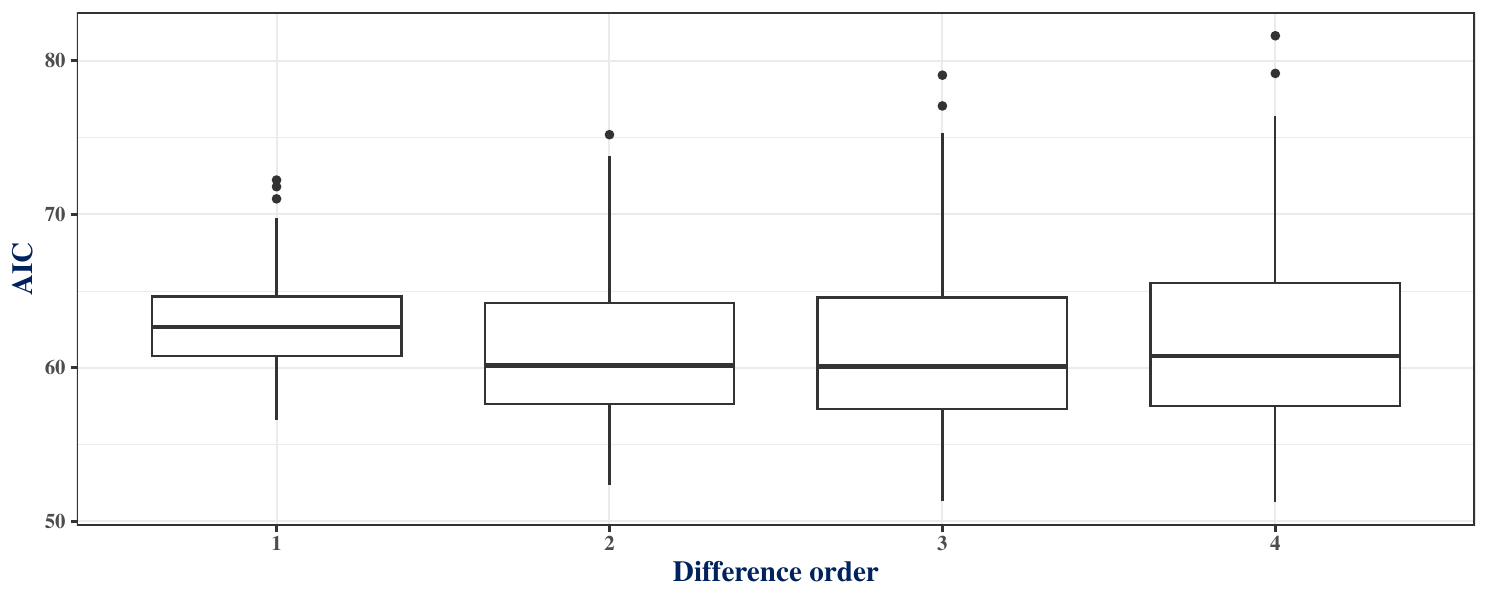}

}

\caption{\label{fig-order-1d}AIC values for different penalization
orders in WH smoothing, based on 100 replicates of the annuity portfolio
(100,000 individuals).}

\end{figure}

\begin{figure}

\centering{

\includegraphics[width=1\linewidth,height=\textheight,keepaspectratio]{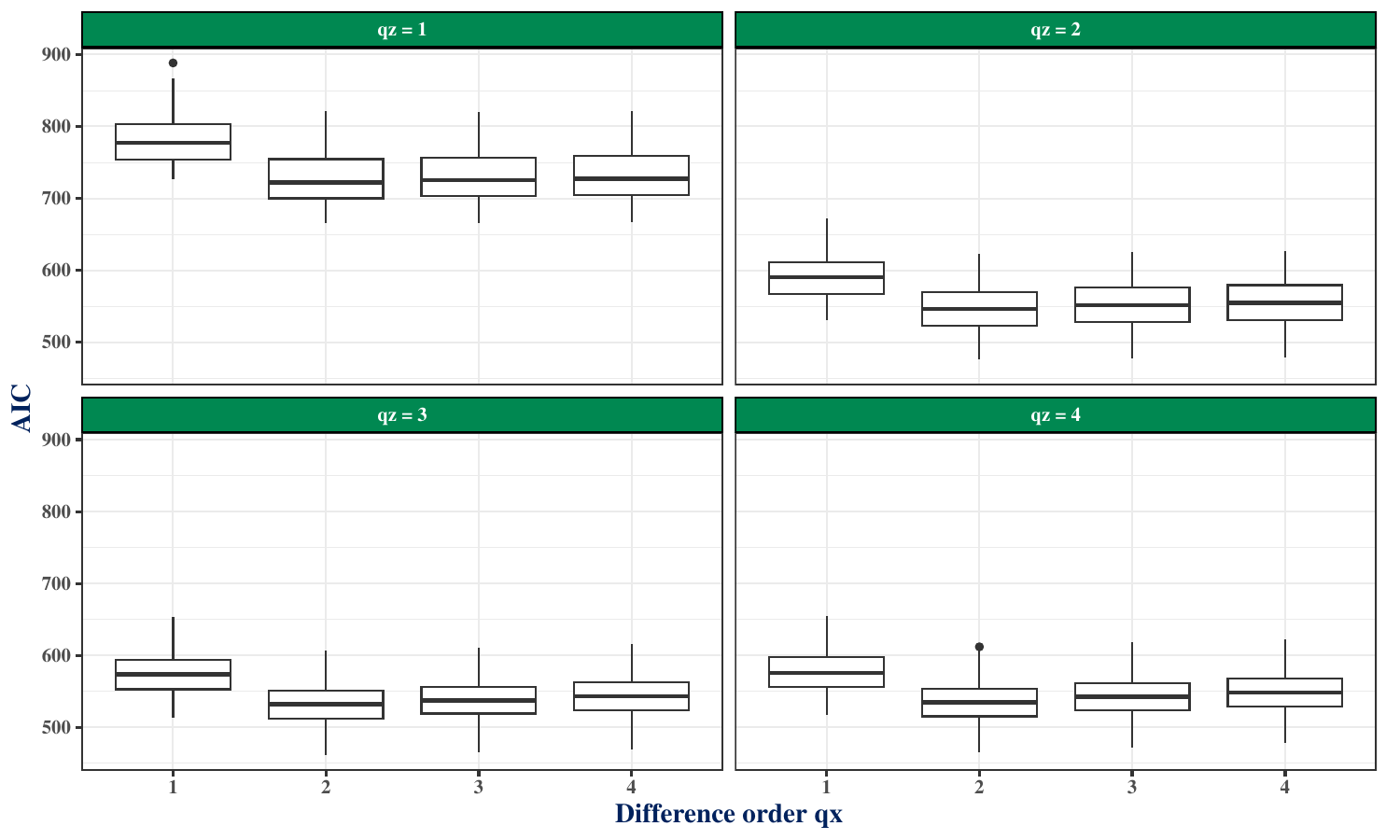}

}

\caption{\label{fig-order-2d}AIC values for combinations of difference
matrix orders along age and duration in WH smoothing, based on 100
replicates of the LTC portfolio (100,000 individuals).}

\end{figure}

It is possible to treat the choice of \(q\) as a model selection
problem. REML is not suitable in this context, as it assumes an equal
fixed effect structure, however standard criteria such as AIC and GCV
can be used instead. Relying on AIC and the 100 replicate datasets
described in Section 6, we found that:

\begin{itemize}
\item
  For the annuity datasets, Figure~\ref{fig-order-1d} shows that
  second-order differences (\(q = 2\)) provides the best results.
\item
  For the duration dimension in LTC datasets, Figure~\ref{fig-order-2d}
  shows that, regarding the age dimension, second-order differences
  (\(q = 2\)) is still the best option while for the duration dimension,
  AIC improves significantly when moving from first- to second-order
  penalization, and then slightly improves further for higher orders.
  This suggests that higher-order prior may be relevant for the duration
  dimension in this particular dataset.
\end{itemize}

Still, the gains from increasing the order beyond 2 are limited and may
not outweigh the risks of erratic extrapolation. Therefore, second-order
differences remain a pragmatic and robust default.


\begin{CSLReferences}{1}{0}
\bibitem[\citeproctext]{ref-akaike1973}
Akaike, Hirotsugu. 1973. {``Information Theory and an Extension of the
Maximum Likelihood Principle.''} In \emph{2nd International Symposium on
Information Theory, 1973}.

\bibitem[\citeproctext]{ref-anderssen1974time}
Anderssen, RS, and Peter Bloomfield. 1974. {``A Time Series Approach to
Numerical Differentiation.''} \emph{Technometrics} 16 (1): 69--75.

\bibitem[\citeproctext]{ref-biessy2015multistates}
Biessy, Guillaume. 2015. {``Long-Term Care Insurance: A Multi-State
Semi-{M}arkov Model to Describe the Dependency Process in Elderly
People.''} \emph{Bulletin Français d'Actuariat} 15(29): 41--73.

\bibitem[\citeproctext]{ref-biessy2016phd}
---------. 2016. {``Semi-Markov Modeling of the Loss of Autonomy Among
Elderly People: Application to Long-Term Care Insurance.''} PhD thesis,
Paris Saclay.

\bibitem[\citeproctext]{ref-bohlmann1899ausgleichungsproblem}
Bohlmann, Georg. 1899. {``Ein Ausgleichungsproblem.''} \emph{Nachrichten
von Der Gesellschaft Der Wissenschaften Zu G{ö}ttingen,
Mathematisch-Physikalische Klasse} 1899: 260--71.

\bibitem[\citeproctext]{ref-brent1973optimize}
Brent, Richard P. 1973. {``Algorithms for Minimization Without
Derivatives, Chap. 4.''} Prentice-Hall, Englewood Cliffs, NJ.

\bibitem[\citeproctext]{ref-brooks1988cross}
Brooks, RJ, M Stone, FY Chan, and LK Chan. 1988. {``Cross-Validatory
Graduation.''} \emph{Insurance: Mathematics and Economics} 7 (1):
59--66.

\bibitem[\citeproctext]{ref-cia2017mortality}
Canadian Institute of Actuaries. 2017. {``Final Report: Canadian
Mortality Improvement Model --- MI-2017.''} Canadian Institute of
Actuaries.
\url{https://www.cia-ica.ca/app/themes/wicket/custom/dl_file.php?fid=14926&p=35281}.

\bibitem[\citeproctext]{ref-carballo2021general}
Carballo, Alba, Maria Durban, Göran Kauermann, and Dae-Jin Lee. 2021.
{``A General Framework for Prediction in Penalized Regression.''}
\emph{Statistical Modelling} 21 (4): 293--312.

\bibitem[\citeproctext]{ref-carballo2021prediction}
Carballo, Alba, Maria Durban, and Dae-Jin Lee. 2021. {``Out-of-Sample
Prediction in Multidimensional p-Spline Models.''} \emph{Mathematics} 9
(15): 1761.

\bibitem[\citeproctext]{ref-clifford1977nonidentifiability}
Clifford, Peter. 1977. {``Nonidentifiability in Stochastic Models of
Illness and Death.''} \emph{Proceedings of the National Academy of
Sciences} 74 (4): 1338--40.

\bibitem[\citeproctext]{ref-cornea2017explicit}
Cornea-Madeira, Adriana. 2017. {``The Explicit Formula for the
Hodrick-Prescott Filter in a Finite Sample.''} \emph{Review of Economics
and Statistics} 99 (2): 314--18.

\bibitem[\citeproctext]{ref-currie2013smoothing}
Currie, Iain D. 2013. {``Smoothing Constrained Generalized Linear Models
with an Application to the Lee-Carter Model.''} \emph{Statistical
Modelling} 13 (1): 69--93.

\bibitem[\citeproctext]{ref-currie2004smoothing}
Currie, Iain D, Maria Durban, and Paul HC Eilers. 2004. {``Smoothing and
Forecasting Mortality Rates.''} \emph{Statistical Modelling} 4 (4):
279--98.

\bibitem[\citeproctext]{ref-currie2006glam}
---------. 2006. {``Generalized Linear Array Models with Applications to
Multidimensional Smoothing.''} \emph{Journal of the Royal Statistical
Society: Series B (Statistical Methodology)} 68 (2): 259--80.

\bibitem[\citeproctext]{ref-delwarde2007smoothing}
Delwarde, Antoine, Michel Denuit, and Paul Eilers. 2007. {``Smoothing
the Lee--Carter and Poisson Log-Bilinear Models for Mortality
Forecasting: A Penalized Log-Likelihood Approach.''} \emph{Statistical
Modelling} 7 (1): 29--48.

\bibitem[\citeproctext]{ref-demmler1975oscillation}
Demmler, A, and Christian Reinsch. 1975. {``Oscillation Matrices with
Spline Smoothing.''} \emph{Numerische Mathematik} 24 (5): 375--82.

\bibitem[\citeproctext]{ref-dempster1977EM}
Dempster, Arthur P, Nan M Laird, and Donald B Rubin. 1977. {``Maximum
Likelihood from Incomplete Data via the EM Algorithm.''} \emph{Journal
of the Royal Statistical Society: Series B (Methodological)} 39 (1):
1--22.

\bibitem[\citeproctext]{ref-eilersmarx1996}
Eilers, Paul H. C., and Brian D. Marx. 1996. {``Flexible Smoothing with
\(B\)-Splines and Penalties.''} \emph{Statistical Science} 11 (2):
89--102.

\bibitem[\citeproctext]{ref-fellner1986}
Fellner, William H. 1986. {``Robust Estimation of Variance
Components.''} \emph{Technometrics} 28 (1): 51--60.

\bibitem[\citeproctext]{ref-fix1951illnessdeath}
Fix, Evelyn, and Jerzy Neyman. 1951. {``A Simple Stochastic Model of
Recovery, Relapse, Death and Loss of Patients.''} \emph{Human Biology}
23 (3): 205--41.

\bibitem[\citeproctext]{ref-giesecke1981use}
Giesecke, Lee, and Defense Manpower Data Center. 1981. {``Use of the
Chi-Square Statistic to Set Whittaker-Henderson Smoothing
Coefficients.''} Smoothing.

\bibitem[\citeproctext]{ref-golub2013matrix}
Golub, Gene H, and Charles F Van Loan. 2013. \emph{Matrix Computations}.
JHU press.

\bibitem[\citeproctext]{ref-gschlossl2011risk}
Gschlössl, Susanne, Pascal Schoenmaekers, and Michel Denuit. 2011.
{``Risk Classification in Life Insurance: Methodology and Case Study.''}
\emph{European Actuarial Journal} 1: 23--41.

\bibitem[\citeproctext]{ref-gu1992gcv}
Gu, Chong. 1992. {``Cross-Validating Non-Gaussian Data.''} \emph{Journal
of Computational and Graphical Statistics} 1 (2): 169--79.

\bibitem[\citeproctext]{ref-gu2013smoothing}
---------. 2013. \emph{Smoothing Spline ANOVA Models}. Vol. 297.
Springer.

\bibitem[\citeproctext]{ref-hastie1990generalized}
Hastie, Trevor J, and Robert J Tibshirani. 1990. \emph{Generalized
Additive Models}. Vol. 43. CRC press.

\bibitem[\citeproctext]{ref-henderson1924new}
Henderson, Robert. 1924. {``A New Method of Graduation.''}
\emph{Transactions of the Actuarial Society of America} 25: 29--40.

\bibitem[\citeproctext]{ref-hodrick1997postwar}
Hodrick, Robert J, and Edward C Prescott. 1997. {``Postwar US Business
Cycles: An Empirical Investigation.''} \emph{Journal of Money, Credit,
and Banking}, 1--16.

\bibitem[\citeproctext]{ref-hoem1971point}
Hoem, Jan M. 1971. {``Point Estimation of Forces of Transition in
Demographic Models.''} \emph{Journal of the Royal Statistical Society:
Series B (Methodological)} 33 (2): 275--89.

\bibitem[\citeproctext]{ref-kauermann2005note}
Kauermann, Göran. 2005. {``A Note on Smoothing Parameter Selection for
Penalized Spline Smoothing.''} \emph{Journal of Statistical Planning and
Inference} 127 (1-2): 53--69.

\bibitem[\citeproctext]{ref-knorr1984multidimensional}
Knorr, Frank E. 1984. {``Multidimensional Whittaker-Henderson
Graduation.''} \emph{Transactions of Society of Actuaries} 36: 213--55.

\bibitem[\citeproctext]{ref-krivobokova2008fast}
Krivobokova, Tatyana, Ciprian M Crainiceanu, and Göran Kauermann. 2008.
{``Fast Adaptive Penalized Splines.''} \emph{Journal of Computational
and Graphical Statistics} 17 (1): 1--20.

\bibitem[\citeproctext]{ref-lee2011}
Lee, Dae-Jin, and Maria Durban. 2011. {``P-Spline ANOVA-Type Interaction
Models for Spatio-Temporal Smoothing.''} \emph{Statistical Modelling} 11
(1): 49--69.

\bibitem[\citeproctext]{ref-marra2012}
Marra, Giampiero, and Simon N Wood. 2012. {``Coverage Properties of
Confidence Intervals for Generalized Additive Model Components.''}
\emph{Scandinavian Journal of Statistics} 39 (1): 53--74.

\bibitem[\citeproctext]{ref-nelder1965optim}
Nelder, John Ashworth, and Roger Mead. 1965. {``A Simplex Method for
Function Minimization.''} \emph{The Computer Journal} 7 (4): 308--13.

\bibitem[\citeproctext]{ref-nelder1972glm}
Nelder, John Ashworth, and Robert WM Wedderburn. 1972. {``Generalized
Linear Models.''} \emph{Journal of the Royal Statistical Society: Series
A (General)} 135 (3): 370--84.

\bibitem[\citeproctext]{ref-patterson1971reml}
Patterson, H. D., and R. Thompson. 1971. {``Recovery of Inter-Block
Information When Block Sizes Are Unequal.''} \emph{Biometrika} 58:
545--54.

\bibitem[\citeproctext]{ref-R2025}
R Core Team. 2025. \emph{R: A Language and Environment for Statistical
Computing}. Vienna, Austria: R Foundation for Statistical Computing.
\url{https://www.R-project.org/}.

\bibitem[\citeproctext]{ref-reinsch1967smoothing}
Reinsch, Christian H. 1967. {``Smoothing by Spline Functions.''}
\emph{Numerische Mathematik} 10 (3): 177--83.

\bibitem[\citeproctext]{ref-reiss2009smoothing}
Reiss, Philip T, and R Todd Ogden. 2009. {``Smoothing Parameter
Selection for a Class of Semiparametric Linear Models.''} \emph{Journal
of the Royal Statistical Society: Series B (Statistical Methodology)} 71
(2): 505--23.

\bibitem[\citeproctext]{ref-rodriguez2015sap}
Rodriguez-Alvarez, Maria Xosé, Dae-Jin Lee, Thomas Kneib, Maria Durban,
and Paul Eilers. 2015. {``Fast Smoothing Parameter Separation in
Multidimensional Generalized p-Splines: The {SAP} Algorithm.''}
\emph{Statistics and Computing} 25 (5): 941--57.

\bibitem[\citeproctext]{ref-rodriguez2019estimation}
Rodríguez-Álvarez, María Xosé, Maria Durban, Dae-Jin Lee, and Paul HC
Eilers. 2019. {``On the Estimation of Variance Parameters in
Non-Standard Generalised Linear Mixed Models: Application to Penalised
Smoothing.''} \emph{Statistics and Computing} 29: 483--500.

\bibitem[\citeproctext]{ref-ruppert2000adaptive}
Ruppert, David, and Raymond J Carroll. 2000. {``Theory \& Methods:
Spatially-Adaptive Penalties for Spline Fitting.''} \emph{Australian \&
New Zealand Journal of Statistics} 42 (2): 205--23.

\bibitem[\citeproctext]{ref-schall1991estimation}
Schall, Robert. 1991. {``Estimation in Generalized Linear Models with
Random Effects.''} \emph{Biometrika} 78 (4): 719--27.

\bibitem[\citeproctext]{ref-soa2018practitioner}
Society of Actuaries. 2018. {``A Practitioner's Guide to Statistical
Mortality Graduation.''}
\url{https://www.soa.org/globalassets/assets/files/resources/tables-calcs-tools/2018-stat-mort-graduation.pdf}.

\bibitem[\citeproctext]{ref-taylor1992bayesian}
Taylor, Greg. 1992. {``A Bayesian Interpretation of
Whittaker---Henderson Graduation.''} \emph{Insurance: Mathematics and
Economics} 11 (1): 7--16.

\bibitem[\citeproctext]{ref-verrall1993state}
Verrall, RJ. 1993. {``A State Space Formulation of Whittaker Graduation,
with Extensions.''} \emph{Insurance: Mathematics and Economics} 13 (1):
7--14.

\bibitem[\citeproctext]{ref-wahba1980gcv}
Wahba, Grace. 1980. \emph{Spline Bases, Regularization, and Generalized
Cross Validation for Solving Approximation Problems with Large
Quantities of Noisy Data}. University of Wisconsin.

\bibitem[\citeproctext]{ref-wahba1985comparison}
---------. 1985. {``A Comparison of GCV and GML for Choosing the
Smoothing Parameter in the Generalized Spline Smoothing Problem.''}
\emph{The Annals of Statistics}, 1378--1402.

\bibitem[\citeproctext]{ref-weinert2007efficient}
Weinert, Howard L. 2007. {``Efficient Computation for
Whittaker--Henderson Smoothing.''} \emph{Computational Statistics \&
Data Analysis} 52 (2): 959--74.

\bibitem[\citeproctext]{ref-whittaker1923new}
Whittaker, Edmund Taylor. 1923. {``On a New Method of Graduation.''}
\emph{Proceedings of the Edinburgh Mathematical Society} 41: 63--75.

\bibitem[\citeproctext]{ref-wood2011reml}
Wood, Simon N. 2011. {``Fast Stable Restricted Maximum Likelihood and
Marginal Likelihood Estimation of Semiparametric Generalized Linear
Models.''} \emph{Journal of the Royal Statistical Society: Series B
(Statistical Methodology)} 73 (1): 3--36.

\bibitem[\citeproctext]{ref-wood2017generalized}
---------. 2017. \emph{Generalized Additive Models: An Introduction with
r}. 2nd ed. chapman; hall/CRC.

\bibitem[\citeproctext]{ref-wood2017fellner}
Wood, Simon N, and Matteo Fasiolo. 2017. {``A Generalized Fellner-Schall
Method for Smoothing Parameter Optimization with Application to Tweedie
Location, Scale and Shape Models.''} \emph{Biometrics} 73 (4): 1071--81.

\bibitem[\citeproctext]{ref-wood2017blacksmoke}
Wood, Simon N, Zheyuan Li, Gavin Shaddick, and Nicole H Augustin. 2017.
{``Generalized Additive Models for Gigadata: Modeling the UK Black Smoke
Network Daily Data.''} \emph{Journal of the American Statistical
Association} 112 (519): 1199--1210.

\end{CSLReferences}
\end{document}